Chemical Equilibrium Calculations for Bulk Silicate Earth Material at High Temperatures

Bruce Fegley, Jr.[1*], Katharina Lodders[1] and Nathan S. Jacobson[2]

[1]Planetary Chemistry Laboratory, Dept. of Earth & Planetary Sciences and McDonnell Center for the Space Sciences, Washington University, St Louis MO 63130, USA

[2]Materials and Structures Division, NASA Glenn Research Center/HX5, 21000 Brookpark Road, Cleveland, OH 44135 USA

*Corresponding author: Email: bfegley@wustl.edu

**Abstract**

The chemical equilibrium distribution of 69 elements between gas and melt is modeled for bulk silicate Earth (BSE) material over a wide P – T range (1000 – 4500 K, $10^{-6}$ – $10^{2}$ bar). The upper pressure end of this range may occur during lunar formation in the aftermath of a Giant Impact on the proto-Earth. The lower pressures may occur during evaporation from molten silicates on achondritic parent bodies. The virial equation of state shows silicate vapor behaves ideally in the P - T range studied. The BSE melt is modeled as a non-ideal solution and the effects of different activity coefficients and ideal solution are studied. The results presented are 50% condensation temperatures, major gas species of each element, and the pressure and temperature dependent oxygen fugacity ($fO_2$) of dry and wet BSE material. The dry BSE model has no water because it excludes hydrogen; it also excludes the volatile elements (C, N, F, Cl, Br, I, S, Se, Te). The wet BSE model has water because it includes hydrogen; it also includes the other volatiles. Some key conclusions include the following: (1) much higher condensation temperatures in silicate vapor than in solar composition gas at the same total pressure due to the higher metallicity and higher oxygen fugacity of silicate vapor (cf. Fegley et al. 2020), (2) a different condensation sequence in silicate vapor than in solar composition gas, (3) good agreement between different activity coefficient models except for the alkali elements, which show the largest differences between models, (4) agreement, where overlap exists, with prior published silicate vapor condensation calculations (e.g., Canup et al. 2015, Lock et al. 2018, Wang et al. 2019), (5) condensation of Re, Mo, W, Ru, Os oxides instead of metals over the entire P – T range, (6) a stability field for Ni-rich metal as reported by Lock et al. (2018), (7) agreement between ideal solution (from this work and from Lock et al. 2018) and real solution condensation temperatures for elements with minor deviations from ideality in the oxide melt, (8) similar 50% condensation temperatures, within a few degrees, in the dry and wet BSE models for the major elements Al, Ca, Fe, Mg, Si, and the minor elements Co, Cr, Li, Mn, Ti, V, and (9) much lower 50% condensation temperatures for elements such as B, Cu, K, Na, Pb, Rb, which form halide, hydroxide, sulfide, selenide, telluride and oxyhalide gases. The latter results are preliminary because the solubilities and activities of volatile elements in silicate melts are not well known, but must be considered for the correct equilibrium distribution, 50% condensation temperatures and mass balance of halide (F, Cl, Br, I), hydrogen, sulfur, selenium and tellurium bearing species between silicate melt and vapor.

## 1. Introduction

The chemistry of hot silicate vapor is important for several reasons yet is little known. For example, the ongoing debate about nebular versus planetary mechanisms responsible for meteoritic and terrestrial abundances of the moderately volatile elements (MVE) and Earth's volatility trend (e.g., Braukmüller et al. 2019, Fegley et al. 2020) dates back over 45 years to the MVE condensation calculations of Wai and Wasson (1977) and the discussion of lunar and terrestrial siderophile and volatile elemental abundances by Ringwood and Kesson (1977). Isotopic evidence for MVE depletion during planetary formation, and experimental and theoretical studies of major element and MVE evaporative loss from meteorite parent bodies, the Moon, and Mars after dissipation of the solar nebula are attracting increased attention (Canup et al. 2015, Lock et al. 2018, Jiang et al. 2019, Sossi et al. 2019, 2020, Ivanov et al. 2022, Neuman et al. 2022, Tian et al. 2021, Zhang et al. 2021).

Other topics of current interest are the chemistry of silicate vapor atmospheres of hot rocky exoplanets and the chemistry of impact-generated silicate melt – vapor debris disks in our solar system and in exoplanetary systems (Visscher and Fegley 2013, Lupu et al. 2014, Lock et al. 2018). Condensation and/or evaporation of elements from silicate melt – vapor is important in all these cases. Geochemists have been forced to use condensation temperatures from solar composition gas as a proxy for element volatility in silicate vapor (Lodders 2003). But there is no reason to expect the same condensation temperatures because hydrogen dominates solar composition gas while oxygen is the major element in silicate vapor. This work models the chemical equilibrium distribution of 69 elements between gas and melt for bulk silicate Earth material over a wide P – T range (1000 – 4500 K, $10^{-6}$ – $10^{2}$ bar) and gives condensation temperatures for the 69 elements studied and the major gas(es) of each element.

This paper is organized as follows:

Section 2 describes computational methods and data sources and is divided into subsections that give an overview of the computations, discuss the CONDOR and FactSage codes, describe the thermodynamic data sources for pure materials, explain details of molten oxide solution modeling in the CONDOR code, present background information on activities and activity coefficients, discuss sources for activity coefficients used in the calculations, and use the virial equation of state to check ideality of bulk silicate Earth vapor under the pressure and temperature conditions in the calculations.

Section 3 describes computational results starting with the condensation curve for molten BSE material over the $10^{-8}$ to $10^{2}$ bar range. Next is a discussion of condensation of the major, minor, and trace elements in BSE material and effects of variable activity coefficients upon condensation temperatures.

Section 4 discusses condensation temperatures in silicate vapor and solar composition material and differences between computed and actual condensation and evaporation temperatures due to evaporation (condensation) coefficients. Section 5 summarizes the work.

Preliminary results of this work were published earlier in two abstracts and one paper (Fegley and Lodders 2017, 2021, Fegley et al. 2020).

## 2. Computational Methods and Data Sources

**Overview**. Different sets of chemical equilibrium calculations for BSE material were done using the CONDOR code at Washington University and the FactSage code at the NASA Glenn Research Center. The FactSage code was also used to compute some of the activity coefficients used in the CONDOR code. The CONDOR code is described and used in several prior papers (e.g., Lodders and Fegley 1993, Fegley and Lodders 1994, Lodders 2003). CONDOR is a mass-balance, mass-action code that computes gas phase and gas – condensed phase chemical equilibrium as a function of temperature, pressure, elemental abundances, and thermodynamic data. CONDOR includes thermal ionization and non-ideal molten oxide solutions. Validation examples and references to other papers using the code are at the Planetary Chemistry Laboratory website (http://solarsystem.wustl.edu/research/condor/).

The CONDOR code was used to compute 50% condensation temperatures (i.e., half of an element in the gas and half in the melt), gas phase and gas – melt chemical equilibria from $10^{-6}$ to $10^{2}$ bar total pressure for 69 elements: Li, Na, K, Rb, Cs; Be, Mg, Ca, Sr, Ba; Sc, Y; La, Ce, Pr, Nd, Sm, Eu, Gd, Tb, Dy, Ho, Er, Tm, Yb, Lu; Ti, Zr, Hf; V, Nb, Ta; Cr, Mo, W; Mn, Re; Fe, Ru, Os; Co, Rh, Ir; Ni, Pd, Pt; Cu, Ag, Au; Zn, Cd, Hg; B, Al, Ga, In, Tl; Si, Ge, Sn, Pb; P, As, Sb, Bi; O; Th, U, Pu in a dry model of BSE composition based on Palme and O'Neill (2014) and Hudson et al (1989) for the $^{244}Pu/^{238}U$ ratio of 0.0068. Calculations were done with 1 K temperature steps and ≤0.5 log P pressure steps. The dry model excludes the volatile elements (H, C, N, F, Cl, Br, I, S, Se, Te), which will be included in a separate set of calculations called the wet model. This was done because the solubilities and activities of volatile elements in silicate melts are not well known, but must be considered because hydrogen, carbon, nitrogen, sulfur, selenium, tellurium and halogens are soluble (to varying extent) in silicate melts. Preliminary results of the wet model calculations for a few elements, whose chemistry is similar in the dry and wet models, are presented here. Elements that are not considered are the noble gases (He, Ne, Ar, Kr, Xe), and the following radioactive elements: Tc (43) Pm (61), Po (84), At (85), Rn (86), Fr (87), Ra (88), Ac (89), Pa (91), and Np (93). Plutonium (94) is considered because $^{244}Pu$ is a short-lived (82 Ma half-life) radionuclide present in the early solar system and it is important for Pu – I – Xe dating of meteorites and of the Earth – Moon system (Hudson et al. 1989, Avice and Marty 2014). Table 1 summarizes compositions of the dry and wet models.

The FactSage thermochemical modeling software package (Bale et al. 2002, 2009, 2016) was used to calculate thermodynamic activities and condensation temperatures in solution using the modified quasi-chemical model. The liquid/glass solution phase of the FToxid-Slag Database was used. This includes Al, As, B, Ca, Co, Cr, Cu, Fe, Ge, K, Li, Mg, Mn, Na, Ni, P, Pb, Si, Sn, Ti, Zn + dilute S and F (Miscibility gap at high $SiO_2$). The core system of this phase is $Al_2O_3$-CaO-FeO-$Fe_2O_3$-MgO-$SiO_2$ and "has been optimized from 25°C to above the liquidus temperatures at all compositions and oxygen partial pressures" (Bale et al. 2009). Numerous other oxides are included in this database, and it is optimized for industrial slags such as an iron blast furnace slag or a copper smelting slag. The database is extrapolated/interpolated to a separate set of conditions in this study. A comparison of compositions of industrial slags, molten BSE, and molten continental crust shows there is some overlap (Table 3 of Fegley et al. 2020). Compositionally the extrapolation/interpolation is reasonable. For example, Fegley et al. (2016) compared activities at 2000 K for $SiO_2$, MgO, CaO, $Al_2O_3$, and FeO computed from the MELTS and the FactSage codes for molten BSE and molten continental crust and found agreement, except for alumina where FactSage gave a($Al_2O_3$) about 100 times smaller than the value from MELTS, which does not include a $MgAl_2O_4$ liquid component. Extrapolations in temperature and pressure are more extreme in the present work. However, as discussed later, the agreement between activity coefficients from FactSage and other models suggests these extrapolations/interpolations are reasonable.

Thermodynamic activities of the melt components are part of the equilibrium solution parameters in FactSage. Model parameters are from many assessments of experimental data and standard solution models, as described in publications on the database (Bale et al. 2002, 2009, 2016). Condensation temperatures were calculated in the "Equilib" module. This is the FactSage free energy minimizer, which calculates the equilibrium products under a given set of conditions. Using the "Formation Target Phase" option, the module steps through temperature until the selected phase (Slag in our case) first appears.

The CONDOR and FactSage codes were used to calculate the condensation curve for the six most abundant oxides (thus the seven most abundant elements) in the BSE according to the compilation of Palme and O'Neill (2014). These six oxides are MgO, $SiO_2$, FeO, CaO, $Al_2O_3$, and $Na_2O$, which are the bold-face entries in Table 1. CONDOR code calculations using all elements included in the dry BSE model give the same condensation temperatures as the seven-element (six oxide) model. This occurs because the additional elements are less abundant than those in the seven-element model and do not significantly alter the oxygen fugacity. The results of the calculations and comparisons to prior work are in the Results and Discussion sections.

**Thermodynamic data sources for pure materials**. The calculations for dry BSE material use thermodynamic data for the elements, and their oxides (e.g., the Li-bearing gases in the dry model are Li, $Li^+$, $Li^-$, $Li_2$, $Li_2^+$, $Li_3^+$, LiNa, LiNaO, LiK, LiRb, $LiBO_2$, LiO, $Li_2O$, $Li_2O_2$ and the Li-bearing condensed phase in the code is $Li_2O$). The wet BSE model calculations also use thermodynamic data for any gases formed with the volatiles H, C, N, F, Cl, Br, I, S, Se, and Te (e.g., the additional Li-bearing gases in the wet model are LiH, $LiBeF_3$, LiN, LiNO, $LiNO_2$, $LiNO_3$, LiOH, $LiOH^+$, $Li_2(OH)_2$, $Li_2SO_4$, LiF, $Li_2F_2$, $Li_3F_3$, LiFO, LiCl, $Li_2Cl_2$, $Li_3Cl_3$, LiClO, $Li_2ClF$, LiBr, $Li_2Br_2$, $Li_3Br_3$, LiI, $Li_2I_2$, and $Li_3I_3$.

The data for the compounds included in the calculations are those tabulated in standard compilations, e.g., the NIST-JANAF Tables (Chase 1998), the US Geological Survey tables (Robie and Hemingway 1995), Schick (1966), and the FactSage and IVTAN code databases described elsewhere (Bale et al. 2002, 2009, 2016, Belov et al. 1999, Glushko et al. 1982, Gurvich et al. 1989-1996). The IVTAN thermodynamic database is the Russian counterpart of the NIST-JANAF Tables and is from the Joint Institute for High Temperatures of the Russian Academy of Sciences. It has the advantage of continual updates that include newer experimental studies that are not in the NIST-JANAF Tables, e.g., Aristova et al (2018) on solid and liquid uranium dioxide. The published tables give book length discussions of data evaluation methods with extensive literature citations (Gurvich et al. 1989-1996). Additional data sources are given in the text as needed when discussing particular compounds.

**Molten oxide solution modeling.** Magnesium exemplifies an element condensing as a single oxide into a non-ideal multicomponent oxide melt and copper exemplifies a multivalent element that can condense as two oxides into a non-ideal multicomponent oxide melt. Ideal gas chemical equilibria are used because as discussed later, the virial equation of state shows silicate vapor is ideal within ≤2% in the P – T range considered.

*Condensation as a single oxide*. The silicate vapor - melt equilibrium for magnesium is

$$Mg\ (gas) + ½\ O_2\ (gas) = MgO\ (melt) \quad (1)$$

The equilibrium constant $K_1$ for reaction (1) is

$$K_1 = \frac{a_{MgO}}{P_{Mg} P_{O_2}^{1/2}}$$

The partial pressure of gaseous Mg in equilibrium with the melt is $P_{Mg}$, $a_{MgO}$ is the thermodynamic activity of *liquid* MgO dissolved in the oxide melt, and $P_{O_2}^{1/2}$ is the square root of the $O_2$ gas partial pressure (not $P_O$, which is completely different). Hence the equilibrium partial pressure of Mg over the oxide melt is

$$P_{Mg} = \frac{a_{MgO}}{K_1 P_{O_2}^{1/2}}$$

Now consider how the equilibrium partial pressure of Mg over the melt relates to the total amount of magnesium in the system, denoted $S_{Mg}$ for sum magnesium. The total amount $S_{Mg}$ is comprised of magnesium in the gas phase (denoted $S_G$ for sum gas) and magnesium condensed as liquid MgO in the oxide melt (denoted $S_C$ for sum condensed)

$$S_{Mg} = S_C(Mg) + S_G(Mg)$$

The fraction of magnesium condensed in the melt is $\alpha_{Mg}$,

$$\alpha_{Mg} = \frac{S_C(Mg)}{S_{Mg}}$$

and the fraction of total magnesium in the gas is thus $(1 - \alpha_{Mg})$,

$$(1 - \alpha_{Mg}) = \frac{S_G(Mg)}{S_{Mg}}$$

Monatomic Mg and all other Mg-bearing gases, in decreasing order of abundance, in dry silicate vapor are in the $S_G$ term,

$$S_G(Mg) = P_{Mg} + P_{Mg^+} + P_{Mg^{2+}} + P_{MgO} + 2P_{Mg_2}$$

The mass balance sum is more complex in wet silicate vapor, e.g., at 3415 K (the 50% condensation temperature at one bar total pressure), the nine most abundant Mg-bearing gases in decreasing order of abundance are

$$S_G(Mg) = P_{Mg} + P_{MgO} + P_{MgOH} + P_{MgF} + P_{MgCl} + P_{Mg^+} + P_{MgH} + P_{MgS} + P_{Mg(OH)_2} + \cdots$$

The entire mass balance sum contains 28 gases, but Mg and MgO are the two major ones, and all other gases combined are <0.1% of total gaseous magnesium.

The variable $\beta_{Mg}$ is the fraction of total gaseous magnesium present as monatomic Mg (gas); this is

$$\beta_{Mg} = \frac{P_{Mg}}{S_G(Mg)}$$

Using $\beta_{Mg}$ to substitute into the expression for $K_1$ gives

$$K_1 = \frac{a_{MgO}}{P_{Mg}P_{O_2}^{1/2}} = \frac{a_{MgO}}{\beta_{Mg} \cdot S_G(Mg) \cdot P_{O_2}^{1/2}}$$

Using the definition of $(1 - \alpha_{Mg})$ – the fraction of total Mg in the gas – one can write

$$\left(1 - \alpha_{Mg}\right) = \frac{a_{MgO}}{K_1 \cdot \beta_{Mg} \cdot S_G(Mg) \cdot P_{O_2}^{1/2}} \frac{1}{S_{Mg}}$$

The code calculates gas chemistry for a given MgO activity and $O_2$ partial pressure, which means that $S_G$ (Mg) is computed. This is one of the iterative procedures in the code.

The MgO activity controls the amount of MgO condensed into the oxide melt because the MgO activity is the product of the MgO mole fraction (X) and activity coefficient ($\gamma$),

$$a_{MgO} = X_{MgO} \cdot \gamma_{MgO}$$

The activity coefficient is unity ($\gamma = 1$) for an ideal solution. Negative deviations from ideality mean $\gamma < 1$ and positive deviations from ideality mean $\gamma > 1$. Calculation of activity coefficients is described in another section below.

The MgO mole fraction is the condensed moles of MgO ($N_{MgO}$) divided by the sum of the moles of all other species (all oxides in this case) in the melt,

$$X_{MgO} = \frac{N_{MgO}}{N_{MgO} + N_{SiO_2} + N_{Al_2O_3} + N_{CaO} + N_{Na_2O} + \cdots}$$

This is equivalent to

$$X_{MgO} = \frac{\alpha_{Mg} S_{Mg}}{\alpha_{Mg} S_{Mg} + \sum v_{i \neq Mg}^{-1} \alpha_i S_i}$$

The summation term in the denominator includes all cationic elements (Al, Ca, Si, Fe, ...) other than Mg in the melt. The moles of the other cationic elements are expressed by their fraction condensed $\alpha_i$ times their respective total abundance $S_i$ with $\nu_i$ being the stoichiometric coefficient of the cation in the oxide formula, e.g., for $Al_2O_3$, $\nu_{Al} = 2$ for mass balance. The condition that all mole fractions sum to unity is another constraint used by the code, i.e.,

$$X_{MgO} + X_{SiO_2} + X_{Al_2O_3} + X_{Na_2O} + \cdots = 1$$

Rewriting the $X_{MgO}$ equation to explicitly include the MgO activity coefficient and to solve for $1/S_{Mg}$,

$$\frac{1}{S_{Mg}} = \frac{\alpha_{Mg}(1 - X_{MgO})}{X_{MgO} \sum v_{i \neq Mg}^{-1} \alpha_i S_i} = \frac{\gamma_{MgO}\, \alpha_{Mg}(1 - X_{MgO})}{a_{MgO} \sum v_{i \neq Mg}^{-1} \alpha_i S_i}$$

Substituting this equation into the prior one for $(1 - \alpha_{Mg})$, the fraction of Mg in the gas,

$$(1 - \alpha_{Mg}) = \frac{S_G(Mg)}{S_{Mg}} = \frac{\gamma_{MgO}\, \alpha_{Mg}(1 - X_{MgO})}{a_{MgO} \sum v_i^{-1} \alpha_i S_i} \cdot S_G(Mg)$$

This equation shows the connection between gas phase and melt chemistry and shows how the activity coefficient of MgO affects its chemistry. Going back to the mole fraction sum shows that the activity coefficient of any oxide also affects the chemistry of all other oxides

$$X_{MgO} + X_{SiO_2} + X_{Al_2O_3} + X_{Na_2O} + \cdots = 1$$

$$\frac{a_{MgO}}{\gamma_{MgO}} + \frac{a_{SiO_2}}{\gamma_{SiO_2}} + \frac{a_{Al_2O_3}}{\gamma_{Al_2O_3}} + \frac{a_{Na_2O}}{\gamma_{Na_2O}} \cdots = 1$$

All chemical equilibrium calculations must preserve mass balance; thus, the effects of varying the activity coefficient of a single melt constituent upon the chemistry of all other melt constituents is always apparent in *all* calculations of multicomponent melt chemistry. Experimental studies of the effects of $SiO_2$ and $Al_2O_3$, the two most abundant acidic oxides in melts, upon solubilities of alkali oxides $M_2O$ (M = Na, K, Rb) in the CaO − $Al_2O_3$ − $SiO_2$ (CAS) and the CAS − MgO (CMAS) systems by Borisov (2009) demonstrate this interdependence, and show

$$\left(\frac{\partial a_{M_2O}}{\partial X_{SiO_2}}\right)_{P,T} > \left(\frac{\partial a_{M_2O}}{\partial X_{Al_2O_3}}\right)_{P,T}$$

in ultramafic melts, but the reverse in felsic melts.

This interdependence is a consequence of the Gibbs − Duhem equation, which constrains the covariance of thermodynamic potentials, volume, entropy, activities, and activity coefficients, of the components in *any* solution (see chapter 34 of Pitzer and Brewer 1961)

$$X_1 dln\gamma_1 + X_2 dln\gamma_2 + X_3 dln\gamma_3 + \cdots = 0$$

This equation shows the activity coefficient of a species in a ternary or multicomponent solution can increase or decrease upon change of concentration of another species in the solution. Figures 4-7 of Borisov (2009), Figure 31 of Kubaschewski and Alcock (1979), and Figure 34-1 of Pitzer and Brewer (1961) illustrate this point. In practical terms the Gibbs − Duhem equation means that changing the activity coefficient for a single abundant species can affect the fraction condensed of other species with similar or smaller abundances. Conversely, changes in activity coefficients for trace species (e.g., CdO) have insignificant effects on condensation of more abundant oxides (e.g., MgO).

Rearranging the (1 − $\alpha_{Mg}$) equation again yields the "key" equation for the fraction of magnesium condensed,

$$\alpha_{Mg} = \frac{a_{MgO} \sum v_{i \neq Mg}^{-1} \alpha_i \, S_i}{\gamma_{MgO} \, S_G(Mg) \left(1 - X_{MgO}\right) + a_{MgO} \sum v_{i \neq Mg}^{-1} \alpha_i S_i}$$

Generalizing, the "key" equation for any metal (metalloid) M that dissolves in a multicomponent melt as a single constituent, e.g., Si as $SiO_2$, Al as $Al_2O_3$, Na as $Na_2O$, and so on,

$$\alpha_M = \frac{a_{M_{v(M)}O_x} \sum v_i^{-1} \alpha_i \, S_i}{\gamma_{M_{v(M)}O_x} \, S_{G(M)} \, \left(1 - X_{M_{v(M)}O_x}\right) + a_{M_{v(M)}O_x} \sum v_i^{-1} \alpha_i S_i} \; with \; (i \neq M)$$

The code solves this equation iteratively for each element considered in the system. The coupled gas chemistry and melt composition link the equations. The required inputs for the calculations are temperature, total pressure, elemental abundances of the system, and activity coefficients as well as the usual basic thermodynamic properties of the elements and their gaseous, liquid, and solid compounds.

If only tiny amounts of element M condense into a melt such that $X_{M_{v(M)}O_x} \ll 1$, then

$$\left(1 - X_{M_{v(M)}O_x}\right) \approx 1$$

And the expression for $\alpha_M$ simplifies to

$$\alpha_M \approx \frac{a_{M_{v(M)}O_x} \sum v_i^{-1} \alpha_i \, S_i}{\gamma_{M_{v(M)}O_x} \, S_{G(M)} + a_{M_{v(M)}O_x} \sum v_i^{-1} \alpha_i S_i} \; with \; (i \neq M)$$

This equation is a useful approximation for quick estimates of trace element condensation into a liquid host phase made of one or more major elements, e.g., a melt in the CMAS system made of CaO, MgO, $Al_2O_3$, and $SiO_2$; and a trace element such as zinc.

*Condensation as two oxides with different oxidation states*. Copper exemplifies multivalent elements that may exist in two different oxidation states with different activity coefficients in oxide melts (Paul 1990). These multivalent elements include Ti, V, Nb, Cr, Mo, W, Mn, Fe, Cu, In, Tl, Ge, Sn, Pb, As, Sb, Bi, Ce, Pr, Eu, U, and Pu. The basic variables are the sum of copper in the gas $S_G$(Cu), the sum of condensed copper $S_C$(Cu), and total copper $S_{Cu}$, which is the sum of $S_G$(Cu) plus $S_C$(Cu). In dry silicate vapor, $S_G$(Cu) has only six terms; in decreasing order of abundance at the 50% condensation temperature (3127 K at 1 bar total P),

$$S_G(Cu) = P_{Cu} + P_{CuO} + 2P_{Cu_2} + P_{Cu^-} + P_{Cu^+} + P_{Cu^{2+}}$$

However, total gaseous copper is almost 100% monatomic Cu with percent levels of CuO and trace amounts of all other Cu gases at temperatures where copper condenses into the oxide melt. The $S_G$(Cu) expression contains 18 species in wet silicate vapor because Cu-bearing halide, hydride, selenide, and sulfide gases are present. At the 50% condensation temperature of 2025 K (1 bar $P_{Total}$) the seven most abundant Cu gases (decreasing order of abundance) in $S_G$(Cu) are

$$S_G(Cu) = P_{Cu} + P_{CuS} + P_{CuH} + P_{CuCl} + P_{CuF} + P_{CuSe} + 2P_{Cu_2}$$

Monatomic Cu again dominates but is only 62% of $S_G$(Cu) with CuS (20%) and CuH (16%) making up most of the rest. Of course, the gas phase speciation varies with P and T and this distribution at 2025 K and 1 bar total P is only illustrative. Holzheid and Lodders (2001) found no evidence for sulfidic dissolution of copper, and it was not considered in the wet model calculations.

The sum of condensed copper is the number of moles (N) of liquid CuO plus liquid $Cu_2O$,

$$S_C(Cu) = N_C(Cu) = N_{CuO} + 2N_{Cu_2O}$$

The fraction condensed $\alpha_{Cu}$ of total copper is

$$\alpha_{Cu} = \frac{S_C(Cu)}{S_{Cu}} = \frac{N_C(Cu)}{S_{Cu}} = \frac{N_{CuO} + 2N_{Cu_2O}}{S_{Cu}}$$

And the fraction of total copper left in the gas is

$$(1 - \alpha_{Cu}) = \frac{S_G(Cu)}{S_{Cu}} = \frac{P_{Cu} + P_{CuO} + \cdots}{S_{Cu}} \cong \frac{a_{Cu_2O}}{K_{Cu} a_{CuO}} \frac{1}{S_{Cu}}$$

If Cu is the only important gas, then $S_G$(Cu) = $P_{Cu}$, otherwise other gases also must be considered as done above. The coexistence of CuO and $Cu_2O$ dissolved in the melt regulates $P_{Cu}$ via

CuO (melt) + Cu (gas) = $Cu_2O$ (melt)

$$K_{Cu} = \frac{a_{Cu_2O}}{a_{CuO}} \frac{1}{P_{Cu}}$$

The mole fractions of CuO and $Cu_2O$ in the melt are

$$X_{Cu_2O} = \frac{N_{Cu_2O}}{N_{Cu_2O} + N_{CuO} + N_{MgO} + N_{SiO_2} + N_{Al_2O_3} + \cdots} = \frac{N_{Cu_2O}}{\sum v_i^{-1} \alpha_i S_i}$$

$$X_{CuO} = \frac{N_{CuO}}{N_{Cu_2O} + N_{CuO} + N_{MgO} + N_{SiO_2} + N_{Al_2O_3} + \cdots} = \frac{N_{CuO}}{\sum v_i^{-1} \alpha_i S_i}$$

The summation term includes all melt constituents (including CuO and $Cu_2O$).

Because the activities of CuO and $Cu_2O$ are the product of their mole fractions and activity coefficients, one can rewrite the equation for $\alpha_{Cu}$ as follows,

$$\alpha_{Cu} S_{Cu} = 2N_{Cu_2O} + N_{CuO} = 2X_{Cu_2O} \sum v_i^{-1} \alpha_i S_i + X_{CuO} \sum v_i^{-1} \alpha_i S_i$$

Rearranging to solve for $(S_{Cu})^{-1}$,

$$\frac{1}{S_{Cu}} = \frac{\alpha_{Cu}}{\left(2X_{Cu_2O} + X_{CuO}\right) \sum v_i^{-1} \alpha_i S_i} = \frac{\alpha_{Cu}}{\left(2\frac{a_{Cu_2O}}{\gamma_{Cu_2O}} + \frac{a_{CuO}}{\gamma_{CuO}}\right) \sum v_i^{-1} \alpha_i S_i}$$

Inserting this into the equation for $(1 - \alpha_{Cu})$ and rearranging gives,

$$\alpha_{Cu} = \frac{\left(2a_{Cu_2O}\gamma_{Cu_2O} + a_{CuO}\gamma_{CuO}\right) \sum v_i^{-1} \alpha_i S_i}{\gamma_{Cu_2O}\gamma_{CuO} S_{G(Cu)} + \left(2a_{Cu_2O}\gamma_{CuO} + a_{CuO}\gamma_{Cu_2O}\right) \cdot \sum v_i^{-1} \alpha_i S_i}$$

The generalized equation for an oxide pair of metal (metalloid) M with formulas $M_{vm1}O_{vo1}$ and $M_{vm2}O_{vo2}$ and an exchange equilibrium between gas and melt analogous to that for copper is

$$v_{o1}\, M_{vm2}O_{vo2}(melt) + uM(gas) = v_{o2} M_{vm1} O_{vo1}(melt)$$

$$u = (v_{m1} v_{o2} - v_{m2} v_{o1})$$

The generalized version of the $\alpha_{Cu}$ equation is

$$\alpha_M = \frac{\left(v_{m1} a_{M_{vm1}O_{vo1}} \gamma_{M_{vm1}O_{vo1}} + a_{M_{vm2}O_{vo2}} \gamma_{M_{vm2}O_{vo2}}\right) \sum v_i^{-1} \alpha_i S_i}{\gamma_{M_{vm1}O_{so1}} \gamma_{M_{vm2}O_{vo2}} S_{G(M)} + \left(v_{m1} a_{M_{vm1}O_{vo1}} \gamma_{M_{vm2}O_{vo2}} + a_{M_{vm2}O_{vo2}} \gamma_{M_{vm1}O_{vo1}}\right) \sum v_i^{-1} \alpha_i S_i}$$

In principle, elements with two or more valence states can form various oxide pairs depending on oxygen fugacity and temperature. For example, Mn can form oxides in five oxidation states (2+, 3+, 4+, 6+, and 7+). However, there is a maximum of two stable coexisting oxides with different valence states that can simultaneously dissolve in the melt, such as $Ce_2O_3/CeO_2$, $CrO/Cr_2O_3$, $CuO/Cu_2O$, $EuO/Eu_2O_3$, $FeO/Fe_2O_3$, TiO/ $Ti_2O_3$ or $Ti_2O_3$/ $TiO_2$. If an element has more than two valence states, the oxide pairs with the highest thermochemical activity are usually the ones that dissolve; these oxides always have neighboring valence states (e.g., 3+ and 4+ but not 3+ and 5+ if 4+ exists). As a corollary, the oxide pair dissolving will also be the one that gives the maximum amount of element condensed. The activities of these oxides can be calculated (provided reliable thermodynamic data exists), but there are no tabulated data for some oxides such as GeO (s,liq) and data for other liquid oxides with high melting points are estimates of uncertain reliability such as CuO (liq). The CONDOR code computes activities of the following liquid oxides of multivalent elements: $CeO_2$, $Ce_2O_3$; CuO, $Cu_2O$; EuO, $Eu_2O_3$; FeO, $Fe_2O_3$, $Fe_3O_4$; MnO, $Mn_3O_4$; $MoO_2$, $MoO_3$; NbO, $NbO_2$, $Nb_2O_5$; SnO, $SnO_2$; TiO, $Ti_2O_3$, $Ti_3O_5$, $Ti_4O_7$, $TiO_2$; $Tl_2O$, $Tl_2O_3$; and VO, $V_2O_3$, $V_2O_4$, $V_2O_5$. These calculations are discussed later.

**Activities and activity coefficients**. The Raoultian activity ($a_i$) of a solid or liquid evaporating to an ideal gas is the ratio of its partial pressure ($P_i$) divided by its saturated vapor pressure ($P_i°$) at the same temperature,

$$a_i = \frac{P_i}{P_i^o}$$

The activity of a pure solid or liquid is unity, while the activity of a species in a solid or liquid solution is proportional to its concentration in the solution. The proportionality constant is the activity coefficient ($\gamma_i$) which is the ratio of activity ($a_i$) divided by concentration ($X_i$)

$$\gamma_i = \frac{a_i}{X_i}$$

An ideal solution has $\gamma = 1$ and non-ideal solutions have $\gamma < 1$ (negative deviations from ideality) or $\gamma > 1$ (positive deviations from ideality). Compound formation between two oxides, such as $MgSiO_3$ in the MgO - $SiO_2$ system, requires $\gamma < 1$ while unmixing into two liquids such as in the NaCl - albite system (Koster van Groos and Wyllie 1969) requires $\gamma > 1$. Activity coefficients < 1 give higher condensation temperatures than for ideal solutions. Activity coefficients > 1 give lower condensation temperatures than for ideal solutions.

**Temperature dependence and extrapolation of activity coefficients.** Activity coefficients trend toward unity (ideality) with increasing temperature and the general equation describing the temperature dependence of activity coefficients (Gaskell 1973) is

$$\left[\frac{\partial ln\gamma_i}{\partial(1/T)}\right]_{P,n} = \frac{\overline{H}_i^E}{R}$$

The subscripts P and n (the number of moles) denote differentiation at constant pressure and constant composition and the enthalpy term is the *excess* partial molal enthalpy of mixing for any species in a solution. Converting to base 10 logs and rearranging gives an equation like that below for CaO,

$$log_{10}\gamma_{CaO}(MAGMA) = -0.144 - \frac{4656.5}{T}$$

The temperature dependence of activity coefficients for a regular solution is (Charles 1967),

$$\left[\frac{\partial ln\gamma_i}{\partial(1/T)}\right]_{P,n} = \frac{\overline{H}_i^E}{R} = Tln\gamma_{i,T} = T \cdot 2.303 \cdot log\gamma_{i,T}$$

In this case the intercept term is zero and the temperature dependence is an equation like

$$log_{10}\gamma_{MnO}(BanYa) = \frac{672.096}{T}$$

Table 2 lists these two equations and all other activity coefficient equations used in this work.

**Oxide activity coefficients for bulk silicate Earth melts.** There are no experimental measurements of oxide activity coefficients in molten BSE material, i.e., peridotite, at super-liquidus temperatures. This is not surprising given the complex composition and high liquidus temperature of dry peridotite at one bar pressure (~2000 K, Takahashi 1986).

The CONDOR code calculations use activity coefficients from six primary sources: (1) Ban-Ya's (1993) implementation of Lumsden's (1961) quadratic solution model for molten oxides, (2) the FactSage code, a modified quasi-chemical model, (3) the MAGMA code, which is an associated solution model (Fegley and Cameron 1987, Schaefer and Fegley 2004), (4) regular solution extrapolation of a single activity coefficient value to higher temperature, e.g., see the discussions of Ag, Cd, Cu, Ge, and In, (5) the MELTS code (Ghiorso and Sack 1995, Ghiorso et al. 2002) used for K, (6) consideration of optical basicity and ionic potential in periodic table groups, e.g., see the later discussions of Li, Rb, Cs, Mo, W, Ag, Cd, Cu, Ga, and Tl. Table 2 lists coefficients for the two-term equations giving $log_{10}$ values of activity coefficients used in the CONDOR code. As Table 2 shows, sources (1), (2), and (4) supplied most of the nominal activity coefficient values. The Lumsden – Ban-Ya quadratic solution model is described next. Details about individual activity coefficient estimates and the dependence of 50% condensation temperatures upon activity coefficients are in the Results section.

**Molten oxide quadratic solution mode**l. Lumsden (1961) developed a regular solution model "to calculate the thermodynamic properties of complex molten slags from those of the component binary systems. The essential features of this problem can be demonstrated by a ternary system." The two key points of his model are (1) a single interaction parameter for each cation pair and (2) regular mixing of cations (e.g., $Fe^{2+}$, $Fe^{3+}$, $Si^{4+}$) in an oxygen ion ($O^{2-}$) matrix. This is a variant of Temkin's ionic mixing model, which postulates ideal mixing of cations with other cations and ideal mixing of anions with other anions in fused salts and metallurgical slags (Temkin 1945, Samarin et al. 1945). The two equations that define the activity coefficient for a cation in the regular solution model are

$$\bar{G}_i^E = \bar{H}_i^E = RTln\gamma_i$$

$$RTln\gamma_i = \sum_j \alpha_{ij} N_j^2 + \sum_j \sum_k (\alpha_{ij} + \alpha_{ik} - \alpha_{jk}) N_j N_k$$

The excess partial molal Gibbs energy, and excess partial molal enthalpy of component "i" are $\bar{G}_i^E$ and $\bar{H}_i^E$, respectively, the interaction energy between cations "i" and "j" is $\alpha_{ij}$, and $N_i$ is the cation mole fraction, i.e., $N_{SiO_2}, N_{NaO_{0.5}}$, and so on. Although quadratic regular solution models neglect ternary and other higher order terms (Grover 1977, Berman 1983, Berman and Brown 1984) they are often used, e.g., see Ghiorso et al. (1983), Ghiorso and Sack (1995).

Lumsden (1961) demonstrated the regular solution model fit data for $Pb^{2+}O^{2-} – Bi_2^{3+}O_3^{2-}$ melts and applied it to $FeO – Fe_2O_3 – SiO_2$ slags. Later work by other metallurgists applied Lumsden's model to multicomponent melts made of various combinations of $Al_2O_3$, CaO, $Cu_2O$, FeO, $Fe_2O_3$, MgO, MnO, $Na_2O$, $SiO_2$, $TiO_2$, and ZnO (e.g., Richards and Thorne 1961, Somerville et al. 1973, Martin and Bell 1974, Altman 1978, Ban-Ya and Shim 1982, Ban-Ya et al. 1984, Nagabayashi et al. 1989, Ban-Ya 1993). These papers show agreement between experimental data and quadratic solution modeling. The CONDOR calculations use the interaction energies and temperature-dependent equations in Tables 1-3 of Ban-Ya (1993) to compute activity coefficients for $Al_2O_3$, CaO, FeO, MgO, MnO, $Na_2O$, $P_2O_5$, $SiO_2$, and $TiO_2$ in BSE melt.

**Ideality of BSE vapor**. An important question is whether silicate vapor is ideal or non-ideal under the pressure (P) and temperature (T) conditions used in the calculations. Corrections for intermolecular forces between neutral atoms and molecules in a gas become significant as the gas density approaches that of the liquid and the ideal gas equation no longer applies. The compression factor Z quantifies the non-ideality and is the ratio of the $PV_m$ product or the $P/\rho_m$ quotient (where $V_m$ = molar volume, and $\rho_m = 1/V_m$ = molar density) to RT:

$$\frac{PV_m}{RT} = \frac{1}{RT} \cdot \frac{P}{\rho_m} = Z = 1 + \frac{B(T)}{V_m} + \frac{C(T)}{V_m^2} + \frac{D(T)}{V_m^3} + \cdots$$

The compression factor Z is unity for an ideal gas (i.e., with attractive and repulsive forces in balance), is greater than unity for an incompressible real gas (repulsive forces are dominant) and is less than unity for a highly compressible real gas (attractive forces are dominant).

The equation above is the virial equation of state (EOS), which has a sound theoretical foundation (Hirschfelder et al. 1964, Mason and Spurling 1969). The B(T), C(T), and D(T) in the virial EOS are the second, third and fourth virial coefficients (henceforth virials) respectively. The first virial is unity and is the ideal gas term. The second virial B arises from interactions between two gas molecules. The third virial C arises from interactions between three gas molecules, and the fourth virial D from interactions of four gas molecules and so on. The virial coefficients for reactive atoms and radicals such as atomic Fe, O, Na, and SiO can be computed from their interatomic (or intermolecular) potentials. Thus, one can rigorously compute fugacity coefficients for these reactive species – something that is not possible with empirical EOS such as the van der Waals or Redlich – Kwong equations.

The second virial is easily, the third virial less easily and the fourth virial with difficulty computed from intermolecular potentials. A common potential is the Lennard-Jones (12-6) potential energy function φ(r) with two force constants σ and ε/k,

$$\varphi(r) = 4\epsilon\left[\left(\frac{\sigma}{r}\right)^{12} - \left(\frac{\sigma}{r}\right)^{6}\right]$$

The distance between two gas molecules (or atoms) is r, σ is the distance at which $\varphi(r) = 0$, ε is the maximum depth of the potential energy well, and ε/k is epsilon divided by Boltzmann's constant and has units of temperature. Hirschfelder et al. (1964) gives equations and tabular data for computing B(T) and C(T) for the Lennard-Jones (12-6) potential. These were used to get the numerical values of B(T) in Table 3. The maximum T and P in the calculations are 5000 K, 100 bar corresponding to $\rho_m \simeq 2.4\times10^{-4}$ mol cm$^{-3}$ (about six times denser than air at room temperature). This low density allows simplification of the virial EOS to the equation below.

$$\frac{PV_m}{RT} = 1 + \frac{B(T)}{V_m}$$

The corresponding equation below gives fugacity coefficients ϕ (see Chapter 16 of Pitzer and Brewer 1961).

$$ln\left(\frac{f}{P}\right) = ln\phi = \frac{B}{RT}P$$

Table 3 gives values for the second virials and fugacity coefficients for several major gases in BSE vapor at 5000 K and 100 bar total pressure. The least ideal gases in Table 3 are sodium and silicon dioxide which are more compressible than an ideal gas (2% and 1% respectively). The other gases for which second virials are known have fugacity coefficients closer to unity. At 1500 K and 100 bar total pressure $O_2$ is the major gas with much smaller amounts of metallic gases. Under these conditions $B(O_2)$ = 25.00 cm$^3$ mol$^{-1}$ and the $O_2$ fugacity coefficient ϕ of 1.02 is 2% less compressible than an ideal gas. Monatomic Hg vapor (0.15%) is the second most abundant gas with *B*(Hg) = −26.15 cm$^3$ mol$^{-1}$. The mercury fugacity coefficient of ϕ = 0.98 is 2% more compressible than an ideal gas. These small deviations from ideality are neglected in the chemical equilibrium calculations.

Many gases in silicate vapor have negative second virials and display negative deviations from ideality. This behavior follows directly from the characteristic temperature dependence of the second virial coefficient *B* (Mason and Spurling 1969). Values of the second virial *B* are negative below the Boyle temperature, zero at the Boyle temperature, and positive above the Boyle temperature of a gas. The Boyle temperature is defined by the equation below.

$$\left(\frac{\partial Z}{\partial \rho}\right)_T = 0\ as\ \rho \to 0$$

The Boyle temperature, $T_B$, is greater than the critical temperature, $T_{Crit}$. Oxygen is a major gas in silicate vapor, and it has $T_B$ = 406 K, $T_{Crit}$ = 154.58 K, and $T_B/T_{crit}$ = 2.63. This is the same as the $T_B/T_{crit}$ ratio for a Lennard-Jones fluid (Apfelbaum and Vorob'ev 2009). The critical temperatures are even higher for other species in silicate vapor; 1763±15 K for Hg (Franck and Hensel 1966), and 2497±15 K for Na (Ohse et al. 1985). As mentioned earlier (Fegley et al. 2020), "the critical temperatures of silica, iron, and silicates are unknown. Estimated critical temperatures for silica range from 4,700 to 13,500 K (Melosh 2007)." More recent estimates of critical temperatures for silica and other minerals are 5200 - 6200 K for silica $Si_{1.1}O_2$ - $Si_{1.4}O_2$ (Connolly 2016), 5500 - 6000 K for $NaAlSi_3O_8$ and 5000 - 5500 K for $KAlSi_3O_8$ (Kobsch and Caracas 2020), 6240±200 K for $Mg_2SiO_4$ from ab initio calculations (Townsend et al 2020), 7928 K for elemental Fe using a corresponding states method (Apfelbaum and Vorob'ev 2015), and 9250±700 K for Fe from rapid pulse heating experiments Pottlacher and Jäger (1996). Using the $T_C/T_B$ ratio of 0.38 for a Lennard-Jones fluid (Apfelbaum and Vorob'ev 2009), the $T_C$ estimates suggest $T_B$ values > 10,000 K for constituents of silicate vapor and thus negative deviations from ideality.

## 3. Results

**BSE melt condensation curve**. The condensation curve is the highest temperature at which melt is stable and is the boundary between unsaturated silicate vapor and condensate plus vapor. The BSE melt condensation curve was calculated (using the six-oxide model for the bulk silicate Earth, boldface entries in Table 1) with the CONDOR code at Washington University and with the FactSage code at NASA Glenn. The six-oxide model makes up more than 99% of the BSE and is the same model used in Fegley et al. (2016). As stated earlier, CONDOR code calculations using all elements in the dry BSE model give the same condensation temperatures for BSE melt as the six-oxide model. In particular, the absence of $TiO_2$ (0.14% molar abundance in the BSE) and other less abundant refractory oxides found in Ca, Al-rich inclusions in meteorites does not change the results. The $TiO_2$ content of the initially condensed BSE melt is small because $TiO_2$ is more volatile than either $Al_2O_3$ or CaO in BSE melt as discussed later.

Figure 1 displays the condensation curve for molten bulk silicate Earth (BSE) material from $10^{-8}$ to $10^2$ bar and compares the results from the CONDOR and FactSage codes with results from the GRAINS code (Lock et al. 2018). Table 4 lists the condensation temperatures plotted in Figure 1. Linear least squares equations for computing total pressure $P_T$ over the melt as a function of temperature [f(T)] or the converse T=f($P_T$) along the (dry) molten BSE condensation curve from $10^{-8}$ to $10^2$ bar are

$$logP_T = 8.564 - \frac{31415.9}{T}$$

$$\frac{10{,}000}{T} = 2.726 - 0.31823logP_T$$

The corresponding equations for the wet molten BSE condensation curve are

$$logP_T = 8.638 - \frac{31503.7}{T}$$

$$\frac{10,000}{T} = 2.742 - 0.31727logP_T$$

The results from the CONDOR and FactSage codes agree within 38 K up to one bar with increasing disagreement at higher pressures reaching 278 K (6%) at 100 bar. FactSage gives higher condensation temperatures than CONDOR at all pressures. Both codes used the same model bulk composition and both codes are well validated as described earlier. The differences in the results are due to the large extrapolation in temperatures and pressures from one bar and temperatures <2000 K where the solution parameters are optimized for use.

Table 4 and Figure 1 also compare CONDOR and FactSage code results to those from the GRAINS code (Lock et al. 2018), which is a Gibbs energy minimization code developed by Petaev (2009). Petaev (personal communication, 2021) kindly supplied us with tabular data for the 20 element ideal solution calculations discussed in Lock et al. (2018) and earlier in several abstracts at the Lunar and Planetary Science Conference (Petaev et al. 2014, 2015, 2016). The agreement with CONDOR and FactSage results is good overall, except at the highest temperatures and pressures studied, e.g., differences are 313 K (6%) at 100 bar between CONDOR and GRAINS. Again, the differences are due to large extrapolations in thermodynamic data.

The GRAINS code calculations include He and the volatiles H, C, N, S, and Cl that are only in the wet BSE model. Table 4 shows the wet BSE model agrees even better with the GRAINS code results at all pressures. The wet BSE model gives lower condensation temperatures at the same total pressure because the O and $O_2$ partial pressures at a given temperature are lowered by H, OH, CO, $CO_2$, $H_2O$, $H_2$, NO, $S_2$, SO, and other volatile-bearing gases. As a result, reactions of Al and Ca with O or $O_2$ forming $Al_2O_3$ and CaO in the molten oxide only reach equilibrium at a lower temperature where the $fO_2$ of wet BSE material is the same as the $fO_2$ of dry BSE material. The bottom row in Table 4 shows the $fO_2$ values along the dry and wet BSE condensation curves, which we found equal within ≤5% at the same total pressure.

The major conclusion from *all* results in Figure 1 is the extremely high temperatures for silicate melt stability in a system with silicate bulk composition in the absence (or with only very small amounts) of hydrogen. At a given pressure and temperature the oxygen fugacity and rocky element abundances (i.e., the metallicity) are significantly higher in a silicate bulk composition system than in solar composition, i.e., $H_2$ and He dominated, material.

Although the CONDOR, FactSage, and GRAINS codes use different activity coefficient models derived from data at <2000 K and have different algorithms, BSE melt condensation temperatures from all three codes agree (within the small differences mentioned above). The highest temperatures involve the largest extrapolations in activity coefficient models and larger differences between the three models are not surprising. Nevertheless, the fact that BSE melt condensation temperatures from the three different codes agree within ≤10% over eight orders of magnitude in pressure and a factor of 2.4 in temperature is noteworthy given the large extrapolations involved in the different activity coefficient models and thermodynamic data.

This good agreement between condensation temperatures exists because the $RTln\gamma_i$ term and differences between $RTln\gamma_i$ terms from different activity coefficient models are smaller than the enthalpy (ΔH°) and entropy (ΔS°) of condensation. We illustrate this using the condensation of CaO, one of the two most refractory oxides in the initially condensed BSE melt:

$$\mathrm{Ca(gas)}+\frac{1}{2}\mathrm{O_2(gas)}=\mathrm{CaO(melt)}$$

The reaction isotherm gives the Gibbs energy change for CaO condensation

$$\Delta G_{CaO} = \Delta G°_{CaO} + RTln\frac{a_{CaO}}{P_{Ca}f_{O_2}^{1/2}}$$

At equilibrium between the gas and melt, ΔG = 0

$$\Delta G_{CaO} = 0 = \Delta G°_{CaO} + RTln\frac{a_{CaO}}{P_{Ca}f_{O_2}^{1/2}}$$

Rewriting the reaction isotherm gives

$$\Delta G°_{CaO} = -RTln\frac{a_{CaO}}{P_{Ca}f_{O_2}^{1/2}} = -RTlnK_{eq}$$

This equation shows the relationship between the standard Gibbs energy change, which is at one bar pressure, and the reaction quotient, which is equal to the equilibrium constant $K_{eq}$.

The standard Gibbs energy change is also given by the Gibbs - Helmholtz equation

$$\Delta G°_{CaO} = \Delta H°_{CaO} - T\Delta S°_{CaO}$$

Rewriting the reaction quotient by substituting for activity with the equation a = γX and using the definition of the partial molal ideal entropy of mixing,

$$\overline{\Delta S}_i^M = -RlnX_i$$

After substituting back into the reaction isotherm and grouping terms,

$$[\Delta H°_{CaO} + RTln\gamma_{CaO}] - T[\Delta S°_{CaO} + \overline{\Delta S}_{CaO}^M] = RT\left(lnP_{Ca} + \frac{1}{2}lnf_{O_2}\right)$$

Over the 2000 – 4500 K range, JANAF (or IVTAN) data show typical ΔH° values for the CaO condensation reaction are −710 to −670 kJ $mol^{-1}$ versus typical $RTln(\gamma_{CaO})$ values of −145±5 kJ $mol^{-1}$, about 17% of the [enthalpy term]. Typical ΔS° values are −170 to −155 J $mol^{-1}$ $K^{-1}$ (2000 – 4500 K) versus typical $RTlnX_{CaO}$ values of 29 J $mol^{-1}$ $K^{-1}$, about 17% of the [entropy term]. The energetic differences between two different activity coefficient models are smaller than these between the Ban-Ya regular solution model and ideal solution. Thus, the CaO condensation temperatures are closer as seen in Figure 1. Table 4 and Figure 1 show the net effect of these differences; the CONDOR and FactSage calculations used two different non-ideal melt models while that of Lock et al. (2018) used an ideal melt model. All three sets of calculations give the same result to first order because the ΔH° and ΔS° for condensation of pure liquid oxides are larger than the ΔH° and ΔS° terms due to formation of non-ideal oxide melts. Ebel (2005) and Ebel and Grossman (2005) reached a similar conclusion in their calculations.

**Melt composition**. The composition of the initially condensed melt varies with total pressure as shown in Figure 2. The melt is mainly $Al_2O_3$ (more abundant) and CaO (less abundant) at lower total pressures with increasing amounts of CaO, MgO and other oxides and decreasing amounts of $Al_2O_3$ and CaO with increasing pressure. As MgO becomes the major oxide in the melt, the amounts of $Al_2O_3$ and CaO are diluted by the more abundant MgO. Table 5 compares melt compositions at $10^{-3}$, 10, and 100 bars calculated by the CONDOR and FactSage codes. The FactSage code shows the same general trend with total pressure, but the overall agreement is only fair. However, the two codes show the same key point; namely the initially condensed melt at any total pressure is more refractory with less $SiO_2$, FeO, and $Na_2O$ than the BSE composition in Table 1.

**Condensation chemistry of major, minor, and trace elements in dry BSE material.**

**Oxygen**. Oxygen, the most abundant element in BSE material, is considered first. Figures 3a (dry BSE model) and 3b (wet BSE model) show logarithmic contours of constant oxygen fugacity ($fO_2$) over the 1500 – 4500 K and $10^{-6}$ to $10^2$ bar total pressure range. The thicker dark green curves in Figures 3a and 3b show 50%, 99%, and 99.99% (only in Figure 3a) condensation of total oxygen into the BSE melt, which causes the curvature and fine structure of the log $fO_2$ contours. Grid calculations with ΔT = 4 K and $\Delta(\log_{10}P)$ = 0.10 or 0.16 gave the same fine structure and took several weeks computer runtime because of the complexity of the calculations. The most abundant oxides in BSE material (MgO, $SiO_2$, FeO) condense into the BSE melt within a small temperature range in the vicinity of the oxygen 50% condensation curve and the amount of oxygen left in the gas decreases markedly. The CONDOR chemical equilibrium calculations give the equilibrium $fO_2$ and the fraction of oxygen condensed into the melt.

A comparison of the shape and range of the log $fO_2$ contours for the dry and wet BSE models is instructive. Above the 50% condensation temperature of oxygen, i.e., to the right of the 50% curve, the $fO_2$ contours have the same shape and about the same value at any P, T point. The small difference between the dry and wet $fO_2$ values leads to slightly lower condensation temperatures for wet BSE melt at the same total pressure ($P_T$) as seen in Table 4.

Below the 50% condensation temperature of oxygen, to the left of the 50% curve, the shape, and values of the $fO_2$ contours are different. In the dry BSE model, the $fO_2$ contours go through a shallow minimum near the 50% condensed curve and then flatten out parallel to isobars. The $fO_2$ is independent of temperature in the low T, high P region at the top left of Figure 3a. In the wet BSE model (Figure 3b) $fO_2$ contours go through a deeper minimum between 50% and 99% condensed and then curve upward with $fO_2$ decreasing steeply with decreasing temperature on isobars and with increasing pressure on isotherms. At the lowest temperatures and highest pressures in the dry BSE model (top left in Figure 3a), almost no rocky elements are in the gas; the $O_2$ fugacity equals the total gas pressure to good approximation. In the same P, T range, speciation in the wet BSE model (in Figure 3b) is completely different with $fO_2$ contours down to $10^{-16}$ bar. The major oxygen – bearing gases are CO and $H_2O$ with the $fO_2$ controlled by the water gas reaction

$$H_2 + CO_2 = H_2O + CO$$

Although $O_2$ does not appear in the water gas reaction, the $H_2/H_2O$ and $CO/CO_2$ ratios each control $fO_2$ and give identical $fO_2$ values.

As temperature is isobarically increased in the dry BSE model, i.e., moving from left to right along horizontals in Figure 3a, rocky elements vaporize from the melt into the gas and reduce the $O_2$ abundance in the gas. The more volatile elements such as Au, Ag, Cd, Cu, Zn, and As vaporize at lower temperatures but are less abundant than major elements such as Si, Fe, Mg, and Ti. Thus, log $fO_2$ contours initially show small deviations from isobars. But as temperature continues increasing, more of the major elements go into the gas and the $O_2$ mole fraction decreases. Several factors including formation of metal (metalloid) oxide gases, thermal dissociation of $O_2$ to O atoms, and eventually thermal ionization of all elements lead to large decreases in the $O_2$ abundance at any given pressure. The $fO_2$ contours slope upward with respect to the isobars they originally followed, e.g., compare the log $P_{Total} = -1$ isobar and the log $fO_2 = -1$ contour in Figure 3a. Deviations between isobars and $fO_2$ contours increase with decreasing pressure and increasing temperature. At 4500 K and $10^{-6}$ bar, $fO_2$ is reduced to $1.5 \times 10^{-14}$ bar by the effects mentioned earlier. The equilibrium partial pressures of the major gases are O ($4.4 \times 10^{-7}$ bar), $Mg^+$ ($1.4 \times 10^{-7}$ bar), $Si^+$ ($7.4 \times 10^{-8}$ bar), Si ($4.6 \times 10^{-8}$ bar), and $Fe^+$ ($1.6 \times 10^{-8}$ bar). Figures 6 – 9 of Fegley et al. (2020) shows these changes at one bar total pressure for Si, Mg, Fe, and O in bulk Earth vapor.

**Major Elements**. The composition of the condensed BSE melt changes with temperature because more refractory elements condense at higher temperatures and more volatile elements condense at lower temperatures along isobars. Figures 4, 8a, 9a, 10a, 11a, 12a, and 12b and Tables 6 – 11 show the 50% condensation curves from CONDOR code calculations for Ca, Al, Mg, Si, Fe, Ti, Mn, Cr, Na, and K as a function of total pressure from $10^{-6}$ to $10^{2}$ bar in the dry BSE model. Table 6 also shows 50% condensation temperatures for Ca, Al, Mg, Si, and Fe in the wet BSE model using the Ban-Ya activity coefficients. Except for nickel, discussed later, the major elements are the most abundant elements in Table 1. The different symbols denote different activity coefficient models used in the calculations. The black curves are fits to the nominal activity coefficient model. The graphs show ideal solution 50% condensation temperatures from $10^{-4}$ to $10^{2}$ bar calculated with the GRAINS code by Lock et al (2018). The graphs also show 50% condensation temperatures at one bar total pressure for elements listed in Table 4 of Ivanov et al. (2022), which are the only tabular values in their paper. We also did ideal solution calculations with the CONDOR code. Figure labels show major gases in the region above the curves where each element is >50% in the gas, and oxide(s) dissolved in the BSE melt below the curve where each element is >50% in the melt. Figures 5, 6, 7, 8b, 9b, and 10b show 50% condensation curves superimposed on logarithmic grids of the ratios of major gases for each element.
**Calcium**. Figure 4a shows the "50%" condensation curve for calcium, which coincides with the curve in Figure 1. Calcium is more than 50% condensed where melt initially forms; at $10^{-6}$ bar Ca is 68% condensed and at $10^{2}$ bar Ca is 98% condensed. The CONDOR code calculations use activity coefficients from the Ban-Ya model, the MAGMA code, and the FactSage code (see Table 2). The MAGMA activity coefficients are linear least squares fits to $\log_{10}\gamma(CaO)$ values versus 1/T where the $\gamma(CaO)$ values were calculated for a nine-oxide model (MgO, $SiO_2$, FeO, CaO, $Al_2O_3$, $Na_2O$, $TiO_2$, $K_2O$, ZnO) of the BSE composition used by Fegley et al. (2016) over the 2000 – 5500 K range. All activity coefficients from the MAGMA code in Table 2 are from the same set of calculations. The FactSage activity coefficients are linear least squares fits to $\log_{10}\gamma(CaO)$ values versus 1/T for the six oxide BSE composition (Table 1) over the 2000 – 3000 K range and were previously used in Fegley et al. (2016). All activity coefficients from the FactSage code in Table 2 are from the same set of calculations except for $As_2O_3$, $B_2O_3$, $GeO_2$, $Li_2O$, PbO, SnO, $SnO_2$, and ZnO done from 1800 – 2200 K. Figure 4a and Table 6 show the condensation temperatures for the three sets of activity coefficients are similar. There is a 94 K range at $10^{-6}$ bar between the highest and lowest values (5% difference) decreasing with increasing pressure to a 10 K (0.2%) difference at 100 bar. Table 6 also shows condensation temperatures from the dry and wet BSE models agree within 1 – 57 K over the $10^{-6}$ – $10^2$ bar range. The ideal solution condensation temperatures, either from Lock et al. (2018) or those computed in this work using the CONDOR code, are lower, as they should be, because the activity coefficient equations for CaO give negative deviations from ideality in the BSE melt. The ideal solution values are a lower limit to the "50%" condensation temperature for CaO (liquid) dissolved in the dry BSE melt. Ideal solution condensation temperatures are useful because they help disentangle volatility and activity coefficient contributions to condensation temperatures, e.g., for the alkalis, alkaline earths, and the Group 13 - 15 elements.

The two major Ca-bearing gases in the dry and wet models are Ca and CaO, in agreement with Knudsen effusion mass spectrometry (KEMS) measurements of molten lunar basalts (DeMaria et al. 1971, Markova et al. 1986) and CaO-bearing melts (Stolyarova et al. 1991, Shornikov et al. 1997, Shornikov 2020). The other Ca-bearing gases included in the calculations ($Ca_2$, $Ca^+$, $Ca^{2+}$) are much less abundant along the condensation curve.

Figure 5a shows the thick red colored CaO "50%" condensation curve superimposed on a logarithmic grid of CaO/Ca molar ratio contours over the 1500 – 4500 K and $10^{-6}$ to $10^2$ bar range. Calcium oxide is the major Ca-bearing gas at low T and high P (top left), Ca is the major Ca-bearing gas at high T and low P (bottom right), and the two gases have equal abundances on the zero contour. The irregularities in the zero contour and others in its vicinity arise from condensation of oxygen into the melt (see Figure 3). The distribution of calcium between monatomic Ca gas and CaO gas depends upon temperature and $fO_2$ via

$$\mathrm{Ca(gas)} + \frac{1}{2}\mathrm{O_2(gas)} = \mathrm{CaO(gas)}$$

The equilibrium constant for this reaction is

$$K_{eq} = \frac{P_{CaO}}{P_{Ca}} \cdot \frac{1}{f_{O_2}^{1/2}} = \frac{X_{CaO}}{X_{Ca}} \cdot \frac{1}{f_{O_2}^{1/2}}$$

Rearranging, taking logs, and using IVTAN data for log $K_{eq}$,

$$log_{10}\left(\frac{X_{CaO}}{X_{Ca}}\right) = \frac{1}{2}log_{10}f_{O_2} + \frac{6474.2}{T} - 1.650$$

The equation shows the CaO/Ca ratio increases with the square root of oxygen fugacity and with decreasing temperature.

**Aluminum**. Figure 4b shows the "50%" condensation curve for aluminum, which coincides with the curve in Figure 1. Aluminum, like Ca, is >50% condensed where melt forms. At $10^{-6}$ bar Al is 98% condensed and at $10^2$ bar Al is 79% condensed. CONDOR code calculations used activity coefficients from the Ban-Ya model, the MAGMA code, and the FactSage code (see Table 2). Table 6 lists the "50%" condensation temperatures for the three sets of activity coefficients and for ideal solution calculations done in this work and by Lock et al (2018). The range between the highest and lowest condensation temperatures from the three activity coefficient models varies from 2 K (0.1%) at $10^{-6}$ bar to 19 K (0.4%) at 100 bar. Figure 2 shows the $Al_2O_3$ content of the melt increases with decreasing pressure and as this occurs the ideal solution model becomes more realistic and all condensation temperatures approach the same value as seen in Figure 4b.

Aluminum gas chemistry is complex. Depending on P and T, monatomic Al or AlO is the major Al-bearing gas along the condensation curve; their relative abundances (Al/AlO) vary from 0.99 to 1 (2159 K, $10^{-6}$ bar) to 0.18 to 1 (4894 K, $10^2$ bar). Either $Al_2O$, $AlO_2$, or $Al_2O_2$ is the next most abundant gas with $Al_2O$ favored at lower pressure and $AlO_2$ or $Al_2O_2$ favored at higher pressure. The other Al-bearing gases in the calculations ($Al^+$, $Al^{2+}$, $Al^-$, $Al_2$, $AlO^+$, $Al_2O_3$) are less abundant.

Figure 6 shows the thick red colored alumina condensation curve superimposed on logarithmic grids of the AlO/Al and $AlO/AlO_2$ molar ratios over the 1500 – 4500 K and $10^{-6}$ to $10^2$ bar range. Figures 6a and 6b show the same trend: lower T and higher P favors the more complex gas in each pair (i.e., AlO vs Al) and higher T and lower P favors the less complex gas in each pair (i.e., AlO vs $AlO_2$). Consistent with our results, the KEMS data of DeMaria et al. (1971) and Markova et al. (1986) show Al > AlO > $Al_2O$ over molten lunar basalts up to 2500 K at pressures to the left of the red curve in Figure 6.

**Magnesium**. Figure 4c shows the 50% condensation curve for magnesium computed using the three sets of MgO activity coefficients from Table 2, and comparison with the Lock et al. (2018) ideal solution values. The dry and wet condensation temperatures are the same within 2 K. The 50% condensation temperatures for the three sets of activity coefficients agree very well with each other and with ideal solution calculations at lower pressures but diverge into two groups with increasing pressure. The Ban-Ya and MAGMA results form one cluster whereas the FactSage and two sets of ideal solution calculations form another cluster (see Table 6). The difference is 260 K (6%) at 100 bar total pressure. Magnesium oxide shows negative deviations from ideality in the BSE melt. Figure 4c and Table 6 show the ideal solution condensation temperatures are lower than those from the activity coefficient models, as they should be. Our ideal solution 50% condensation temperatures are slightly lower than those of Lock et al (2018). The small disagreements could be due to different data sources or extrapolation of data from lower to higher temperatures.

The chemical equilibrium calculations include Mg, $Mg^+$, $Mg^{2+}$, $Mg_2$, and MgO; but only Mg and MgO are important along the condensation curve as seen in Figure 5b. Monatomic Mg is dominant at high temperature and low pressure (lower right) while MgO is dominant at low temperature and high pressure (upper left); the two gases have comparable abundances along the condensation curve at the highest P and T (upper right corner) in Figure 5b. KEMS studies of molten lunar basalts (DeMaria et al. 1971, Markova et al. 1986), molten olivine (Costa et al. 2017), and molten MgO-$SiO_2$ (Kambayashi and Kato 1984) show Mg is the major Mg-bearing gas in agreement with our low-pressure results. A comparison of Figures 5a and 5b shows CaO gas is more stable than MgO gas at a given temperature and pressure, i.e., the CaO/Ca log zero contour is at higher T than the Mg/MgO contour at any P.

**Silicon**. Figure 4d shows the 50% condensation curve for silicon computed using $SiO_2$ activity coefficients from Table 2 and comparison with the Lock et al. (2018) ideal solution values. Table 6 shows the dry and wet 50% condensation temperatures are identical within 1 – 7 K. As seen for magnesium in Figure 4c, the 50% condensation temperatures for silicon agree well with one another and with ideal solution calculations at low pressures but disagree with increasing pressure; the difference is 274 K (6%) at 100 bar total pressure (see Table 6). The condensation temperatures using activity coefficients from Ban-Ya, FactSage, and ideal solution are closer to one another than those using MAGMA code activity coefficients. The latter give higher condensation temperatures because the associated solution model in MAGMA computes $\gamma(SiO_2) < 1$. Molten $SiO_2$ has activity coefficients near unity with either positive or negative deviations from ideality (Ryerson 1985, Turkdogan 1983) and has a small partial molal enthalpy of solution that can be endothermic or exothermic (Morishita et al. 2004, Wilding and Navrotsky 1998) depending on temperature and melt composition.

Figure 7 shows the Si condensation curve imposed on graphs of SiO/$SiO_2$ molar ratios and SiO/Si molar ratios. The calculations include Si, $Si^+$, $Si^{2+}$, $Si^-$, $Si_2$, $Si_3$, SiO, and $SiO_2$; but SiO and $SiO_2$ are the only important gases and SiO is always dominant along the condensation curve in the dry and wet models. Figure 7b shows SiO is orders of magnitude more abundant than Si in the vicinity of the condensation curve. Unlike calcium, magnesium, and aluminum, monatomic silicon vapor is never important for gas – melt equilibria. Several KEMS studies show SiO is the major Si-bearing gas over molten silicates at low pressures (e.g., DeMaria et al. 1971, Markova et al. 1986, Stolyarova et al. 1996, Costa et al. 2017)

**Iron**. CONDOR calculations included FeO, $Fe_3O_4$, and $Fe_2O_3$ with FeO having the largest activity. The FeO/$FeO_{1.5}$ ratio along the 50% condensation curve varies from 35 (1818 K, $10^{-6}$ bar) to 540 (4599 K, $10^2$ bar). FactSage calculations give smaller FeO/$FeO_{1.5}$ ratios, e.g., 30 (3300 K, 1 bar) vs. 228 (3284 K, 1 bar) than CONDOR. The differences arise from differences in thermodynamic data for the pure liquid oxides and their solution parameters, which are magnified because of the exponential dependence of equilibrium constants on $\Delta G°$ values.

Figure 8a shows the 50% condensation curve for FeO computed using FeO activity coefficients from Table 2 and comparison with the Lock et al. (2018) ideal solution values. The small temperature coefficients of $\gamma$(FeO) values in Table 2 are consistent with the small temperature dependence of FeO activity coefficients in CMAS melts (Holzheid et al. 1997).

Table 6 and Figure 8a show the 50% condensation temperatures for FeO agree well with one another and with ideal solution calculations at low pressures but increasing disagreement with increasing pressure, reaching 230 K (5%) at 100 bar. Condensation temperatures from the MAGMA code are always higher because the associated solution model in MAGMA computes $\gamma$(FeO) <1 versus $\gamma$(FeO) ≥1 from ideal solution and the Ban-Ya and FactSage models. However, Figures 7 and 8 of Schaefer and Fegley (2004) display good agreement between MAGMA code calculations and experimental measurements for the amounts of FeO left in vaporization residues as a function of the fraction vaporized for two chondritic compositions (Hashimoto 1983, Wang et al. 2001) and an Apollo 16 lunar basalt (Markova et al 1986). Schaefer and Fegley (2004) found good agreement for all other oxides in both the experiments and MAGMA code ($SiO_2$, MgO, CaO, $Al_2O_3$, and $TiO_2$).

Figure 8b displays the FeO 50% condensation curve superimposed on logarithmic contours of the Fe/FeO gas molar ratios. Monatomic Fe is the major gas along the condensation curve and Fe/FeO ratios vary from ≈10 at lower pressures to ≈2 at higher pressures in the dry and wet models. The calculations also include $Fe^+$, $Fe^{2+}$, and $Fe^-$, which are insignificant. Gas – melt chemical equilibrium calculations for a Type B1 CAI composition (Schaefer and Fegley 2004) and for chondrule melts (Ebel 2005) give similar results. KEMS studies of saturated vapor over molten silicates show Fe is the major Fe-bearing gas and FeO is less abundant (Dhima et al. 1986, Markova et al. 1986, Costa et al. 2017).

**Titanium**. CONDOR calculations included TiO, $Ti_2O_3$, $Ti_3O_5$, $Ti_4O_7$, and $TiO_2$, which has the largest activity. The $TiO_2/TiO_{1.5}$ ratio varies from 6 (1882 K, $10^{-6}$ bar) to 0.35 (4778 K, $10^2$ bar) along the 50% condensation curve. FactSage calculations give different values for the $TiO_2/TiO_{1.5}$ ratio, e.g., 33 vs. 0.75 (CONDOR) at 3433 K, 1 bar. The melting point, enthalpy of fusion, and liquid heat capacity of titanium oxides are different in different databases (e.g., JANAF and IVTAN); titanium condensation was modeled as dissolution of $TiO_2$ (liq) in the BSE melt.

Figure 9a shows 50% condensation calculations using the $TiO_2$ activity coefficients in Table 2 and comparison with the Lock et al. (2018) ideal solution values. Table 7 and Figure 9a show good agreement between the real solution $TiO_2$ 50% condensation temperatures within 110 – 150 K (about 5%) over the $10^{-6}$ to $10^2$ bar range. The MAGMA code results are always higher because the associated solution model in MAGMA computes $\gamma(TiO_2)$ <1 but the FactSage $\gamma(TiO_2)$ values shift from positive to negative deviations from ideality with increasing temperature instead of trending toward ideality as do the Ban-Ya and MAGMA code values.

Figure 9b displays the $TiO_2$ 50% condensation curve superimposed on logarithmic contours of the $TiO_2$/TiO gas molar ratios. Titanium dioxide is the major gas along the condensation curve, TiO is less abundant, and Ti, $Ti^+$, $Ti^{2+}$, and $Ti^-$ are insignificant. DeMaria et al. (1971) found TiO ≈ $TiO_2$ and Markova et al. (1986) found TiO ≳ $TiO_2$ in their KEMS studies; however, these two studies are in the region where $TiO/TiO_2$ ≥1 at lower P and higher T than the thick red curve in Figure 9b.

**Manganese**. CONDOR calculations included MnO, $MnO_2$, and $Mn_3O_4$ with MnO having the largest activity. FactSage calculations included $Mn_2O_3$ (but not $Mn_3O_4$). Neither IVTAN, JANAF, nor Robie and Hemingway (1995) give data for $Mn_2O_3$ liquid and it is not considered in the CONDOR code calculations. Figure 10a is the 50% condensation curve for MnO dissolving in the BSE melt.

The black curve is the Ban-Ya (1993) model. Figure 10a and Table 8 show good agreement between the Ban-Ya, FactSage, and ideal solution models. Figure 10b overlays the MnO 50% condensation curve on logarithmic contours of constant Mn/MnO gas molar ratios. Manganese monoxide is the major gas along the condensation curve at higher pressure and Mn is the major gas at lower pressure. The KEMS results of Zaitsev and Mogutnov (1995) show monatomic Mn is the major Mn-bearing gas over $MnO-SiO_2$ melts in the 1369-1817 K range at low pressures in agreement with Figure 10b.

**Chromium**. CONDOR calculations included $Cr_2O_3$ but not CrO because there are no tabulated data for CrO (s, liq) in IVTAN, JANAF, or Robie and Hemingway (1995). Toker et al (1991) and Pretorius and Muan (1992) give data for pure CrO (liquid) and its activity coefficients in CaO – $Al_2O_3$ – $CrO_x$ – $SiO_2$ melts, but including CrO (s, liq) in the CONDOR code is beyond the scope of this work and chromium condensation was computed as $Cr_2O_3$ dissolution in BSE melt as done by Lock et al. (2018).

Figure 11a is the 50% condensation curve for $Cr_2O_3$ dissolving in the BSE melt based on FactSage $\gamma(Cr_2O_3)$ values in Table 2. Table 9 shows excellent agreement between the 50% condensation temperatures in the dry and wet BSE models and good agreement with ideal solution results within 1 – 100 K (≤2%) from $10^{-6}$ to $10^{2}$ bar total pressure. Ivanov et al. (2022) used the FactSage code and give a 50% condensation temperature of 3346 K at one bar total pressure. Using the FactSage code, which includes CrO (liquid), we obtained a similar value of 3358 K at one bar total pressure and a molar ratio of $\simeq$0.3 for $CrO_{1.5}$/CrO at the 50% condensation temperature. This is 162 K higher than the 50% condensation temperature 3196 K computed by CONDOR using FactSage activity coefficients for $Cr_2O_3$.

Chromium gas chemistry is complex and Cr, $Cr^+$, CrO, $CrO_2$, and $CrO_3$ are included in the calculations. The four most abundant Cr-bearing gases along the condensation curve are $CrO_2 \gtrsim$ CrO > Cr > $CrO_3$. This is the same sequence as Grimley et al (1961) found over $Cr_2O_3$ under oxidizing conditions in his KEMS study.

**Sodium**. Figure 12a shows the 50% condensation curve for $Na_2O$ dissolving in the BSE melt based on the Ban-Ya model and comparison with similar calculations using activity coefficients from Sossi et al (2019), the FactSage and MAGMA code models (see Table 2) and our ideal solution values. Figure 12a and Table 10 show 50% condensation temperatures from the four activity coefficient models agree within 45 – 190 K (1–11%) with 50% condensation temperatures from the MAGMA code values intermediate between those from FactSage (high) and Ban-Ya (low) and better agreement at higher pressures. The ideal solution condensation temperatures from Lock et al (2018) and this work are significantly lower than the real solution values because $Na_2O$ has large negative deviations from ideality in binary, ternary, and more complex silicate melts (Charles 1967, Gaskell 1989, Itoh and Yokokawa 1984, Lambotte and Chartrand 2013, Mathieu et al. 2008, 2011, Sossi et al. 2019, Zaitsev et al 1999).

The chemical equilibrium calculations include Na, $Na^+$, $Na^-$, $Na_2$, $NaBO_2$, NaK, NaO, $Na_2O$, and $Na_2O_2$ but only Na and NaO are important. Monatomic Na is the major gas along the condensation curve with Na/NaO ratios increasing from $\simeq$ 3 at 100 bar to $\simeq 10^4$ at $10^{-6}$ bar. The KEMS studies of molten basalt vaporization by DeMaria et al. (1971) and Markova et al. (1986) show monatomic Na is the major sodium gas in agreement with our calculations.

**Potassium**. Figure 12b shows the 50% condensation curve for $K_2O$ dissolving in the BSE melt based on the MELTS activity coefficients and comparisons with similar calculations using the three other activity coefficient models in Table 2, our ideal solution values, and results from Wang et al. (2019). Figure 12b and Table 11 show 50% condensation temperatures from the four activity coefficient models agree within 133 – 385 K with the FactSage values intermediate between the higher MAGMA and lower values from our regular solution fit to the 1673 K point of Sossi et al. (2019). The disagreement increases with increasing total pressure. The ideal solution temperatures of Lock et al (2018) and this work are much lower than those for the real solution models as expected from the large negative deviations from ideality displayed by $K_2O$ in binary, ternary, and higher order silicate melts (Charles 1967, Hastie et al 1982, Sossi et al 2019, Yazhenskikh et al 2011, Zaitsev et al 2000a,b). Figure 12b also shows good agreement between the present work and the results of Wang et al. (2019).

The chemical equilibrium calculations include K, $K^+$, $K_2$, $KBO_2$, NaK, KO, $K_2O$, and $K_2O_2$ but only K, $K^+$, and KO are important. Monatomic K gas is the major gas along the condensation curve with $K^+$, and KO the second and third most important gases. The KEMS studies of molten basalt vaporization by DeMaria et al. (1971) and Markova et al. (1986) show monatomic K is the major potassium gas in agreement with our calculations.

**Other elements.** The remaining elements in the dry BSE model are systematically considered starting with alkalis in group 1 and ending with actinides (Th, U, Pu). As discussed earlier, we exclude H, C, N, F, Cl, Br, I, S, Se, and Te from consideration because their solubilities and activity coefficients in molten silicates are not well known, but must be considered for their correct equilibrium distribution, 50% condensation temperatures, and mass balance. This is beyond the scope of this paper but is planned for subsequent calculations for a wet BSE model.

The gas phase speciation over the BSE melt for many elements is similar to that in the saturated vapor over their respective pure oxides, e.g., Ca, Al, Mg, Si, Fe, Ti, Mn, Cr, Na, and K. However, there are significant changes for elements such as As, B, Cs, Ga, In, Li, Mo, P, Re, Sb, Tl, V, and W. For example, the saturated vapor over pure oxides of As, B, Li is dominated by gaseous $As_4O_6$, $B_2O_3$, and $Li_2O$, but over the BSE melt the dominant gases of As, Ga, and Li are AsO, Ga, and Li, respectively. These changes occur because the gas chemistry of all elements is coupled and (in particular) because the $fO_2$ of the BSE melt is superimposed onto the gas phase equilibria of individual minor and trace elements. This is described in the lithium section below.

Activity coefficient models or estimates were used for several elements (Li, Rb, Cs; V, Mo, W) but ideal solution calculations were done for other elements where estimates of activity coefficients in BSE melt were not done for various reasons (Be, Sr, Ba; Sc, Y, La; the rare earth elements REE; Zr, Hf; Nb, Ta; Re; Ru, Os; Rh, Ir; Pd, Pt, Au, Hg; Th, U, Pu). Most of the latter elements are lithophile refractory or siderophile trace elements (Kornacki and Fegley 1986).

**Group 1 (Li, Rb, Cs)**. Sodium and potassium were discussed earlier; lithium, rubidium, and cesium are considered now. Several measurements show negative deviations from ideality increasing from $Li_2O$ to $Na_2O$ to $K_2O$; a smaller number of measurements show the trend continues from $K_2O$ to $Rb_2O$ to $Cs_2O$. This is seen from alkali oxide activities derived from phase diagrams for $Li_2O$-$SiO_2$, $Na_2O$-$SiO_2$ and $K_2O$-$SiO_2$ melts (Charles 1967), Langmuir vaporization measurements of alkali oxide activities (Li, Na, K, Rb) in basaltic melt (Sossi et al. 2019), the partial molal enthalpy of solution for $SiO_2$ in Li, Na, K, and Cs oxide - silica binary melts (Morishita et al. 2004), and the position of silica liquidus curves in binary alkali oxide - silica melts (Rey 1948). Increasing optical basicity (Duffy 2004) and decreasing cationic potential (Z/r), which relate to electronegativity and metal - oxygen bond strengths, suggest increasingly negative deviations from ideality downward in group 1 from Li to Cs. Smooth, often linear correlations between optical basicity and log $\gamma$(oxide) in melts exist for CaO, MgO, $SiO_2$, $Al_2O_3$, $Na_2O$ and $P_2O_5$ (Duffy et al 1978, Gaskell 1989, Beckett 2002, Mathieu et al. 2011).

**Lithium condensation**. Figure 11b and Table 12 compare the 50% condensation temperatures for $Li_2O$ computed using the four sets of activity coefficients in Table 2 and ideal solution. $Li_2O$, like $Na_2O$ and $K_2O$, has negative deviations from ideality in binary melts with $SiO_2$ although the database is smaller (Charles 1967, Shakhmatkin and Vedishcheva 1994). Sossi et al (2019) report positive deviations from ideality, which is consistent with the known immiscibility in some regions of the $Li_2O$ - $SiO_2$ - $Al_2O_3$ system (e.g., Konar et al 2018). Nassarella and Fruehan (1992) measured $\gamma(Li_2O)$ =0.05 at 1573 K in a CaO - $Al_2O_3$ - $CaF_2$ - $Li_2O$ melt. Sossi et al (2019) measured two different $\gamma(Li_2O)$ values at 1673 K: one for $fO_2 = 10^{-8}$ bar and one in air. The large spread of condensation temperatures shows the poor state of knowledge about $Li_2O$ activity coefficients in multicomponent silicate melts.

**Lithium Gas Chemistry**. Monatomic Li and LiO are the two major Li-bearing gases and Li (g) is the major species along the condensation curve. KEMS studies of $Li_2SiO_3$, $Li_2TiO_3$, and $Li_2ZrO_3$ vaporization (Asano and Nakagawa 1989, Kato et al. 1993, Nakagawa et al. 1982) show monatomic Li and the less abundant LiO are the major lithium gases in agreement with the calculations for saturated vapor over BSE melt. But this is different than in saturated vapor over pure $Li_2O$ where $Li_2O \gtrsim Li > LiO$ with minor amounts of $Li_3O$ and $Li_2O_2$ (Table 1, Sossi and Fegley 2018). The difference arises because the $fO_2$ of saturated $Li_2O$ vapor is different than that over the BSE melt and because the partial pressure of $Li_2O$ (g) is independent of oxygen fugacity.

The major vaporization reactions, equilibrium constants, and gas partial pressures are

$$Li_2O\ (s,liq) = Li_2O\ (g)$$

$$K_{Li_2O} = \frac{P_{Li_2O}}{a_{Li_2O(s,liq)}}$$

$$P_{Li_2O} = K_{Li_2O} \cdot a_{Li_2O(s,liq)}$$

$$Li_2O\ (s,liq) = 2\ Li\ (g) + ½\ O_2\ (g)$$

$$K_{Li} = \frac{P_{Li}^2 f_{O_2}^{1/2}}{a_{Li_2O(s,liq)}}$$

$$P_{Li} = \frac{\left(K_{Li} \cdot a_{Li_2O(s,liq)}\right)^{1/2}}{f_{O_2}^{1/4}}$$

$$Li_2O\ (s,liq) = LiO\ (g) + Li\ (g)$$

$$K_{LiO} = \frac{P_{LiO}P_{Li}}{a_{Li_2O(s,liq)}}$$

$$P_{LiO} = \frac{K_{LiO} \cdot a_{Li_2O(s,liq)}}{P_{Li}} = \frac{K_{LiO} \cdot f_{O_2}^{1/4}}{K_{Li}^{1/2}}$$

Some thermal dissociation of $O_2$ occurs via

$$O_2\ (g) = 2O\ (g)$$

$$K_O = \frac{P_O^2}{P_{O_2}} = \frac{P_O^2}{f_{O_2}}$$

$$P_O = \left(K_O \cdot f_{O_2}\right)^{1/2}$$

Neglecting the minor gases $Li_3O$, $Li_2O_2$, and ions does not alter the conclusions.

A comparison of the $Li_2O$, Li, and LiO equilibrium partial pressures in saturated $Li_2O$ vapor (denoted by the subscript "pure") and in saturated BSE melt vapor (denoted by the subscript "BSE") is instructive. Starting with $Li_2O$ gas,

$$\frac{P_{Li2O,BSE}}{P_{Li2O,pure}} = \frac{X_{Li2O,BSE} \cdot \gamma_{Li2O,BSE}}{a_{Li2O,pure}} = X_{Li2O,BSE} \cdot \gamma_{Li2O,BSE}$$

The $Li_2O$ partial pressure over the BSE melt is independent of the $fO_2$ of melt vapor but is proportional to the $Li_2O$ mole fraction in the melt, which is about $6 \times 10^{-6}$ (see Table 1). Unless the $Li_2O$ activity coefficient is >>1, the $Li_2O$ activity in the melt is <<1 and the $Li_2O$ partial pressure is lower than that over pure $Li_2O$ which has an activity of unity.

The Li and LiO equilibrium partial pressure ratios are

$$\frac{P_{Li,BSE}}{P_{Li,pure}} = \left(X_{Li2O,BSE} \cdot \gamma_{Li2O,BSE}\right)^{1/2} \left(\frac{f_{O_2,pure}}{f_{O_2,BSE}}\right)^{1/4}$$

$$\frac{P_{LiO,BSE}}{P_{LiO,pure}} = \left(X_{Li2O,BSE} \cdot \gamma_{Li2O,BSE}\right)^{1/2} \left(\frac{f_{O_2,BSE}}{f_{O_2,pure}}\right)^{1/4}$$

The partial pressure ratios of monatomic Li and LiO are both proportional to the square root of the $Li_2O$ activity in the BSE melt. However, their dependence on oxygen fugacity is different. At constant temperature, $fO_{2,pure}$ over pure $Li_2O$ is constant while $fO_{2,BSE}$ for the BSE melt can be variable. For the case of $fO_{2,BSE} > fO_{2,pure}$, the LiO partial pressure ratio decreases (proportional to the 4th root of the $fO_2$ ratio (BSE/pure), whereas the monatomic Li gas partial pressure ratio increases by the same factor. Conversely for $fO_{2,BSE} < fO_{2,pure}$, the LiO partial pressure ratio increases and that of monatomic Li decreases.

The oxygen fugacity $fO_{2,pure}$ in saturated vapor over pure $Li_2O$ is fixed at any temperature by stoichiometry and is

$$f_{O2,pure} = \frac{1}{4}\left(P_{Li,pure} - P_{LiO,pure} - 2\,P_{O,pure}\right).$$

However, $fO_{2,BSE}$ is not controlled by trace $Li_2O$ evaporation but instead comes from evaporation and dissociation of major oxides such as $SiO_2$ from the melt. There is no simple expression for $fO_{2,BSE}$ because several evaporating oxides can contribute oxygen to the vapor phase and non-linear coupled equations for all gas species equilibria must be solved simultaneously.

**Rubidium**. The activity coefficient data of Sossi et al. (2019) for $Rb_2O$ in molten ferrobasalt seem to be the only data for $\gamma(Rb_2O)$ in multicomponent silicate melts. The nominal value for $\gamma(Rb_2O)$ is a regular solution extrapolation of the 1673 K point of Sossi et al (2019) for a synthetic ferrobasalt. This regular solution equation was used because a linear least square fit of log $\gamma$ vs. 1/T to all three Rb data points in Table 4 of Sossi et al. (2019) gave increasingly negative deviations from ideality with increasing temperature, which is unreasonable.

Initially we used the published $Rb_2SiO_3 – SiO_2$ phase diagram of Alekseeva (1963) to estimate an equation for $\gamma(Rb_2O)$ from the eutectic reaction between $Rb_2Si_4O_9$, $Rb_2O$ (melt), and $SiO_2$ (quartz), and derived log $\gamma(Rb_2O) = -15{,}200/T$ using data from Knacke et al. (1991). However, Lapshin et al. (2006a) claim the phase diagram is incorrect because they crystallized (and identified via X-ray diffraction) $Rb_6Si_{10}O_{23}$ from glass with the composition $Rb_2Si_4O_9$. These two silicates have similar Rb/Si molar ratios of 0.25 ($Rb_2Si_4O_9$) versus 0.30 ($Rb_6Si_{10}O_{23}$). Kracek (1933) published the liquidus curve for the $Rb_2Si_2O_5 – SiO_2$ part of the phase diagram and he does not find the existence of a tetrasilicate. We do not use either our thermodynamic analysis of the $Rb_2SiO_3 – SiO_2$ phase diagram or that of Kim and Sanders (1991).

Figure 12c and Table 13 compare the Rb 50% condensation temperatures for our real and ideal solution models. The ideal solution is a strict lower limit although the nominal activity coefficient equation may not give an upper limit to the 50% condensation temperatures if the "true" $Rb_2O$ activity coefficients are even smaller, i.e., larger negative deviations from ideality.

The chemical equilibrium calculations considered nine Rb-bearing gases (Rb, $Rb^+$, $Rb^-$, $Rb_2$, RbK, $RbBO_2$, RbO, $Rb_2O$, and $Rb_2O_2$) but only Rb, $Rb^+$, and RbO are significant with Rb more important. Rubidium gas chemistry in BSE material is like that in the saturated vapor over liquid $Rb_2O$ because Rb is the major gas in both cases (see Table 1 of Sossi and Fegley 2018). A KEMS study of $Rb_2O – B_2O_3 – SiO_2$ melt vaporization shows Rb is the major gas (Lopatin et al 2007) for $Rb_2O/B_2O_3$ ratios > 1. Molecular gases such as $Rb_2O$ and $Rb_2O_2$ are less important in BSE vapor as expected from the discussion of $Li_2O$ vaporization.

**Cesium**. Little is known about thermodynamic properties of $Cs_2O$-bearing silicate melts. Borisov (2009) measured relative activity coefficients of alkali oxides in aluminosilicate melts and found $\gamma(Cs_2O)$ just below $\gamma(Rb_2O)$ and a few times smaller than $\gamma(Na_2O)$. The existence of two Cs-Na silicates in the $Cs_2O – SiO_2 – Na_2O$ system ($CsNaSiO_3$ and $CsNaSi_2O_5$) means there are negative deviations from ideality between $Na_2O$ and $Cs_2O$ in the ternary system (Bennour et al. 1996a). Bennour et al (1996b, 1999) studied mixing in $Cs_2O$ - $SiO_2$ - $Na_2O$ melts by KEMS and successfully modeled their data using the interaction parameters $W_{Cs-Si}$ = −232,500 J mol$^{-1}$, and $W_{Cs-Na}$ = 0 J mol$^{-1}$. The $W_{Cs-Si}$ parameter corresponds to log $\gamma(Cs_2O)$ = −12,144/T.

The situation with the $Cs_2O$ – $SiO_2$ phase diagram of Alekseeva (1966) is like that above for the $Rb_2SiO_3$ – $SiO_2$ phase diagram. Initially we estimated $\gamma(Cs_2O)$ from the eutectic reaction between $Cs_2Si_4O_9$, $Cs_2O$ (melt), and $SiO_2$ (quartz), and derived log $\gamma(Cs_2O)$ = −13,700/T with data from Knacke et al. (1991). Then we found Lapshin et al. (2006b) who claim the $Cs_2O$ – $SiO_2$ phase diagram of Alekseeva (1966) is incorrect because they crystallized (and identified) $Cs_6Si_{10}O_{23}$ from glass with the composition $Cs_2Si_4O_9$. Suzuki et al. (2020) synthesized and measured the X-ray diffraction pattern of $Cs_2Si_4O_9$, found some discrepancies with the XRD pattern of Alekseeva (1966), but used the sample for low temperature heat capacity measurements and published data for $Cs_2Si_4O_9$ thermal properties from 0 – 300 K. Cordfunke and Ouweltjes (1993) published the standard enthalpy of formation for $Cs_2Si_4O_9$ at 298 K. Whether or not $Cs_2Si_4O_9$ exists, we do not use either our thermodynamic analysis of the $Cs_2O$ – $SiO_2$ phase diagram or that of Kim and Sanders (1991) in our work.

Cesium is an instance where analogy to molten salts (of elements in the same group of the periodic table) helped to estimate the activity coefficient. Molten Cs and Na fluorides were used because of (1) the similar VI coordinate radii of oxide (140 pm) and fluoride (133 pm) ions (Shannon 1976), (2) the oxide $O^{2-}$ and fluoride $F^{1-}$ ions are isoelectronic, and (3) ideal (or nearly so) Temkin mixing of fluoride ions with oxide ions is documented for PbO-$PbF_2$, (Jeffes and Sridhar 1968) CaO-$CaF_2$, and MgO-$MgF_2$ (Kim et al. 2012). Lumsden (1966) derived $\log_{10} \gamma^{o}$ = 175/T for CsF dilute solution in molten NaF; a small positive deviation from ideality. Based on the work of Borisov (2009) suggesting $\gamma(Cs_2O) < \gamma(Na_2O)$, Bennour (1996b, 1999) suggesting $\gamma(Cs_2O) = \gamma(Na_2O)$, and Lumsden (1966) suggesting $\gamma(Cs_2O) > \gamma(Na_2O)$, we adopted an equation setting log $\gamma(Cs_2O)$ = log $\gamma(Na_2O)$ from FactSage (see Table 2).This equation gives $\gamma(Cs_2O) < \gamma(Na_2O)$ from the Ban-Ya (1993) equation for log $\gamma(Na_2O)$ in Table 2. Table 13 and Figure 12d show the 50% condensation temperatures and condensation curve for $Cs_2O$ using the two estimated activity coefficient equations and comparison to ideal solution. The ideal solution 50% condensation temperatures at $10^{-6}$ to $10^{-2}$ bar for cesium are below the solidus temperature of peridotite; the same is true for Li at $10^{-6}$ bar and for Na, K, and Rb at $10^{-6}$ to $10^{-4}$ bar.

Nine Cs-bearing gases are included in the chemical equilibrium calculations: Cs, $Cs^+$, $Cs^-$, $Cs_2$, $CsBO_2$, CsO, $Cs_2O$, $Cs_2O^+$, and $Cs_2O_2$, but only Cs, $Cs^+$, and CsO are significant with Cs dominant along the condensation curve. The KEMS study of crystalline cesium aluminosilicates by Odoj and Hilpert (1980a, b) showed Cs is the major Cs-bearing gas. Cesium gas chemistry in BSE material is different than the speciation in saturated vapor over liquid $Cs_2O$ in which $Cs_2O$ is the major gas (Table 1, Sossi and Fegley 2018). The difference is due to the same reasons that discussed above for lithium.

**Group 2 (Alkaline Earths Be, Mg, Ca, Sr, Ba)**. Table 6 and Figures 4a, 4c, and 5 give the results for Ca and Mg gas chemistry and real and ideal solution 50% condensation temperatures into BSE melt. Table 14 summarizes ideal solution 50% condensation temperatures for all alkaline earth oxides dissolved in BSE melt and shows BeO is the most refractory alkaline earth oxide in BSE melt assuming ideal solution.

The existence of $Be_2SiO_4$ (phenakite), $BeAl_2O_4$ (chrysoberyl), and $Al_2Be_3Si_6O_{18}$ (beryl) suggests $\gamma(BeO) < 1$. We did not try to constrain the BeO activity coefficient using periodic trends because beryllia (BeO) does not have the NaCl crystal structure as do the other alkaline earth oxides. It has the wurtzite structure like zincite (ZnO). Chemically, Be is more like Al than the alkaline earths, and its optical basicity ($\Lambda = 0.48$) is closer to that of $Al_2O_3$ ($\Lambda = 0.60$) than to other alkaline earth oxides (Lebouteiller and Courtine 1998). The optical basicity of BeO suggests amphoteric behavior in melts (Duffy 1993, Navrotsky 1994).

Negative deviations from ideality are expected for SrO and BaO in BSE melt because of (1) the existence of Sr and Ba aluminates, aluminosilicates, and titanates, (2) KEMS measurements showing negative deviations from ideality for SrO and BaO in binary and ternary silicate melts (Lopatin et al. 2006, Stolyarova and Lopatin 2005, Tyurnina et al. 2005), and (3) thermodynamic analyses of the SrO – $SiO_2$ and BaO – $SiO_2$ phase diagrams of Eskola (1922). However, the activity coefficients from the KEMS studies, the phase diagrams, and FactSage disagree enough that we did not use them and did ideal solution calculations for BaO and SrO.

The major alkaline earth gases in BSE vapor are Be ≈ BeO, Mg, Ca, Sr ≈ SrO, and BaO, which are the same as those in the saturated vapor over the pure alkaline earth monoxides (see Table 1, Sossi and Fegley 2018). Note that Be does not follow the trend of decreasing M/MO gas ratios from Mg to Ba. The twelve Be-bearing gases in the chemical equilibrium calculations are Be, $Be^+$, $Be_2$, $(BeO)_{1-6}$, $Be_2O$, $BeBO_2$, and $BeB_2O_4$. Monatomic Be and BeO are the two major gases along the 50% condensation curve; at one bar total pressure Be is 54.4%, BeO is 45.6%, and $BeBO_2$ is $7\times10^{-4}$% of all Be in the gas. The BeO gas polymers are unimportant. The BeO/Be ratio increases along the condensation curve to 2.1 at 4894 K, 100 bar pressure. The major Mg and Ca gases were discussed earlier. The major Sr-bearing and Ba-bearing gases along the SrO and BaO 50% condensation curves are Sr and SrO, and BaO. The SrO/Sr ratio increases from 0.40 to 4.2 and the BaO/Ba ratio decreases from 3,400 to 28 from $10^{-6}$ to $10^2$ bar total pressure. The approximate equality of SrO and Sr and the dominance of BaO in BSE vapor are also seen in saturated vapor over solid SrO (1575 - 2100 K) and solid BaO (1200 - 1917 K).

**Group 3 (Sc, Y)**. Yttrium is similar to light rare earth elements (REE) and we mention studies relevant to REE below. Table 15 gives 50% condensation temperatures for ideal solution of $Sc_2O_3$ and $Y_2O_3$ in BSE melt. These temperatures are higher than the oxide 50% condensation temperatures in solar composition gas at the same total pressure (Kornacki and Fegley 1986, Lodders 2003). The necessary data for real solution calculations are unavailable. Evidence for negative deviation from ideality includes double oxide formation between group 3 oxides and CaO, phase diagrams for group 3 oxides with CaO, formation of group 3 and REE silicates, and KEMS studies of $Sc_2O_3$, $Y_2O_3$, and $La_2O_3$ in oxide melts (e.g., Costa et al 2019, Jacobson 1989, Jacobson et al. 2001, Stolyarova and Semenov 1994, Stolyarova et al. 2020, Zhi et al. 2018). Evidence for positive deviations from ideality includes high-temperature liquid unmixing in the $La_2O_3$ - $SiO_2$ and $Y_2O_3$ - $SiO_2$ binary systems (Hageman and Oonk 1986). The major gases of Sc and Y in BSE vapor are ScO and YO, which are the same as the major gases in saturated vapor over pure group 3 sesquioxides (Sossi and Fegley 2018).

**Rare Earth Elements (REE, La to Lu).** Table 16 lists ideal solution 50% condensation temperatures for the REE, which are higher than the oxide 50% condensation temperatures in solar composition gas at the same total pressure (Kornacki and Fegley 1986, Lodders 2003). The necessary data for calculations are unavailable but non-ideal solution of the REE in BSE melt is expected because REE form compounds with the major oxides in BSE melt, e.g., oxyapatite compounds with CaO and $SiO_2$ (Costa et al 2019), aluminates such as $LaAlO_3$, and titanates such as $CeTi^{3+}O_3$. Different activity coefficients are expected for $Ce_2O_3$ and $CeO_2$ and for EuO and $Eu_2O_3$ dissolved in BSE melt because different activity coefficients are derived for Ce and Eu oxides dissolved in silicate melts at much lower temperatures (Johnston 1965, Morris et al 1974, Cachia et al. 2006, Burnham et al 2015), but there are insufficient data to do calculations.

**Group 4 (Ti, Zr, Hf)**. Titanium chemistry was discussed earlier, and Table 7 and Figure 9 give 50% condensation temperatures for non-ideal dissolution of $TiO_2$ in the BSE melt, a comparison to ideal solution results, and shows the major Ti-bearing gases. Table 15 compares ideal solution 50% condensation temperatures for $TiO_2$, $ZrO_2$, and $HfO_2$ in BSE melt. Zirconia and hafnia are more refractory than titania and are more than 50% condensed as soon as BSE melt forms. Figure 13a shows the ideal solution 50% condensation curves for $ZrO_2$ and $HfO_2$ from $10^{-6}$ to $10^2$ bar total pressure. The major gases are $ZrO_2$ > ZrO and HfO > $HfO_2$; monatomic Zr and Hf gases are insignificant. The $HfO_2$ (g) data have an uncertainty of ±30 kJ mol$^{-1}$ and are of low accuracy (6-F class of the IVTAN database). The $HfO_2$ condensation calculations are preliminary because of the uncertain $HfO_2$ (g) thermodynamic data.

The necessary interaction coefficients to model non-ideal solutions of $ZrO_2$ and $HfO_2$ in BSE melt do not exist. However, negative deviations from ideality are expected for $ZrO_2$ and $HfO_2$ in BSE melt based on existing calorimetry. Linard et al (2008) measured $\Delta H^\circ_{soln}$ = −20±3 kJ mol$^{-1}$ at infinite dilution for $HfO_2$ solution in $Na_2O \cdot 3SiO_2$ melt at 1733 K. Costa et al (2018) found $\Delta H^\circ_{soln}$ = −14.07±3.35 kJ mol$^{-1}$ for solution of 7 mol% $Y_2O_3$ - 93mol% $ZrO_2$ in a CMAS melt at 1723 K. Both results indicate negative deviations from ideality but neither melt has identical composition to BSE melt. In contrast, Linard et al (2008) found $\Delta H^\circ_{soln}$ = +25.26 kJ mol$^{-1}$ for $TiO_2$ solution in $Na_2O \cdot 3SiO_2$ melt at 1731 K. An endothermic $\Delta H^\circ_{soln}$ implies positive deviations from ideality, as seen for FactSage $\gamma(TiO_2)$ values at lower temperatures in BSE melt.

**Group 5 (V, Nb, Ta)**. Table 17 gives 50% condensation temperatures for V (real and ideal solution), and ideal solution values for Nb, and Ta. The dioxides and monoxides are the major gases for V, Nb, and Ta. The $MO_2$/MO ratios decrease from $10^{-6}$ to $10^2$ bar along the 50% condensation curves; from 10,750 to 34 for $VO_2$/VO, from 23,800 to 23 for $NbO_2$/NbO, and from 46,200 to 140 for $TaO_2$/TaO. The following solid (liquid) oxides are in the condensation calculations: VO, $V_2O_3$, $V_2O_4$, and $V_2O_5$ for V, NbO, $NbO_2$, and $Nb_2O_5$ for Nb, and $Ta_2O_5$ for Ta. Tantalum pentoxide $Ta_2O_5$ is the only stable solid oxide of tantalum (Garg et al 1996, Jacob et al 2009). Table 2 shows the real solution vanadium condensation calculations use the estimates $\gamma$(VO) = $\gamma$(FeO) and $\gamma(V_2O_3)$ = $\gamma(Ti_2O_3)$. Ideal solution condensation was used for the other V, Nb, and Ta oxides in the absence of reliable estimates. The existence of stable vanadates, niobates, and tantalates (Kubaschewski 1972, Jacob and Rajput 2014) implies negative deviations from ideality but the amphoteric V and Nb oxides and acidic Ta oxides may have positive interaction parameters ($\alpha$) with $SiO_2$, like $Fe_2O_3$ and $TiO_2$ (amphoteric) and $P_2O_5$ (acidic) with $SiO_2$ (Ban-Ya 1993).

**Group 6 (Cr, Mo, W)**. Chromium chemistry in BSE material was discussed earlier (see Table 9 and Figure 11a). The oxides $MoO_2$, $MoO_3$, and $WO_3$ are included in the calculations and their activity coefficients are estimated from basicity and ionic potential comparisons. Tungsten dioxide decomposes before melting and neither JANAF nor IVTAN gives thermodynamic data for $WO_2$ liquid. Only liquid $WO_3$ is considered in the calculations, which use JANAF data (Chase 1998). Neglect of $WO_2$ liquid (not in either JANAF or IVTAN) does not introduce errors because O'Neill et al. (2008) report $W^{6+}$ ($WO_3$) is the dominant tungsten species at $fO_2$ of $10^{-11}$ bar in CMAS melts at 1673 K. An $fO_2$ of $10^{-11}$ bar is five orders of magnitude lower than the lowest $fO_2$ in equilibrium with BSE melts in the P, T range we studied (see Figure 3).

The optical basicity of $MoO_2$ ($\Lambda$ = 0.96) lies between FeO (1.00) and $TiO_2$ (0.75). The ionic potential for $Mo^{4+}$ is 6.15, closer to that of $Ti^{4+}$ (6.61) than to $Fe^{2+}$ (3.28). The geometric mean of FeO and $TiO_2$ activity coefficients was used as the nominal value for $MoO_2$ in Table 2. Optical basicity values show $WO_3$ (0.51) and $MoO_3$ (0.52) are strong acids lying between $SiO_2$ (0.48) and $Al_2O_3$ (0.60). The estimate $\gamma(WO_3) = \gamma(SiO_2)$ was used because the optical basicity of $WO_3$ (0.51) and $SiO_2$ (0.48) are similar and the ionic potentials of $W^{6+}$ and $Si^{4+}$ are identical (10). The ionic potential of $Mo^{6+}$ (10.2) is much closer to that of $Si^{4+}$ (10) than that of $Al^{3+}$ (5.61). Two studies show $\gamma(WO_3)$ is slightly smaller than $\gamma(MoO_3)$. O'Neill et al (2008) found the activity coefficient ratio $\gamma(MoO_3)/\gamma(WO_3) \approx 4$ in CMAS melts at 1673 K. The EMF measurements of Lin and Elliott (1983, 2001) on $Na_2O$ – $MoO_3$ and $Na_2O$ – $WO_3$ melts give $\gamma(MoO_3)/\gamma(WO_3) \approx 2$ at 1200 K. The $MoO_3$ activity coefficient $\gamma(MoO_3)$ is set four times larger than $\gamma(WO_3)$, i.e., a smaller negative deviation from ideality.

Table 18 gives 50% condensation temperatures for real and ideal solution of $Cr_2O_3$, $MoO_2$ + $MoO_3$ and $WO_3$ in BSE melt. The real solution calculations show tungsten is the most refractory and molybdenum is the most volatile of the group 6 elements, except at 100 bar total pressure where Cr is more volatile than Mo. The major Mo-bearing gases are $MoO_3 > MoO_2 > MoO >> Mo$ with $Mo_2O_6 \gtrsim MoO$ at the lowest total pressures. The major W-bearing gases follow a similar pattern: $WO_3 > WO_2 > WO >> W$ with $W_2O_6 \gtrsim WO$ at the lowest total pressures. Condensation of Mo and W oxides dissolved in BSE melt occurs because Mo and W oxide gases are dominant while monatomic Mo and W are trace species. Along the 50% condensation curves from $10^{-6}$ to $10^2$ bar total pressure, the $MoO_3$/Mo ratio varies from $8\times10^{13}$ to $2\times10^3$ and the $WO_3$/W ratio varies from $9\times10^{11}$ to $3\times10^3$. Molybdenum and W gas chemistry in BSE material is different than the speciation in saturated vapor over pure $MoO_2$, $MoO_3$, and $WO_3$ (Table 1 of Sossi and Fegley 2018) for the same reasons as discussed above for lithium.

**Group 7 (Mn, Tc, Re)**, **Group 8 (Fe, Ru, Os)**, **Group 9 (Co, Rh, Ir)**, **Group 10 (Ni, Pd, Pt)**. These four groups are discussed together because their chemistry is similar, especially Mn and Fe; Co and Ni; Re, Ru, Os; and Rh, Ir, Pd, and Pt. The chemistry of Mn and Fe was discussed earlier. Table 6 and Figure 8 show Fe chemistry and Table 8 and Figure 10 show Mn chemistry. Experimental studies indicate noble metals (M) dissolve as oxides in silicate melts via redox sensitive reactions such as

$$M + \frac{x}{4}O_2 = MO_{x/2}(in\ melt)$$

(e.g., O'Neill and Nell 1997, Borisov and Walker 2000 and references therein). As discussed in section 2, activity coefficients of the dissolved metal oxide liquids are needed to compute solubilities. Neither the melting points nor the enthalpies of fusion of noble metal oxides are known and solubility calculations cannot be done. However, 50% condensation temperatures of the stable condensates (either liquid metals or solid metal oxides) were computed to estimate noble metal volatility in BSE material.

**Technetium**. Technetium is unstable and the longest-lived isotopes ($^{97}$Tc, $^{98}$Tc) have half-lives of about 4.2 million years. Technetium chemistry is not considered because insufficient thermodynamic data exist to do so.

**Rhenium**, **Ruthenium, and Osmium**. These three elements are considered together because they form volatile oxide gases and as a result two competing factors affect their chemistry in BSE material. The first factor is their refractory nature, i.e., the low vapor pressure of metallic Re, Ru, and Os at a given temperature. The second factor is their high affinity for oxygen, i.e., the formation of well-known oxide gases such as $Re_2O_7$ (rhenium heptoxide), $RuO_4$ (ruthenium tetroxide), and $OsO_4$ (osmium tetroxide) and solid oxides such as $ReO_2$, $RuO_2$, and $OsO_2$ (Pownceby and O'Neill 1994, O'Neill and Nell 1997). Mass spectrometric and other experimental studies show Re, Ru, and Os form monoxide, dioxide, and trioxide gases (e.g., Hildenbrand and Lau 1992, Nuta et al 2021, Skinner and Searcy 1973, Watson et al. 1991). This complicates their chemistry at the high $fO_2$ of the BSE melt – vapor system at $10^{-6}$ to $10^{2}$ bar total pressure and leads to consideration of multiple oxide gases and solids in the condensation calculations. Rhenium, ruthenium, and osmium condense as dioxide solids (Table 19). The data thermodynamic data used in the calculations are from Barin (1989), Fegley and Palme (1985), O'Neill and Nell (1997), and Pownceby and O'Neill (1994).

Initially the high volatility of Re, Ru, and Os in dry BSE material was surprising. However, Jacobson et al. (2001) found Re oxidized rapidly at 600 - 1400 °C (873 - 1673 K) in $O_2$–Ar gas mixtures with $fO_2 \geq 3\times10^{-4}$ bar. Borisov and Jones (1999) found Re wire loops in gas mixing furnaces evaporate within a few hours at 1673 K at the $fO_2$ of the quartz - fayalite - magnetite (QFM) oxygen fugacity buffer ($fO_2 = 3.6\times10^{-7}$ bar). Watson et al. (1991) reported that Os metal coatings on satellite optics decayed in low Earth orbit at O atom partial pressures of about $10^{-10}$ bar. Figure 3 shows the $fO_2$ at 1673 K in the dry BSE model is up to a factor of 5 lower than the total pressure below $10^{-3}$ bar and is about the same as the total pressure from $10^{-3}$ to $10^{2}$ bars.

We also wondered why similar behavior was not seen in earlier calculations about refractory metal nugget formation in Ca,Al-rich inclusions under oxidizing conditions (Fegley and Palme 1985). The reason is that the oxygen fugacities in the dry BSE model are orders of magnitude higher than those in the metal nugget calculations (at the same P and T). For example, the most oxidizing $fO_2$ in Figure 1 of Fegley and Palme (1985) is about $10^{-12.3}$ bar at 1600 K and $10^{-3}$ bar total pressure versus $fO_2 = 10^{-3}$ bar in the dry BSE model at $10^{-3}$ bar total pressure.

**Cobalt and Nickel**. Table 20 and Figure 14 show equilibrium chemistry of Co and Ni in BSE material. The real solution 50% condensation curves are computed using activity coefficients from FactSage (Table 2). Holzheid et al. (1997) give similar values of CoO ($\gamma = 1.51\pm0.28$) and NiO ($\gamma = 2.70\pm0.52$) activity coefficients in silicate melts over a range of temperatures (1573 - 1873 K) and oxygen fugacities (log $fO_2$ = −3 to +1.5 relative to the $Fe-Fe_{1-x}O$ buffer). The real and ideal 50% condensation curves agree within ≤34 K for both Ni and Co with the real solution curves lying at slightly lower temperature due to the small positive deviations from ideality.
We also find Ni metal condensation in agreement with Lock et al. (2018). This occurs over a P - T range where the $fO_2$ of BSE material is along the Ni - NiO oxygen fugacity buffer.

**Rhodium, Iridium, Palladium, and Platinum.** The condensation temperatures and volatility sequence for the noble metals are different in BSE material than solar composition material. The dry BSE calculations in Table 19 show Pt is the most refractory noble metal in BSE material except at $10^{-6}$ bar where Ir condenses 27 K higher. At $10^{-4}$ bar total pressure, the volatility sequence from most refractory to least refractory in dry and wet BSE material (Table 19), and in solar composition material (Lodders 2003) is

| | |
|---|---|
| Dry BSE | Pt ≳ Ir > Rh > Os > Re >> Au > Pd > Ru |
| Wet BSE | Os > Ir > Re > Ru > Pt ≳ Rh > Pd > Au |
| Solar | Re < Os << Ir < Ru < Pt < Pd << Au |

With increasing pressure, the stable Ir condensate in BSE material changes from metal to $IrO_2$ with a range in which metal plus oxide are stable at $fO_2$ values on the Ir - $IrO_2$ oxygen fugacity buffer (O'Neill and Nell 1997). Depending on total pressure either oxide or monatomic metal gases are the major noble metal gases except gold which is always Au gas in dry BSE material.

**Group 11 (Cu, Ag, Au)**. Gold was discussed with the Pt-group metals.

**Copper**. Figure 14c shows the 50% condensation curve for copper computed using Altman's (1978) regular solution model and comparisons to calculations using regular solution fits to the 1573 K point of Holzheid and Lodders (2001), the 1673 K point of Sossi et al. (2019), the 1923 K point of Wood & Wade (2013), and ideal solution. Table 2 lists the activity coefficient equations for the four regular solution models. Figure 14c and Table 21 show good agreement between the four different activity coefficient models and the ideal solution model. The three major Cu-bearing gases are Cu >> CuO > $Cu_2$ gas in all models. Monatomic copper is the only Cu-bearing gas observed by KEMS studies of $Cu_2O$ evaporation (Table 1, Sossi and Fegley 2018).

The calculations do not include CuO dissolution into BSE melt because there are no thermodynamic data for liquid CuO in the JANAF or FactSage databases. The IVTAN data for liquid CuO at one bar are incorrect, as found during this work. Tenorite (CuO) melts incongruently to solid $Cu_2O$ + melt at one bar $O_2$ pressure. Reported temperatures of this peritectic point and oxygen content of the melt vary slightly; Roberts and Smyth (1921) measured 1101 °C and $X_O$ = 0.399 while Kosenko and Emel'chenko (2001) give 1102 °C and $X_O$ = 0.397. The congruent melting point of CuO is not known and estimates include 1228 °C and 24.7 bar $O_2$, 1278 °C and 1,268 bar $O_2$, and 1347 °C and $10^4$ bar $O_2$ (Schmid 1983, Schramm et al. 2005, Kosenko and Emel'chenko 2001).

**Silver**. Table 22 and Figure 14d show a comparison of 50% condensation temperatures for $Ag_2O$ computed with ideal solution and two different activity coefficient models. Monatomic Ag gas is the major gas in all models. Sossi et al (2019) report $\gamma(AgO_{0.5})$ = 0.67 at 1673 K for solid silver oxide dissolved in a synthetic ferrobasalt melt. (Silver oxide decomposes prior to melting and there are no experimental data for $Ag_2O$ liquid.) The regular solution model in Table 2 based on this point agrees well with the ideal solution model but the second model assuming $Ag_2O$ has the same activity coefficient as $Cu_2O$ gives temperatures about 140 - 500 K lower than the other two models over the $10^{-6}$ - $10^{2}$ bar range. Prior work on silver solubility in melts in the $Ag_2O$ – $B_2O_3$ – $Na_2O$ – $Al_2O_3$ system (Willis and Hennessy 1953, Maekawa et al. 1969, Wakasugi et al. 1997) shows negative deviations from ideality for $Ag_2O$. The existence of $Ag_4SiO_4$ and other silver silicates (Klein and Jansen 2008) indicates negative deviations from ideality in the $Ag_2O$ – $SiO_2$ system. For now, all that can be said is that Ag is more volatile than Cu in BSE material and further measurements of the $Ag_2O$ activity coefficient are needed.

**Group 12 (Zn, Cd, Hg)**.

**Zinc**. Table 23 and Figure 15a show a comparison of 50% condensation temperatures for ZnO computed with ideal solution and the four different activity coefficient models listed in Table 2. The five models agree within ≤72 K. The FactSage and MAGMA code models agree with one another better than with the other two real solution models. This is plausibly due to using BSE composition for these two models versus a CAS eutectic melt (Reyes and Gaskell 1983) and a ferrobasalt melt (Sossi et al. 2019). But all models give 50% condensation temperatures within a few percent of each other. Figure 15a also shows 50% condensation temperatures calculated by Wang et al. (2019) and they agree well with the present work. Monatomic Zn is the major Zn-bearing gas in saturated BSE vapor over the $10^{-6}$ to $10^{2}$ bar range, in agreement with KEMS studies showing ≤ 0.1% ZnO in the saturated vapor over pure ZnO (Wriedt 1987a).

**Cadmium**. Table 24 and Figure 15b compare 50% condensation temperatures for CdO computed with ideal and real solution models. The real solution model has negative deviations from ideality and always lies above the ideal solution model. Sossi et al. (2019) report $\gamma$(CdO) = 0.15 at 1400 °C for CdO dissolved in a synthetic ferrobasalt melt. Assuming regular solution behavior, their datum yields log $\gamma$(CdO) = −1378.5/T. Two arguments support a negative deviation from ideality for CdO. The larger optical basicity (1.12 for CdO vs 0.92 for ZnO) and smaller ionic potential (2.10 for $Cd^{2+}$ vs 2.70 for $Zn^{2+}$) suggest CdO is a stronger base than ZnO with larger negative interaction parameters. Second, the $CdF_2$ – $CaF_2$ cigar-shaped phase diagram (O'Horo and White 1971) would have a different topology if $\gamma(CdF_2)$ was significantly different than $\gamma(CaF_2)$ and would look like those for the $CdF_2$ – $ZnF_2$ or $CdF_2$ – $PbF_2$ systems. Using the fluoride as a proxy for the oxide suggests CdO, like CaO, should display negative deviations from ideality in oxide melts. Monatomic Cd is the major Cd-bearing gas in agreement with vaporization studies done by KEMS and other methods (Wriedt 1987b).

**Mercury**. Mercury stays in the gas as Hg (g) and does not condense into BSE melt. Condensation temperatures were computed using extrapolations of $\log_{10}$ a(HgO) vs. 1/T from the higher temperature CONDOR code results. Mercury (II) oxide condenses at higher temperature than Hg and the temperatures for initial condensation of HgO ($a_{HgO}$ = 1) are:

| log $P_T$ | −6 | −4 | −2 | 0 | 2 |
|---|---|---|---|---|---|
| T/K | 339 | 393 | 468 | 579 | 758 |

Mercury is the most volatile element in BSE material and in solar composition material.

**Group 13** (B, Al, Ga, In, Tl). Aluminum was discussed earlier; see the results in Table 7 and Figures 4b and 6. Group 13 elements, except for thallium, are more stable as sesquioxides $M_2O_3$ than suboxides $M_2O$, and are acidic or amphoteric in oxide melts. Boric oxide $B_2O_3$ is a strong acid and lead borate ($2PbO \cdot B_2O_3$) is a solvent used for high-temperature molten oxide calorimetry (Navrotsky 1977). Optical basicity $\Lambda$ increases, acidity decreases, and cation ionic potential Z/r decreases down group 13; the values are

| | $B_2O_3$ | $Al_2O_3$ | $Ga_2O_3$ | $In_2O_3$ |
|---|---|---|---|---|
| $\Lambda$ | 0.42 | 0.60 | 0.71 | 1.25 |
| Z/r | 11.1 | 5.56 | 4.84 | 3.70 |

(Duffy 2004, 2005, Shannon 1976). This trend suggests larger negative interaction parameters and larger negative deviations from ideality down the group, which is seen from the activity coefficients of 0.28 – 0.37 (1573 – 1773 K) for $AlO_{1.5}$ and 0.02 (1923 K) for $InO_{1.5}$ in CMAS melts (Table 2, Sossi and Fegley 2018). However, the activity coefficients of Group 13 oxides in silicate melts are poorly known and need more study.

**Boron**. Table 25 and Figure 16a show 50% condensation temperatures for boron as $B_2O_3$ for real and ideal solution models. The black line in Figure 16a is computed with the FactSage activity coefficient in Table 2; the ideal solution condensation temperatures are much lower as seen in Figure 16a and Table 25. Boron gas chemistry is complex. The major gases are alkali metaborates ($MBO_2$ with M = Na, K, Rb, Cs) > $BO_2$ > BO > $B_2O_3$ >> B over the $10^{-6}$ to $10^2$ bar range. Lithium metaborate is not an important gas because lithium is fully condensed in the BSE melt at temperatures where boron is 50% condensed.

**Gallium**. Table 25 and Figure 16b show 50% condensation temperatures for $Ga_2O_3$ for real and ideal solution in BSE melt. The existence of gallium analogs of albite ($NaGaSiO_3$) and anorthite ($CaAlGaSi_2O_8$) and metal – silicate partitioning experiments indicate $Ga_2O_3$ displays negative deviations from ideality in silicate melts (Goldsmith 1950, Capobianco et al 1999) despite positive deviations from ideality in the $Ga_2O_3$ - $SiO_2$ binary (Glasser 1959). The logarithm of the $Ga_2O_3$ activity coefficient log $\gamma(Ga_2O_3)$ is interpolated from the logarithms of activity coefficient values for $Al_2O_3$ and $In_2O_3$ as a function of optical basicity (Table 2). The major Ga-bearing gases in saturated BSE vapor are Ga > GaO > $Ga_2O$ over the $10^{-6}$ to $10^2$ bar range, while $Ga_2O$ is the major Ga-bearing gas in the saturated vapor over pure $Ga_2O_3$ (Table 1, Sossi and Fegley 2018). The different speciation arises for the same reasons discussed above for lithium.

**Indium**. Table 25 and Figure 16c show 50% condensation temperatures for ideal and real solution of indium in BSE melt. CONDOR calculations used two real solution models; one is a regular solution fit to the 1923 K point of Wood and Wade (2013) and the other one sets the $In_2O_3$ activity coefficient equal to the FactSage value for $Al_2O_3$ (see Table 2). The differences between the 50% condensation temperatures from the two real solution models are 13 – 62 K while the ideal solution temperatures are 150 – 640 K lower than the real solution values. The major In-bearing gases in saturated BSE vapor are In > InO > $In_2O$ over the $10^{-6}$ to $10^2$ bar range. Several KEMS studies show $In_2O$ is the major In-bearing gas in the saturated vapor over pure $In_2O_3$ (Table 1, Sossi and Fegley 2018). The different indium gas speciation arises for the same reasons discussed above for lithium.

**Thallium**. Thallous oxide $Tl_2O$ is the major thallium – bearing oxide in the melt. This is consistent with thallium chemistry in two other ionic solvents – water and molten salts. Monovalent Tl chemistry is qualitatively different than that of the other Group 13 elements, which are trivalent, and its chemistry has similarities to alkali element chemistry (e.g., chapter 9 of Cotton and Wilkinson 1972). For example, boric oxide is a network former while $Tl_2O$ is a network modifier like $K_2O$ or $Ag_2O$. Aluminum, gallium, and indium oxides are all amphoteric and can serve as network formers or modifiers (Reddy et al. 2007). The molten salt chemistry of thallous halides is like that of alkali halides and silver halides (Lumsden 1966). The ionic potential of six-coordinate $Tl^{+1}$ is 0.67 and is closer to that of $Rb^{+1}$ (0.66) than to $K^{+1}$ (0.72) or $Ag^{+1}$ (0.87). Thus, the estimate $\gamma(Tl_2O) = \gamma(Rb_2O)$ was used in the condensation calculations. The ionic potential of $Tl^{3+}$ is closer to $Fe^{3+}$ than $Al^{3+}$ and $\gamma(Tl_2O_3) = \gamma(Fe_2O_3)$ was used in the calculations.

Table 25 and Figure 16d show the results for thallium. The ideal solution 50% condensation temperatures in Table 25 show Tl is the most volatile Group 13 element in BSE material over the $10^{-6}$ to $10^2$ bar range. The sequence from most volatile (lowest 50% $T_{cond}$) to most refractory (highest 50% $T_{cond}$) is Tl < B < In < Ga < Al. Table 25 shows effects of non-ideal solution on volatility; Tl is the most or 2nd most volatile element but the subsequent volatility sequence changes from Tl < B ≃ In < Ga < Al at $10^{-6}$ bar to In < Tl ≃ Ga < B ≃ Al at $10^2$ bar.

The major Tl-bearing gases in saturated BSE vapor are Tl > TlO > $Tl_2O$ over the $10^{-6}$ to $10^2$ bar range. Knudsen effusion, KEMS, and transpiration studies show $Tl_2O$ is the major Tl-bearing gas in the saturated vapor over pure $Tl_2O$, $Tl_4O_3$, and $Tl_2O_3$ (Table 1, Sossi and Fegley 2018). The different thallium gas speciation in saturated vapor above BSE melt and thallium oxides arises for the same reasons discussed above for lithium.

**Group 14 (C, Si, Ge, Sn, Pb)**. All group 14 elements except carbon, which is not discussed further, are in the dry BSE model. The results for silicon dissolution as $SiO_2$ in the BSE melt and Si gas chemistry in saturated BSE vapor are in Table 6 and Figures 4d and 7.

Silicon, Ge, Sn, and Pb all form stable dioxide solids and Pb also forms a stable monoxide solid. Brewer and Green (1957) showed SiO solid is unstable at all temperatures. Brewer and Zavitsanos (1957) studied the Ge – $GeO_2$ phase diagram by differential thermal analysis and concluded "it is of the simple eutectic type with a miscibility gap for the liquids". They found no evidence for the existence of GeO solid. The KEMS study of Drowart et al (1965b) shows GeO solid is metastable and disproportionates to solid $GeO_2$ + Ge. Platteeuw and Meyer (1956) showed SnO solid is unstable and decomposes into Sn + $SnO_2$ at and above 573 K. Lead monoxide solid is stable at high temperature (Drowart et al., 1965a).

Knudsen effusion mass spectrometry studies show the stable dioxides of Si, Ge, and Sn evaporate incongruently to a mixture of monoxide, dioxide, and monatomic gas with a bulk oxygen/metal ratio of two in the saturated vapor over the pure solid or liquid oxide (Table 1, Sossi and Fegley 2018). Thermogravimetric and KEMS studies show $PbO_2$ decomposes upon heating to lower lead oxides $PbO_x$ (x <2) plus $O_2$ (gas), ultimately forming PbO, which evaporates incongruently to a mixture of PbO, Pb, and $(PbO)_{2-6}$ gases with a bulk Pb/O ratio of unity (Drowart et al. 1965, Otto 1966).

The calculations show monoxides are the major gases for Si, Ge, and Sn in saturated BSE vapor and that the dioxide and monatomic gases are less abundant along the 50% condensation curves. The major Ge-bearing gases are GeO > $GeO_2$ > Ge > $Ge_2O_2$ over the $10^{-6}$ to $10^2$ bar total pressure range. The major Sn-bearing gases are SnO > Sn > $SnO_2$ over the same total pressure range. Lead chemistry is different. The major Pb-bearing gases in saturated BSE vapor are Pb ≳ PbO >> $PbO_2$ over the $10^{-6}$ to $10^2$ bar total pressure range, with Pb a few times more abundant than PbO, which is never the major gas. Table 26 summarizes gas chemistry at $10^{-4}$ bar total pressure for Si, Ge, Sn, and Pb at the temperatures where each element is 50% condensed. The gaseous M/MO, $M/MO_2$, and $MO/MO_2$ molar ratios increase downward in Group 14 from Si to Ge to Sn to Pb. The monatomic and monoxide gases become more important and the dioxide gases less important going down Group 14. Except for PbO (liquid), where PbO polymer gases are important (Drowart et al. 1965), the major gases in saturated BSE vapor are the same as those in saturated vapor over the pure Group 14 oxides.

**Germanium**. Table 27 and Figure 17a show the results for Ge chemistry in BSE material. Literature reports indicate germanium dissolves in CMAS, CMS-FeO, and CAS-FeO silicate melts as $GeO_2$ and/or GeO (Kegler and Holzheid 2011 Mare et al. 2020, Yan and Swinbourne 2003, Shuva et al. 2016). There are no thermodynamic data for formation of GeO (solid, liquid) from its constituent elements in any of the standard data compilations, so only $Ge^{4+}$ ($GeO_2$) was considered. FactSage calculations give $\gamma(GeO_2)$ = 7.4 independent of temperature for the six-component BSE melt, but Sossi et al. (2019) found negative deviations from ideality in a ferrobasalt melt (see Table 2). Magnesia and $SiO_2$ are the two major oxides in BSE melt (Table 1) and the $GeO_2$ – MgO phase diagram (Robbins and Levin 1959) and activity measurements in $GeO_2$ – $SiO_2$ melts (Yoshikawa et al 2004) show negative deviations from ideality at low $GeO_2$ concentrations. Unless there are large positive interaction parameters of $GeO_2$ with the less abundant oxides (FeO, CaO, $Al_2O_3$, $Na_2O$, etc.) in BSE melt negative deviations from ideality are expected for $GeO_2$ dissolved in BSE melt. Table 27 shows the maximum ΔT between the five models ranges from 41 K at $10^{-6}$ bar total pressure to 155 K at 100 bar total pressure. This is about a 3.3% difference across the entire pressure range.

**Tin**. Table 28 and Figure 17b show the results for tin chemistry in BSE material. The 50% condensation temperatures are for dissolution of liquid SnO + $SnO_2$ in BSE melt. Gurvich et al (1991) give thermodynamic data for solid and liquid SnO and $SnO_2$ in the IVTAN database and a discussion of their thermodynamic data assessment and choices. The real solution model uses FactSage values for $\gamma$(SnO) and sets $\gamma(SnO_2) = \gamma(GeO_2)$ from FactSage. Figure 17b shows good agreement between the present work and results from Wang et al. (2019). Carbo-Nover and Richardson (1972) reported ideal solution for SnO in SnO - $SiO_2$ melts (with < 1% $SnO_2$) at 1273 - 1523 K. However, Kozuka et al. (1968) and Grau and Flengas (1976) found $\gamma$(SnO) > 1 in the same system.

**Lead**. Table 28 and Figure 17c show the results for lead chemistry in BSE material. Table 2 lists the activity coefficients for the four real solution models used. Two of these are from Sossi et al. (2019) who assumed either PbO (liquid) vaporization to PbO (gas)

$$\text{PbO (liquid)} = \text{PbO (gas)}$$

or the dissociative congruent vaporization of PbO (liquid) via

$$\text{PbO (liquid)} = \text{Pb (gas)} + \tfrac{1}{2}\,\text{O}_2\text{ (gas)}$$

depending upon the oxygen fugacity in their experiments. Our regular solution fits to their derived activity coefficients at 1673 K are in Table 2. The other real solution models are from FactSage and Wood and Wade (2013).

The condensation temperatures from the four real solution models agree within 8 K over the $10^{-6}$ to $10^{2}$ bar range. The ideal solution condensation temperatures are lower by 10 – 15 K over the same total pressure range. This difference occurs because the real solution models predict PbO has negative deviations from ideality in the BSE melt. Table 28 shows similar condensation temperatures for PbO and SnO + $SnO_2$ dissolved in the BSE melt. Lead is more volatile than tin at lower total pressure but the two are equally volatile at higher total pressure.

**Group 15 (N, P, As, Sb, Bi)**. All group 15 elements except nitrogen are in the dry BSE model. Optical basicities of group 15 oxides vary from strongly acidic to basic down the group: 0.27 ($N_2O_5$), 0.33 ($P_2O_5$), 1.02 ($As_2O_3$), 1.14 ($Sb_2O_3$), and 1.19 ($Bi_2O_3$). The optical basicities suggest less strongly negative deviations from ideality down Group 15 and similar activity coefficients for Sb and Bi oxides.

**Phosphorus**. Table 29 and Figure 18a give the results for three different real solution models of $P_2O_5$ condensation in BSE melt and show the result from Ivanov et al. (2022) for comparison. Table 2 gives the activity coefficient equation for the Ban-Ya (1993) model. The equation for $\gamma(P_2O_5)$ from Suito et al (1981) is

$$log_{10}\gamma_{P_2O_5}(Suito) = -1.01\left(23X_{CaO} + 17X_{MgO} + 8X_{FeO}\right) - \frac{26{,}270}{T} + 11.24$$

Using the BSE melt composition in Table 1 gives the equation listed in Table 2. The differences between calculated 50% condensation temperatures are about 200 K at $10^{-6}$ bar and $10^{2}$ bar but smaller at the intermediate pressures. No ideal solution calculations were done for $P_2O_5$ condensation it displays large negative deviations from ideality in silicate melts.

Phosphorus oxide gas chemistry is complex. Knudsen effusion mass spectrometry shows $P_4O_{10}$ is the major gas and $P_4O_9$ the minor gas in the saturated vapor over commercial “phosphorus pentoxide” at ≤150 °C (Muenow et al. 1970). [Commercial “phosphorus pentoxide” is the metastable hexagonal polymorph, which is composed of $P_4O_{10}$ molecules (Jung and Hudon 2012)]. Muenow et al (1970) found increasing amounts of other phosphorus oxide gases in the saturated vapor with increasing temperature, e.g., $P_4O_{6\text{-}10}$, $P_3O_{4\text{-}7}$, $P_2O_{2\text{-}5}$, and $PO_{1\text{-}3}$. Kazenas and Petrov (1997) confirmed their results. In contrast, the two major P-bearing gases in saturated BSE vapor are $PO_2 \gtrsim PO$ with less abundant gases being $P_2O_{3\text{-}5} \gtrsim P > P_4O_6$. Markova et al. (1986) state that PO and $PO_2$ are the phosphorus gases in the saturated vapor over molten lunar basalts. The different phosphorus gas speciation in saturated vapor above BSE melt and “phosphorus pentoxide” arises for the same reasons discussed above for lithium.

**Arsenic**. Tables 29 and 30 and Figure 18b give the results for $As_2O_3$ condensation in BSE melt. Arsenic condensation was computed as $As_2O_3$ ($AsO_{1.5}$) dissolution in molten silicate and used activity coefficients from FactSage and regular solution fits to the low ($\gamma$ = 0.03) and high ($\gamma$ = 0.56) values at 1573 K for $AsO_{1.5}$ dissolved in $FeO_x$ - CaO - $SiO_2$ copper smelting slags (Chen and Jahanshahi 2010). The equations for the three different real solution models for $As_2O_3$ activity coefficients are in Table 2.

Figure 18b shows the two curves based on Chen and Jahanshahi (2010) lie at much lower temperature than the FactSage curve. This difference could be due to the difference between activity coefficients in BSE melt and the copper smelting slags. The FactSage condensation curve was used in this work. The choice (of $As^{3+}$) is consistent with literature studies, which conclude trivalent As dissolves in molten silicates (Canali et al. 2017, Chen and Jahanshahi 2010, Hidayat et al. 2021, Maciag and Brenan 2020, Shuva et al. 2016b).

Arsenic gas chemistry is complex and Knudsen effusion mass spectrometry (Brittain et al. 1982, Kazenas and Petrov 1997) shows $As_4O_6$ is the major gas in the saturated vapor over pure arsenolite (octahedral $As_2O_3$). In contrast, the major As-bearing gases in saturated BSE vapor are AsO > As with $As_2$ > $As_4$ > $As_4O_6$ much less abundant. The different arsenic gas speciation in saturated vapor above BSE melt and octahedral $As_2O_3$ arises for the same reasons discussed above for lithium.

**Antimony**. Tables 29 and 30 and Figure 18c show 50% condensation temperatures for $Sb_2O_3$ dissolution in BSE melt. Maciag and Brenan (2020) conclude antimony is $Sb^{3+}$ in basaltic melts. Activity coefficient data in Shuva et al. (2016) show negative deviations from ideality increase from $Bi_2O_3$ to $Sb_2O_3$ to $As_2O_3$ at constant T, $fO_2$, and melt composition in $FeO_x$ - CaO - $SiO_2$ copper smelting slags. Whether this trend applies to mafic and ultramafic melts is unknown. The existence of stable Mg and Ca antimonides such as $MgSb_2O_4$ and $CaSb_2O_4$ shows negative deviations from ideality in the MgO – $Sb_2O_3$ and CaO – $Sb_2O_3$ systems (Katayama et al. 1987, 1998). Negative deviations from ideality are assumed for $Sb_2O_3$ dissolution in BSE melt and $\gamma(Sb_2O_3) = \gamma(As_2O_3)$ is estimated in the absence of other data.

Knudsen effusion (Behrens and Rosenblatt 1973) and KEMS (Semenov et al. 1983) show $Sb_4O_6$ is the major gas in the saturated vapor over subliming orthorhombic $Sb_2O_3$ (valentinite). In contrast, the major Sb-bearing gases in saturated BSE vapor are SbO > Sb followed by $Sb_2$ > $Sb_4O_6$ > $Sb_4$ along the Sb 50% condensation curve. The different antimony gas speciation in saturated vapor above BSE melt and octahedral $Sb_2O_3$ arises for the same reasons discussed above for lithium.

**Bismuth**. Tables 29 and 30 and Figure 18d show $Bi_2O_3$ condensation in BSE melt. Three sets of calculations were done: (1) ideal solution, (2) estimating $\gamma(Bi_2O_3) = \gamma(As_2O_3)$ from FactSage, and (3) a regular solution fit to $\gamma(BiO_{1.5}) = 0.4$ at 1573 K in $FeO_x$ - CaO - $SiO_2$ copper smelting slags (Paulina et al. 2013), which is given in Table 2.

Figure 18d shows the large differences between 50% condensation temperatures between the three sets of calculations. There are no intermediate compounds in the MgO – $Bi_2O_3$ system, but the binary systems of $Bi_2O_3$ with $Al_2O_3$, $Na_2O$, and $TiO_2$ contain intermediate compounds (Kargin et al 2008, Jiang et al 2021). Consequently, the activity coefficient equation based on $Bi_2O_3$ dissolution in $FeO_x$ - CaO - $SiO_2$ copper smelting slags (Paulina et al. 2013), may underestimate the negative deviations from ideality for $Bi_2O_3$ dissolved in BSE melt. Based on Shuva et al. (2016) cited above, the "true" $Bi_2O_3$ activity coefficient and 50% condensation temperature should lie between the two sets of curves in Figure 18d.

The major Bi-bearing gases in saturated BSE vapor are Bi >> BiO, which are also the major gases in saturated vapor over $Bi_2O_3$ (Table 1, Sossi and Fegley 2018).

**Groups 16 - 18 (Chalcogens, halogens, noble gases).** Oxygen is the only Group 16 – 18 element in the dry BSE model. Table 31 gives 50% condensation temperatures for oxygen computed using the nominal activity coefficients for oxides in the BSE melt and a comparison to the 50% oxygen condensation temperature from Ivanov et al. (2022) at one bar total pressure. Sulfur, selenium, and tellurium are in the wet BSE model and their condensation chemistry is complicated. For example, the dissolved form of sulfur depends upon oxygen fugacity with sulfide favored at lower $fO_2$ and sulfate favored at higher $fO_2$ (Figure 12 and associated discussion in Zolotov and Fegley 1999). Ivanov et al. (2022) list 2983 K for the 50% condensation temperature of sulfur in BSE melt at one bar total pressure but do not provide details necessary to evaluate their result. The dissolution of selenium and tellurium is also $fO_2$ dependent with reduced selenium and tellurium species favored at lower $fO_2$ and oxidized selenium and tellurium species favored at higher oxygen fugacity. However, the solubilities and activity coefficients of chalcogens in BSE melt are not known and must be estimated using Temkin mixing or another method. The solubilities of halogens in BSE melt are not known and need to be estimated from existing solubility data in basaltic melt at lower temperatures. This work is beyond the scope of this paper.

**Actinides (Th, U, Pu)**. Table 15 and Figure 13b give 50% condensation temperatures for ideal solution of $ThO_2$, $UO_2$, and $PuO_2$ in BSE melt. Uranium is most volatile, and thorium is the most refractory. The major uranium-bearing gases are $UO_3$ > $UO_2$ > UO >> U with $UO_2$ (g) becoming more important with increasing pressure. The major thorium-bearing gases are $ThO_2$ > ThO > Th with $ThO_2$ always dominant. Plutonium gas chemistry is unique. The major Pu-bearing gases are $PuO_2$ > PuO > Pu > $Pu^+$ from $10^{-6}$ to $10^2$ bar total pressure.

**Thermal ionization**. Thermal ionization is included in the chemical equilibrium calculations as discussed earlier. Figure 19 shows logarithmic contours of the electron partial pressure over 1500 – 4500 K and $10^{-6}$ to $10^2$ bar total pressure. Thermal ionization increases electrical conductivity and magnetic fields in the proto-lunar disk, which drive turbulence and bipolar outflows that remove mass from the Earth – Moon system (Gammie et al. 2016). Cesium has the lowest ionization potential (3.89 electron volts) of any naturally occurring element, but calcium is significantly more abundant with a low ionization potential of 6.11 eV making $Ca^+$ the major cation and major electron source in a proto-lunar disk composed of silicate vapor and melt.

## 4. Discussion

**Comparison of condensation temperatures in silicate and solar composition material.** Calculated 50% condensation temperatures for the $10^{-6}$ to $10^2$ bar total pressure range are in the preceding tables and figures. A summary at $10^{-4}$ bar total pressure is in Table 32. Comparison of Table 32 with Table 8 of Lodders (2003) shows condensation temperatures in BSE material are significantly higher than those in solar composition material. Even mercury condenses about 140 K higher in silicate than in solar composition material. Two factors are responsible for the large differences.

The first factor is the absence in dry BSE material of hydrogen and helium, which dilute the partial pressures of rocky element gases in solar composition material. In other words, the metallicity of silicate vapor is significantly higher than that of solar composition material. The hydrogen and helium abundances in wet BSE material are orders of magnitude lower than in solar composition material and the same argument holds.

The second factor is the significantly higher oxygen fugacity of silicate material. This follows from Goldschmidt's (1928) observation that oxygen is the most abundant element in silicates (on an atomic basis). Figure 1 of Fegley et al. (2020) shows the large difference between the $fO_2$ of silica vapor, BSE vapor, and solar composition gas as a function of temperature. The reader can make a similar quantitative comparison using Figure 3a (dry BSE) and Figure 3b (wet BSE) and the equation below, which gives the temperature dependent oxygen fugacity ($fO_2$) of solar composition gas from 300 – 2500 K, which is controlled by reaction of hydrogen with water vapor (p. 33, Lodders and Fegley 2011):

$$2H_2 + O_2 = 2H_2O$$

$$log_{10} f_{O_2}(solar) = 2log\left(\frac{X_{H_2O}}{X_{H_2}}\right) + 5.67 - \frac{25{,}664}{T}$$

Previous chemical equilibrium calculations by Schaefer and Fegley (2010) for H chondritic material give the same conclusion, significantly higher condensation temperatures in H chondritic silicate than in solar composition material (see their Table 4 and Figures 3-5). We do not belabor this point further and refer the reader to prior papers making the same point (Fegley and Cameron 1987, O'Neill 1991, Visscher and Fegley 2013, Canup et al. 2015, Lock et al. 2018, Sossi and Fegley 2018, Sossi et al 2019, Fegley et al. 2020).

**Condensation Coefficients and Non-Equilibrium Condensation (Evaporation)**. At thermodynamic equilibrium condensation and evaporation temperatures are identical. However, not all vapor molecules that strike the surface of a condensed phase (liquid or solid) will equilibrate with the surface and condense (Searcy 1970). The fraction of molecules that strike the surface and condense is defined as the condensation coefficient, $\alpha_c$. Similarly, a vaporization coefficient, $\alpha_v$, is defined as the ratio of measured vapor flux to the flux calculated from the equilibrium pressure. Searcy (1970) notes that in non-equilibrium vaporization the condensation and vaporization coefficients are not strictly equal. In practice, they are probably close.

Measurements of vaporization and condensation coefficients are limited. Most measurements are for vaporization coefficients. They have been reported for some solid and liquid oxides and range from 1 to $10^{-2}$. Schaefer and Fegley (2004) gave a detailed discussion with their Table 10 and we briefly recapitulate some of their major points. Condensation (vaporization) coefficients are less than unity for solid oxides, e.g., Jacobson et al. (2017) report $\alpha_v = (2\rightarrow5) \times 10^{-3}$ for $SiO_2$ (solid) from 1800 - 1950 K. This is about a factor of 2-3 lower than $\alpha_v = (11\text{-}15) \times 10^{-3}$ found by Hashimoto (1990); similar differences occur for $\alpha_v$ values for $\alpha$-$Al_2O_3$ (corundum) listed in Table 10 of Schaefer and Fegley (2004). Burns (1966) found $\alpha_v = 1$ for liquid $Al_2O_3$, $Ga_2O_3$, and $In_2O_3$ at their melting points. However, Hashimoto (1990) reported vaporization coefficients less than unity for liquid silica and liquid $Mg_2SiO_4$. Shornikov (2015) estimated the following values for these solid rock-forming oxides: 0.66±0.15 (CaO), 0.50±0.20 (MgO), 0.33±0.02 ($Al_2O_3$), 0.23±0.02 ($TiO_2$), $(2.2\pm0.3) \times 10^{-2}$ ($SiO_2$), and values of unity for all the liquid oxides.

More experimental data is needed to confirm or refute values of unity for all liquid oxides. The preceding discussion is only for pure solid compounds.

There are limited measurements of vaporization coefficients for specific components of complex melts (e.g., Richter et al. 2002, 2007). These are determined by taking the measured vapor flux from a free surface and comparing it to the calculated equilibrium flux using solution models (e.g., Berman 1983, Berman and Brown 1984). Richter et al. (2002) report vaporization coefficients from CAI-like melts for Mg and SiO ranging from 0.1 to 0.2. There is some correlation with increasing MgO content leading to larger vaporization coefficients. In their calculation of evaporation of forsterite, Tsuchiyama et al. (1999) use a vaporization coefficient of 0.1. The value of the vaporization coefficient for a particular species such as Mg or SiO depends upon the calculated activity coefficient for the parent species in the melt (e.g., MgO, $SiO_2$). As shown in Table 2 and in figures throughout this paper, different melt models give different values for activity coefficients and thus different equilibrium flux values. The "true" values of vaporization coefficients from multicomponent melts are not yet known.

If condensation coefficients for oxide dissolution in BSE melt are less than unity the condensing flux would be reduced, and the "true" 50% condensation temperature would be higher. In the absence of constraints on condensation coefficients we assume they are unity, as done by other authors (Canup et al. 2015, Lock et al. 2018, Wang et al. 2019, Ivanov et al. 2022).

## 5. Summary

The gas – melt chemistry of bulk silicate Earth (BSE) material was modeled using ideal gas, non-ideal melt chemical equilibrium calculations over the 1000 – 4500 K temperature range and $10^{-6}$ to $10^{2}$ bar pressure range. The virial equation of state was used to verify the near ideality of silicate vapor (within 2%) in this pressure – temperature range. The effects of different activity coefficients and ideal solution upon the results were studied. The results presented are 50% condensation temperatures, major gases of each element, and the pressure and temperature dependent oxygen fugacity ($fO_2$) of dry and wet BSE material. The dry BSE model has no water because it excludes hydrogen; it also excludes the volatile elements (C, N, F, Cl, Br, I, S, Se, Te). The wet BSE model has water because it includes hydrogen; it also includes the other volatiles. Key conclusions include the following: (1) much higher condensation temperatures in silicate vapor than in solar composition gas at the same total pressure because of the higher total metallicity and higher oxygen fugacity of silicate vapor, (2) a different condensation sequence in silicate vapor than in solar composition gas, (3) good agreement between different activity coefficient models except for the alkali elements, which show the largest differences between models, (4) agreement, where overlap exists, with published silicate vapor condensation calculations (e.g., Canup et al. 2015, Lock et al. 2018, Wang et al. 2019), (5) condensation of Mo and W oxides instead of metals, (6) a stability field for Ni metal as reported by Lock et al. (2018), (7) agreement between ideal solution (from this work and from Lock et al. 2018) and real solution condensation temperatures for elements with minor deviations from ideality in the oxide melt, (8) similar 50% condensation temperatures, within a few degrees, in the dry and wet BSE models for Al, Ca, Co, Cr, Fe, Li, Mg, Mn, Si, Ti, V, and (9) much lower 50% condensation temperatures for elements such as B, Cu, K, Na, Pb, Rb, which form halide, hydroxide, sulfide, selenide, telluride and oxyhalide gases. The latter results are preliminary because solubilities and activities of volatile elements in silicate melts are not well known, but must be used for the correct equilibrium distribution, 50% condensation temperatures and mass balance of halide (F, Cl, Br, I), hydrogen, sulfur, selenium and tellurium bearing species between silicate melt and vapor.

**Acknowledgments**

Even though we chose a different journal for publication, we thank Sasha Krot for contacting us in October 2020 about authoring a paper in Meteoritics and Planetary Science (MAPS) in memory of John Wasson, whom BF and KL knew for many years. John Wasson was a practitioner of condensation calculations and his study of moderately volatile element (MVE) condensation in the solar nebula (Wai and Wasson 1977) strongly motivated one of us to follow suit (Fegley and Lewis 1980). We thank Misha Petaev for sharing his 50% condensation temperatures with us, helpful discussions, and his formal review. We also thank an anonymous referee, Donna Knaster-Duvall, Astrid Holzheid, Oleg Yakovlev, Herbert Palme, Sara Port, Paolo Sossi, Channon Visscher, and Zhe Zhang for helpful comments. BF and KL thank the staff of the Interlibrary Loan Service of the Washington Universities Libraries for their superb efforts in obtaining references for us. The McDonnell Center for the Space Sciences and NSF Astronomy Program Grant AST-1517541 supported the work at Washington University.

**Figure Captions**

Figure 1. The condensation curve for molten BSE material as a function of total pressure and a comparison of results from the CONDOR, FactSage and GRAINS codes. The curve shows where melt becomes stable; only vapor is present to the left above the curve. A mixture of vapor plus melt is present on and below the curve to the right. Figure 2 illustrates the composition of the melt as a function of total pressure.

Figure 2. The composition of the initially condensed melt as a function of total pressure using the nominal activity coefficients in Table 2. Note the composition axis is logarithmic. Figure 1 gives the melt condensation curve. See the text and Table 5 for a comparison of initial melt compositions from the CONDOR and FactSage codes.

Figure 3. Oxygen equilibrium chemistry in dry (a) and wet (b) BSE material as a function of temperature and total pressure. The curves are logarithmic contours of constant oxygen fugacity ($fO_2$). The dark green curves show the 50%, 99%, and 99.99% (3a only) condensation curves for oxygen. The CONDOR code computes the $O_2$ fugacity; there is no separate $fO_2$ buffer. Kinks in the contours in this figure and subsequent figures are due to size of the P and T steps in the grid. See the text for a detailed discussion of this figure.

Figure 4. Calcium (a), aluminum (b), magnesium (c), and silicon (d) equilibrium chemistry in BSE material as a function of temperature and total pressure. The black curves are the 50% condensation curves. Calcium and aluminum are >50% condensed where melt forms. See the text for a detailed discussion of this figure.

Figure 5. The distribution of calcium (a) and magnesium (b) between monatomic and monoxide gases as a function of temperature and total pressure. The logarithmic contours give the CaO/Ca (5a) and Mg/MgO (5b) molar ratios. The red colored 50% condensation curves for calcium and magnesium are superimposed on the contours.

Figure 6. The distribution of aluminum between monatomic and oxide gases as a function of temperature and total pressure. The red colored 50% condensation curve for aluminum is superimposed on the logarithmic molar contours. Molar ratios of AlO/Al are in (a) and molar ratios of $AlO/AlO_2$ are in (b).

Figure 7. The distribution of silicon between monatomic and oxide gases as a function of temperature and total pressure. The red colored 50% condensation curve for silicon is superimposed on the logarithmic molar contours. Molar ratios of $SiO/SiO_2$ and of SiO/Si are in (a) and (b), respectively.

Figure 8. (a) Iron equilibrium chemistry in BSE material as a function of temperature and total pressure. The black curve is the 50% condensation curve for FeO. (b) The distribution of iron between Fe and FeO gases as a function of temperature and total pressure. The logarithmic contours give the Fe/FeO molar ratios. The red colored 50% condensation curve for iron is superimposed on the contours.

Figure 9. (a) Titanium equilibrium chemistry in BSE material as a function of temperature and total pressure. The black curve is the 50% condensation curve for $TiO_2$. (b) The distribution of titanium between TiO and $TiO_2$ gases as a function of temperature and total pressure. The logarithmic contours give the $TiO_2/TiO$ molar ratios. The red colored 50% condensation curve for titanium is superimposed on the contours. The other Ti-bearing gases included in the calculations are less important (Ti, $Ti^+$, $Ti^{2+}$).

Figure 10. (a) Manganese equilibrium chemistry in BSE material as a function of temperature and total pressure. The black curve is the 50% condensation curve for MnO. (b) The distribution of manganese between Mn and MnO gases as a function of temperature and total pressure. The logarithmic contours give the Mn/MnO molar ratios. The red colored 50% condensation curve for MnO is superimposed on the contours.

Figure 11. (a) Chromium equilibrium chemistry in BSE material as a function of temperature and total pressure. The black curve is the 50% condensation curve for chromium. (b) Lithium equilibrium chemistry in BSE material as a function of temperature and total pressure. The black curve is the 50% condensation curve for lithium.

Figure 12. Equilibrium chemistry of sodium (a), potassium (b), rubidium (c), and cesium (d) in BSE material as a function of temperature and total pressure. The black curves are the 50% condensation curves. See the text for a detailed discussion of this figure.

Figure 13. (a) Equilibrium chemistry of zirconium and hafnium (a) and actinides (b) in BSE material as a function of temperature and total pressure. The black curves are the ideal solution condensation curves. See the text for a detailed discussion of this figure.

Figure 14. Equilibrium chemistry of cobalt (a), nickel (b), copper (c), and silver (d) in BSE material as a function of temperature and total pressure. The black curves are the 50% condensation curves. See the text for a detailed discussion of this figure.

Figure 15. Equilibrium chemistry of zinc (a) and cadmium (b) in BSE material as a function of temperature and total pressure. The black curves are the 50% condensation curves.

Figure 16. Equilibrium chemistry of boron (a), gallium (b), indium (c), and thallium (d) in BSE material as a function of temperature and total pressure. The black curves are the 50% condensation curves. Figure 4b shows the same for aluminum.

Figure 17. Equilibrium chemistry of germanium (a), tin (b), and lead (c) in BSE material as a function of temperature and total pressure. The black curves are the 50% condensation curves. Silicon equilibrium chemistry is in Figure 4d.

Figure 18. Equilibrium chemistry of phosphorus (a), arsenic (b), antimony (c), and bismuth (d) in BSE material as a function of temperature and total pressure. The black curves are the 50% condensation curves.

Figure 19. Logarithmic contours of the electron partial pressure in equilibrium with BSE material as a function of temperature and total pressure. The kinks in the contours are due to size of the P and T steps in the grid.

Table 1. Composition model for the Bulk Silicate Earth (BSE)[a,b]

| Oxide | Mol% | Oxide | Mol% |
|---|---|---|---|
| **MgO** | **47.67** | $Rb_2O$ | $1.8 \times 10^{-5}$ |
| **$SiO_2$** | **39.48** | $La_2O_3$ | $1.3 \times 10^{-5}$ |
| **FeO** | **5.90** | $Dy_2O_3$ | $1.2 \times 10^{-5}$ |
| **CaO** | **3.40** | $Gd_2O_3$ | $9.7 \times 10^{-6}$ |
| **$Al_2O_3$** | **2.30** | $HfO_2$ | $8.8 \times 10^{-6}$ |
| **$Na_2O$** | **0.29** | $Sm_2O_3$ | $7.5 \times 10^{-6}$ |
| NiO | 0.17 | $Er_2O_3$ | $7.3 \times 10^{-6}$ |
| $TiO_2$ | 0.14 | $Yb_2O_3$ | $7.2 \times 10^{-6}$ |
| $Cr_2O_3$ | 0.13 | SnO | $6.2 \times 10^{-6}$ |
| MnO | 0.10 | $Pr_2O_3$ | $4.9 \times 10^{-6}$ |
| $K_2O$ | $1.74 \times 10^{-2}$ | PbO | $4.7 \times 10^{-6}$ |
| CoO | $9.1 \times 10^{-3}$ | $Eu_2O_3$ | $2.9 \times 10^{-6}$ |
| $P_2O_5$ | $7.3 \times 10^{-3}$ | $MoO_3$ | $2.6 \times 10^{-6}$ |
| $V_2O_3$ | $4.4 \times 10^{-3}$ | $Ho_2O_3$ | $2.5 \times 10^{-6}$ |
| ZnO | $4.3 \times 10^{-3}$ | $As_2O_3$ | $2.4 \times 10^{-6}$ |
| SrO | $1.3 \times 10^{-3}$ | $ThO_2$ | $1.9 \times 10^{-6}$ |
| $Sc_2O_3$ | $9.6 \times 10^{-4}$ | $Tb_2O_3$ | $1.8 \times 10^{-6}$ |
| $Cu_2O$ | $8.2 \times 10^{-4}$ | CdO | $1.6 \times 10^{-6}$ |
| $Li_2O$ | $6.0 \times 10^{-4}$ | $Tm_2O_3$ | $1.1 \times 10^{-6}$ |
| $ZrO_2$ | $5.9 \times 10^{-4}$ | $Lu_2O_3$ | $1.1 \times 10^{-6}$ |
| BaO | $2.6 \times 10^{-4}$ | $Ta_2O_5$ | $6.2 \times 10^{-7}$ |
| $Ga_2O_3$ | $1.6 \times 10^{-4}$ | $UO_2$ | $5.0 \times 10^{-7}$ |
| $Y_2O_3$ | $1.2 \times 10^{-4}$ | $In_2O_3$ | $4.1 \times 10^{-7}$ |
| $B_2O_3$ | $6.3 \times 10^{-5}$ | $Cs_2O$ | $3.5 \times 10^{-7}$ |
| $GeO_2$ | $8.6 \times 10^{-5}$ | $WO_3$ | $3.4 \times 10^{-7}$ |
| BeO | $3.6 \times 10^{-5}$ | $Ag_2O$ | $1.5 \times 10^{-7}$ |
| $NbO_2$ | $3.3 \times 10^{-5}$ | $Sb_2O_3$ | $1.2 \times 10^{-7}$ |
| $Ce_2O_3$ | $3.3 \times 10^{-5}$ | $Tl_2O$ | $5.2 \times 10^{-8}$ |
| $Nd_2O_3$ | $2.4 \times 10^{-5}$ | $Bi_2O_3$ | $3.7 \times 10^{-8}$ |

[a]Calculated from Palme and O'Neill (2014).

[b]The molar percentages of volatiles in the wet BSE model are 0.31 $H_2O$, 0.044 $CO_2$, 0.0326 S, $6.9 \times 10^{-3}$ F, $4.4 \times 10^{-3}$ Cl, $4.9 \times 10^{-6}$ Br, $3.0 \times 10^{-7}$ I ,$7.5 \times 10^{-4}$ N, $5.0 \times 10^{-6}$ Se, $4.0 \times 10^{-7}$ Te, and $1.6 \times 10^{-6}$ Hg.

Table 2. Activity coefficients ($\log_{10} \gamma = A + B/T$)

| Oxide | A | B | Notes |
|---|---|---|---|
| $Ag_2O$ | 0 | −582 | Regular solution 1673 K point, Sossi et al. (2019) |
| | 0 | 3582.8 | Set = $\gamma(Cu_2O)$, Sossi et al. (2019) |
| $Al_2O_3$ | 0 | −6253.138 | Ban-Ya (1993) |
| | −0.938 | 720.9 | MAGMA code |
| | −12.515 | 19,111.25 | FactSage |
| $As_2O_3$ | 3.4014 | −26309.08 | FactSage 1800−2200 K, CMAS+FeO |
| | 0 | −4791 | Chen and Jahanshahi (2010), $\gamma$ ($AsO_{1.5}$) = 0.03 at 1573 K |
| | 0 | −396.2 | Chen and Jahanshahi (2010), $\gamma$ ($AsO_{1.5}$) = 0.56 at 1573 K |
| $B_2O_3$ | −7.1517 | 508.158 | FactSage 1800−2200 K, CMAS+FeO |
| $Bi_2O_3$ | 3.4014 | −26309.08 | Set = $\gamma(As_2O_3)$ from FactSage |
| | 0 | −1251.8 | Paulina et al. (2013), regular solution, 1573 K point |
| CaO | −0.246 | −6710.736 | Ban-Ya (1993) |
| | −0.144 | −4656.5 | MAGMA code |
| | 1.30 | −7014.6 | FactSage |
| CdO | −1.23 | 803.1 | Regular solution, liquid standard state, 1473 K point, Sossi et al. (2019) |
| CoO | 0.303 | 0 | FactSage BSE melt, 2000−3000 K mean $\gamma$ = 2.01±0.07 |
| CrO | 0.165 | 601.518 | FactSage |
| $Cr_2O_3$ | −2.318 | 4413.29 | FactSage |
| $Cs_2O$ | −0.083 | −13634.1 | See text, set = $\gamma(Na_2O)$ FactSage |
| | 0 | −12144. | $Cs_2O$-$Na_2O$-$SiO_2$ ternary, Bennour (1999), see text |
| CuO | 4 | 0 | Set = 10,000, see text |
| $Cu_2O$ | 0 | 373.261 | Altman (1978) |
| | 0 | 3144.1 | Regular solution, 1573 K point, Holzheid and Lodders (2001) |
| | 0 | 3582.8 | Regular solution, 1673 K point, Sossi et al. (2019) |
| | 0 | 2094 | Regular solution, 1923 K point, Wood and Wade (2013) |
| FeO | 0.373 | 399.1 | Ban-Ya (1993) |
| | 0.025 | −141.8 | MAGMA code |
| | 0.282 | 49.7 | FactSage |
| $Fe_2O_3$ | −1.2867 | 1336.0 | FactSage |
| $Ga_2O_3$ | 0 | −6314 | Interpolated from log $\gamma$ vs OB using $Al_2O_3$ and $In_2O_3$ |
| $GeO_2$ | 3.669 | −7886 | Sossi et al. (2019) |
| | 0 | −1601 | Regular solution, 1673 K point, Sossi et al. (2019) |
| | 0.8692 | 0 | $\gamma(GeO_2)$ = 7.4, FactSage |
| | −0.010 | −580.4 | $\gamma(GeO_2) = \gamma(SiO_2)$ from FactSage |
| $In_2O_3$ | 0 | −6534.2 | Regular solution, 1923 K point, Wood and Wade (2013) |
| $K_2O$ | 0.8218 | −18774. | pMELTS, BSE 2000-3000 K |
| | −1.64 | −15671.2 | MAGMA code |
| | −1.13 | −16818.8 | FactSage |
| | 0 | −12239 | Regular solution, 1673 K point, Sossi et al. (2019) |
| $Li_2O$ | −3.1647 | −7945.375 | FactSage 1800−2200 K CMAS+FeO |
| | 0 | −2046. | Regular solution, 1573 K point, Nassarella and Fruehan (1992) |
| | 0 | 853.7 | Regular solution, 1673 K, $fO_2 = 10^{-8}$ bar point, Sossi et al. (2019) |
| | 0 | 3973. | Regular solution, 1673 K, $fO_2$ = air point, Sossi et al. (2019) |
| MgO | 0.09574 | −1774.577 | Ban-Ya (1993) |
| | 0.021 | −1800.3 | MAGMA code |
| | 0.565 | −2794.3 | FactSage |
| MnO | 0 | 672.096 | Ban-Ya (1993) |
| | 1.068 | −2787.3 | FactSage |
| $MoO_2$ | 0.187 | 681.0 | Geometric mean of FeO and $TiO_2$, Ban-Ya (1993) |
| $MoO_3$ | 0.7494 | −347.091 | Set $\log \gamma(MoO_3) = \log \gamma(WO_3) + \log(4)$ |
| $Na_2O$ | 1.19437 | −11521.64 | Ban-Ya (1993) |
| | 0.573 | −13034.2 | MAGMA code |
| | −0.083 | −13634.1 | FactSage |
| | 0 | −10039 | Regular solution, 1673 K point, Sossi et al. (2019) |
| NiO | 0.352 | 114.05 | FactSage |
| $P_2O_5$ | −12.0506 | 4076.293 | Ban-Ya (1993) |
| | 1.8919 | −25111.69 | FactSage |
| | 1.75 | −26270 | Suito (1981) |
| PbO | 0.7172 | −1863.825 | FactSage 1800−2200 K |
| | 0 | −1265 | Regular solution, 1923 K point, Wood and Wade (2013) |
| | 0 | −689 | PbO(liq) = Pb(g) + ½$O_2$(g), 1673 K, Sossi et al. (2019) |
| | 0 | −1320 | PbO (liq) = PbO (g), 1673 K, Sossi et al. (2019) |
| $Rb_2O$ | 0 | −13,400 | Regular solution, 1673 K point, Sossi et al. (2019) |
| $Sb_2O_3$ | 3.4014 | −26309.08 | Set = $As_2O_3$, FactSage |
| $SiO_2$ | 0.14730 | −347.091 | Ban-Ya (1993) |
| | −0.320 | −929.6 | MAGMA code |
| | −0.010 | −580.4 | FactSage |
| SnO | 0.470 | −90.4 | FactSage 1800−2200 K CMAS+FeO |
| $SnO_2$ | 0.86923 | 0 | Set = $GeO_2$, FactSage |
| $TiO_2$ | 0 | 963.090 | Ban-Ya (1993) |
| | −0.305 | −964.6 | MAGMA code |
| | −1.407 | 3073.5 | FactSage |
| $Ti_2O_3$ | −2.432 | 289.76 | FactSage |
| $Tl_2O$ | 0 | −13145 | Basicity modified $Rb_2O$ value |
| $Tl_2O_3$ | −1.2867 | 1336.0 | Set = $Fe_2O_3$, FactSage |
| VO | 0.373 | 399.1 | Set = FeO, Ban-Ya (1993) |
| $V_2O_3$ | −1.1319 | −16818.75 | Set = $Ti_2O_3$, FactSage, Pistorius (2006) |
| $WO_3$ | 0.14730 | −347.091 | Set = $SiO_2$, Ban-Ya (1993) |
| ZnO | 1.126 | −1868.7 | FactSage 1800−2200 K, CMAS+FeO |
| | 3.624 | −7061.6 | E1 composition, Reyes and Gaskell (1983) |
| | 4.161 | −7488.1 | 1573 – 1823 K, Sossi et al. (2019) |
| | 0 | −51.65 | MAGMA code, Visscher and Fegley (2013) |

Table 3. Second virials and fugacity coefficients at 5000 K in BSE vapor

| Gas | ε/k | σ (Å) | $b_o$ ($cm^3$) | B(T) | ϕ | Reference |
|---|---|---|---|---|---|---|
| $O_2$ | 118 | 3.460 | 52.26 | 26.97 | 1.006 | HCB |
| O | 81.5 | 2.956 | 32.58 | 16.19 | 1.004* | W62 |
| SiO | 569 | 3.374 | 48.45 | 21.07 | 1.005 | Sv62 |
| $SiO_2$ | 2954 | 3.706 | 64.21 | -59.89 | 0.99 | Sv62 |
| Mg | 1614 | 2.926 | 31.60 | –2.69 | 0.9994 | Sv62 |
| Fe | 3521 | 2.435 | 18.21 | –24.39 | 0.994 | T62 |
| Na | 8471 | 3.0788 | 36.82 | –73.77 | 0.98** | HB87 |
| Hg† | 851 | 2.898 | 30.70 | –26.15† | 0.98† | HCB |

HCB: Hirschfelder et al. (1964). W62: Woolley (1962). Sv62: Svehla (1962). T62: Turkdogan (1962). HB87: Holland and Bilosi (1987).

*Capitelli and Lamanna (1976) considered all electronic states of interacting oxygen atoms and computed $B_{Total}$ = 10.55 $cm^3$ at 5000 K corresponding to ϕ = 1.0025.

**Holland and Bilosi (1987) used the Hulbert – Hirschfelder potential to compute B for Na.

†At 1500 K where Hg is the 2nd most abundant gas with a mole fraction of $1.5\times10^{-3}$.

Table 4. Condensation curve for bulk silicate Earth material

| log P | -6 | -4 | -2 | 0 | 1 | 2 |
|---|---|---|---|---|---|---|
| Dry | 2159 | 2498 | 2960 | 3645 | 4151 | 4894 |
| Wet | 2158 | 2496 | 2949 | 3619 | 4113 | 4837 |
| FactSage | 2161 | 2499 | 2964 | 3683 | 4272 | 5172 |
| GRAINS | 2150 | 2458 | 2923 | 3544 | 3979 | 4581 |
| | | | | | | |
| $fO_2$ (bar)[a] | $4.4\times10^{-8}$ | $7.3\times10^{-6}$ | $1.1\times10^{-3}$ | 0.126 | 1.23 | 11.3 |

[a]The $fO_2$ is identical within ≤5% in the dry and wet BSE models.

Table 5. Composition of initial condensates (mol%)

| | $P = 10^{-3}$ bar | | P = 10 bar | | P = 100 bar | |
|---|---|---|---|---|---|---|
| Oxide | CONDOR[a] | FactSage[b] | CONDOR[a] | FactSage[b] | CONDOR[a] | FactSage[b] |
| CaO | 48. | 21. | 34. | 3. | 24. | 1. |
| MgO | 2. | 2. | 44. | 52. | 59. | 58. |
| $Al_2O_3$ | 50. | 76. | 20. | 16. | 13. | 14. |
| $SiO_2$ | 0.06 | 0.6 | 0.8 | 28. | 1. | 26. |
| FeO | 0.01 | 0.04 | 1. | 0.3 | 3. | 0.4 |
| $Na_2O$ | 1.e-14 | 1.e-7 | 4.e-5 | 1.e-4 | 7.e-6 | 4.e-4 |
| Total | 100.1[c,d] | 99.6[c,d] | 100.0[c] | 99.3[d] | 100.2[c,d] | 99.4[d] |

[a]The calculated condensation temperatures are 2706 K at $10^{-3}$ bar, 4152 K at 10 bar, and 4895 K at 100 bar. [b]The calculated condensation temperatures are 2710 K at $10^{-3}$ bar, 4272 K at 10 bar and 5168 K at 100 bar. [c]Also includes 0.03% $TiO_2$ ($10^{-3}$ bars), and 0.2% $TiO_2$ at 10 and 100 bars. [d]Totals are not 100% due to rounding.

Table 6. Ca, Al, Mg, Si, and Fe 50% condensation temperatures

| log P (bar) | BanYa | BY wet | MAGMA | FactSage | Ideal (Lock) | Ideal (this work) |
|---|---|---|---|---|---|---|
| **Calcium** | | | | | | |
| -6 | 2159 | 2158 | 2090 | 2065 | – | 1880 |
| -4 | 2498 | 2496 | 2461 | 2411 | 2304 | 2216 |
| -2 | 2960 | 2949 | 2956 | 2897 | 2678 | 2700 |
| 0 | 3645 | 3619 | 3640 | 3633 | 3426 | 3448 |
| 2 | 4894 | 4837 | 4893 | 4884 | 4581 | 4758 |
| **Aluminum** | | | | | | |
| -6 | 2159 | 2158 | 2157 | 2159 | – | 2134 |
| -4 | 2498 | 2496 | 2495 | 2498 | 2429 | 2460 |
| -2 | 2960 | 2949 | 2956 | 2961 | 2885 | 2901 |
| 0 | 3645 | 3619 | 3641 | 3648 | 3496 | 3555 |
| 2 | 4894 | 4837 | 4913 | 4901 | 4581 | 4734 |
| **Magnesium** | | | | | | |
| -6 | 1866 | 1865 | 1895 | 1871 | – | 1817 |
| -4 | 2198 | 2197 | 2228 | 2197 | 2157 | 2135 |
| -2 | 2678 | 2677 | 2712 | 2663 | 2602 | 2592 |
| 0 | 3417 | 3415 | 3467 | 3368 | 3330 | 3303 |
| 2 | 4721 | 4719 | 4822 | 4570 | 4581 | 4561 |
| **Silicon** | | | | | | |
| -6 | 1853 | 1852 | 1895 | 1868 | – | 1820 |
| -4 | 2159 | 2157 | 2221 | 2181 | 2124 | 2132 |
| -2 | 2586 | 2585 | 2683 | 2619 | 2573 | 2572 |
| 0 | 3233 | 3230 | 3386 | 3284 | 3235 | 3228 |
| 2 | 4334 | 4327 | 4608 | 4425 | 4352 | 4341 |
| **Iron** | | | | | | |
| -6 | 1818 | 1814 | 1871 | 1832 | – | 1813 |
| -4 | 2138 | 2135 | 2208 | 2148 | 2120 | 2131 |
| -2 | 2581 | 2579 | 2679 | 2597 | 2584 | 2584 |
| 0 | 3284 | 3282 | 3423 | 3318 | 3320 | 3303 |
| 2 | 4599 | 4596 | 4812 | 4657 | 4581 | 4627 |

See text for details.

Table 7. Ti 50% condensation temperatures – different activity coefficient models

| log P (bar) | BanYa | MAGMA | FactSage | Ideal (Lock) | Ideal (this work) |
|---|---|---|---|---|---|
| -6 | 1882 | 1990 | 1896 | – | 1879 |
| -4 | 2213 | 2318 | 2239 | 2129 | 2192 |
| -2 | 2690 | 2803 | 2744 | 2649 | 2650 |
| 0 | 3433 | 3580 | 3558 | 3388 | 3378 |
| 2 | 4778 | 4913 | 4894 | 4581 | 4694 |

See text for details.

Table 8. Mn 50% condensation temperatures – different activity coefficient models

| log P (bar) | BanYa | FactSage | Ideal (Lock) | Ideal (this work) |
|---|---|---|---|---|
| -6 | 1807 | 1835 | – | 1808 |
| -4 | 2134 | 2149 | 2111 | 2127 |
| -2 | 2579 | 2594 | 2573 | 2574 |
| 0 | 3287 | 3286 | 3275 | 3279 |
| 2 | 4597 | 4497 | 4487 | 4563 |

See text for details.

Table 9. Cr 50% condensation temperatures – FactSage and Ideal

| log P (bar) | FactSage | FS (wet) | Ideal (Lock) | Ideal (this work) |
|---|---|---|---|---|
| -6 | 1803 | 1798 | – | 1802 |
| -4 | 2121 | 2115 | 2106 | 2113 |
| -2 | 2554 | 2547 | 2534 | 2540 |
| 0* | 3196 | 3188 | 3154 | 3173 |
| 2 | 4299 | 4290 | 4199 | 4253 |

*Ivanov et al. (2022) give a 50% condensation temperature of 3346 K at 1 bar total pressure. See text for details.

Table 10. Na 50% condensation temperatures – real and ideal solution

| log P (bar) | BanYa | MAGMA | FactSage | Sossi | Ideal (Lock) | Ideal (this work) |
|---|---|---|---|---|---|---|
| -6 | 1477 | 1569 | 1627 | 1488 | – | 1102 |
| -4 | 1820 | 1942 | 2013 | 1853 | 1583 | 1364 |
| -2 | 2373 | 2486 | 2527 | 2436 | 1800 | 1809 |
| 0 | 3154 | 3184 | 3199 | 3176 | 2469 | 2743 |
| 2 | 4276 | 4324 | 4426 | 4330 | 3772 | 4223 |

See text for details.

Table 11. K 50% condensation temperatures – different activity coefficient models

| log P (bar) | MELTS | MAGMA | FactSage | Sossi | Ideal (Lock) | Ideal (this work) |
|---|---|---|---|---|---|---|
| -6 | 1528 | 1574 | 1564 | 1441 | – | 1021 |
| -4 | 1863 | 1954 | 1930 | 1775 | <1500 | 1267 |
| -2 | 2403 | 2550 | 2485 | 2321 | 1752 | 1691 |
| 0 | 3161 | 3339 | 3187 | 3140 | 2297 | 2699 |
| 2 | 4286 | 4648 | 4410 | 4263 | 3807 | 4223 |

See text for details.

Table 12. Li 50% condensation temperatures – real and ideal models

| log P (bar) | FactSage | NF92 | Sossi-A | Sossi-B | Ideal |
|---|---|---|---|---|---|
| -6 | 1793 | 1499 | 1463 | 1447 | 1422 |
| -4 | 2122 | 1819 | 1794 | 1775 | 1694 |
| -2 | 2564 | 2346 | 2326 | 2304 | 2100 |
| 0 | 3242 | 3135 | 3127 | 3119 | 2801 |
| 2 | 4493 | 4238 | 4233 | 4228 | 4205 |

NF-92: Nassaralla and Fruehan (1992). Sossi-A: $LiO_{.5}$ $\gamma$ = 1.8, Sossi-B: $LiO_{.5}$ $\gamma$ = 15.4, both at 1673 K from Sossi et al. (2019).

Table 13. Rb, Cs 50% condensation temperatures – real and Ideal

| | real | | | ideal | |
|---|---|---|---|---|---|
| log P (bar) | Rb | Cs-1 | Cs-2 | Rb | Cs |
| −6 | 1439 | 1416 | 1371 | 1024 | 626 |
| −4 | 1772 | 1746 | 1702 | 1270 | 864 |
| −2 | 2305 | 2270 | 2227 | 1658 | 1426 |
| 0 | 3122 | 3108 | 3088 | 2464 | 2802 |
| 2 | 4230 | 4221 | 4218 | 4172 | 3718 |

Real solution models: Cs-1: $\log \gamma = -0.083 - 13{,}634.1/T$. Cs-2: $\log \gamma = -12{,}144/T$

Table 14. Alkaline Earth 50% condensation temperatures – ideal solution

| log P (bar) | Be | Mg | Ca | Sr | Ba |
|---|---|---|---|---|---|
| -6 | 2101 | 1817 | 1880 | 1866 | 1791 |
| -4 | 2466 | 2135 | 2216 | 2203 | 2124 |
| -2 | 2960 | 2592 | 2700 | 2695 | 2571 |
| 0 | 3645 | 3303 | 3448 | 3473 | 3303 |
| 2 | 4894 | 4561 | 4758 | 4894 | 4794 |

Table 15. Sc, Y, Ti, Zr, Hf, Th, U, Pu 50% condensation temperatures – ideal solution

| log P (bar) | Sc | Y | Ti | Zr | Hf | Th | U | Pu |
|---|---|---|---|---|---|---|---|---|
| -6 | 2054 | 2060 | 1879 | 2159 | 2159 | 2067 | 1805 | 1924 |
| -4 | 2352 | 2350 | 2192 | 2498 | 2498 | 2371 | 2129 | 2253 |
| -2 | 2771 | 2759 | 2650 | 2960 | 2961 | 2806 | 2568 | 2733 |
| 0 | 3444 | 3430 | 3378 | 3645 | 3646 | 3520 | 3275 | 3459 |
| 2 | 4614 | 4578 | 4694 | 4894 | 4900 | 4837 | 4792 | 4685 |

Table 16. REE 50% condensation temperatures – ideal solution

| log P (bar) | La | Ce[a] | Pr | Nd | Sm | Eu[b] | Gd |
|---|---|---|---|---|---|---|---|
| -6 | 1850 | 1808 | 1841 | 1857 | 1864 | 1934 | 1910 |
| -4 | 2147 | 2122 | 2141 | 2159 | 2167 | 2293 | 2207 |
| -2 | 2565 | 2564 | 2562 | 2578 | 2584 | 2825 | 2624 |
| 0 | 3195 | 3229 | 3193 | 3209 | 3214 | 3645 | 3237 |
| 2 | 4270 | 4514 | 4268 | 4285 | 4280 | 4894 | 4289 |

| log P (bar) | Tb | Dy | Ho | Er | Tm | Yb | Lu |
|---|---|---|---|---|---|---|---|
| -6 | 1964 | 1956 | 1952 | 1994 | 2017 | 1958 | 2012 |
| -4 | 2256 | 2242 | 2236 | 2276 | 2288 | 2264 | 2286 |
| -2 | 2690 | 2673 | 2662 | 2694 | 2733 | 2699 | 2694 |
| 0 | 3342 | 3300 | 3277 | 3327 | 3365 | 3341 | 3310 |
| 2 | 4433 | 4350 | 4320 | 4370 | 4384 | 4374 | 4333 |

[a]Ideal solution of $CeO_2$ and $Ce_2O_3$
[b]Ideal solution of $EuO$ and $Eu_2O_3$

Table 17. Group 5 (V, Nb, Ta) 50% condensation temperatures

| log P (bar) | V (real) | V (ideal) | Nb (ideal) | Ta (ideal) |
|---|---|---|---|---|
| -6 | 1729 | 1610 | 1858 | 1828 |
| -4 | 2072 | 2043 | 2181 | 2124 |
| -2 | 2533 | 2532 | 2654 | 2537 |
| 0 | 3190 | 3197 | 3410 | 3160 |
| 2 | 4354 | 4417 | 4825 | 4243 |

Table 18. Group 6 (Cr, Mo, W) 50% condensation temperatures – real and ideal solution

| | Real solution | | | Ideal solution | | |
|---|---|---|---|---|---|---|
| log P (bar) | Cr | Mo | W | Cr | Mo | W |
| -6 | 1803 | 1527 | 1846 | 1802 | 1640 | 1815 |
| -4 | 2121 | 1921 | 2160 | 2113 | 2058 | 2132 |
| -2 | 2554 | 2496 | 2618 | 2540 | 2539 | 2584 |
| 0* | 3196 | 3186 | 3356 | 3173 | 3211 | 3303 |
| 2 | 4299 | 4376 | 4708 | 4253 | 4511 | 4625 |

*Ivanov et al. (2022) report 3346 K (Cr), 3385 K (Mo), and 3337 K (W) 50% condensation temperatures at one bar total pressure.

Table 19. Noble metal 50% condensation temperatures

| Dry BSE | | | | | | | | |
|---|---|---|---|---|---|---|---|---|
| log P (bar) | Re | Ru | Os | Rh | Ir | Pd | Pt | Au |
| -6 | 755 | 1003 | 855 | 1451 | 1491 | 1091 | 1464 | 1100 |
| -4 | 1380 | 1089 | 1480 | 1681 | 1761 | 1197 | 1772 | 1245 |
| -2 | 2301 | 1183 | 2307 | 1743 | 1115 | 1328 | 2080 | 1440 |
| 0 | 3100 | 1268 | 3064 | 1752 | 1211 | 1490 | 2247 | 1713 |
| 2 | 4188 | 1312 | 4236 | 1754 | 1333 | 1701 | 2271 | 2114 |
| Wet BSE | | | | | | | | |
| log P (bar) | Re | Ru | Os | Rh | Ir | Pd | Pt | Au |
| -6 | 1518 | 1498 | 1618 | 1360 | 1633 | 1043 | 1408 | 686 |
| -4 | 1750 | 1679 | 1850 | 1517 | 1807 | 1169 | 1538 | 734 |
| -2 | 2146 | 1957 | 2152 | 1712 | 2039 | 1331 | 1673 | 824 |
| 0 | 2611 | 2301 | 2575 | 1946 | 2347 | 1539 | 1822 | 940 |
| 2 | 3036 | 2718 | 3084 | 2224 | 2739 | 1794 | 1994 | 1094 |

Table 20. Co, Ni 50% condensation temperatures – real and ideal solution

| | Cobalt | | Nickel | |
|---|---|---|---|---|
| log P (bar) | FactSage | ideal | FactSage | ideal |
| -6 | 1781 | 1794 | 1722 | 1755 |
| -4 | 2112 | 2118 | 2051 | 2085 |
| -2 | 2557 | 2560 | 2503 | 2530 |
| 0 | 3225 | 3239 | 3166 | 3181 |
| 2 | 4474 | 4508 | 4262 | 4287 |

Table 21. Copper 50% condensation temperatures – real and ideal solution

| log P (bar) | Cu-1 | Cu-2 | Cu-3 | Cu-4 | Ideal |
|---|---|---|---|---|---|
| -6 | 1452 | 1428 | 1421 | 1441 | 1454 |
| -4 | 1783 | 1759 | 1752 | 1771 | 1786 |
| -2 | 2317 | 2291 | 2285 | 2303 | 2320 |
| 0 | 3127 | 3119 | 3117 | 3122 | 3128 |
| 2 | 4236 | 4231 | 4231 | 4233 | 4236 |

Cu real solution models: (1) Altman (1978). (2) Holzheid and Lodders (2001). (3) Sossi et al (2019). (4) Wood and Wade (2013). See text for details.

Table 22. Silver 50% condensation temperatures – real and ideal solution

| log P (bar) | ideal | Ag-1 | Ag-2 |
|---|---|---|---|
| -6 | 1035 | 1056 | 910 |
| -4 | 1257 | 1283 | 1104 |
| -2 | 1612 | 1645 | 1410 |
| 0 | 2266 | 2315 | 1974 |
| 2 | 3819 | 3876 | 3372 |

Real solution models: (1) $\log \gamma = -582.0/T$ Sossi et al. (2019) 1673 K point. (2) $\log \gamma = \log \gamma(Cu_2O) = 3582.8/T$, Sossi et al. (2019) $Cu_2O$ 1673 K point. See text for details.

Table 23. Zinc 50% condensation temperatures - real and ideal solution

| log P (bar) | Zn-1 | Zn-2 | Zn-3 | Zn-4 | ideal |
|---|---|---|---|---|---|
| -6 | 1453 | 1451 | 1467 | 1463 | 1450 |
| -4 | 1785 | 1787 | 1793 | 1787 | 1787 |
| -2 | 2319 | 2330 | 2309 | 2288 | 2330 |
| 0 | 3129 | 3137 | 3103 | 3065 | 3132 |
| 2 | 4242 | 4259 | 4223 | 4213 | 4256 |

Models:(1) FactSage. (2) MAGMA. (3) Reyes and Gaskell (1983). (4) Sossi et al. (2019).

Table 24. Cd 50% condensation temperatures - real and ideal solution

| log P (bar) | Real | Ideal |
|---|---|---|
| −6 | 1361 | 1262 |
| −4 | 1733 | 1638 |
| −2 | 2296 | 2254 |
| 0 | 3125 | 3120 |
| 2 | 4246 | 4240 |

Table 25. Group 13 (B, Al, Ga, In, Tl) 50% condensation temperatures

| Ideal solution | | | | | |
|---|---|---|---|---|---|
| log P (bar) | B | Al | Ga | In | Tl |
| -6 | 1113 | 2134 | 1615 | 1446 | 924 |
| -4 | 1355 | 2460 | 1891 | 1757 | 1137 |
| -2 | 1756 | 2952 | 2350 | 2205 | 1490 |
| 0 | 2574 | 3632 | 3127 | 2890 | 2209 |
| 2 | 4193 | 4825 | 4228 | 4057 | 4064 |
| Real solution | | | | | |
| log P (bar) | B | Al | Ga | In | Tl |
| -6 | 1447 | 2159 | 1739 | 1473 | 1405 |
| -4 | 1784 | 2498 | 2033 | 1789 | 1729 |
| -2 | 2336 | 2960 | 2452 | 2299 | 2252 |
| 0 | 3147 | 3645 | 3143 | 3070 | 3106 |
| 2 | 4281 | 4894 | 4238 | 4206 | 4223 |

Table 26. Group 14 (Si, Ge, Sn, Pb) gas chemistry at $10^{-4}$ bar for the dry BSE model

| Z | Element | $T_{50}$/K* | Major gas | M/MO | M/$MO_2$ | MO/$MO_2$ |
|---|---|---|---|---|---|---|
| 14 | Si | 2159 | SiO | $2.3\times10^{-8}$ | $3.2\times10^{-7}$ | $1.4\times10^{1}$ |
| 32 | Ge | 1781 | GeO | $2.3\times10^{-7}$ | $7.2\times10^{-3}$ | $3.1\times10^{4}$ |
| 50 | Sn | 1791 | SnO | $4.3\times10^{-4}$ | $5.2\times10^{2}$ | $1.2\times10^{6}$ |
| 82 | Pb | 1782 | Pb | 4 | $7.1\times10^{8}$ | $1.7\times10^{8}$ |

*Nominal values. The $T_{50}$ ranges from different activity coefficient models are 2159-2221 K for Si, 1725-1783 K for Ge, and 1782-1789 K for Pb. See text for details.

Table 27. Ge 50% condensation temperatures - real and ideal solution

| log P (bar) | Ge-1 | Ge-2 | Ge-3 | Ge-4 | Ideal |
|---|---|---|---|---|---|
| -6 | 1464 | 1423 | 1449 | 1456 | 1445 |
| -4 | 1781 | 1725 | 1775 | 1783 | 1768 |
| -2 | 2269 | 2195 | 2296 | 2260 | 2283 |
| 0 | 2960 | 2992 | 3113 | 3118 | 3107 |
| 2 | 4072 | 4212 | 4224 | 4227 | 4222 |

Ge real solution models: (1) Sossi et al. (2019). (2) FactSage. (3) $\gamma(GeO_2) = \gamma(SiO_2)$. (4) Regular solution fit to 1673 K point of Sossi et al. (2019).

Table 28. Sn and Pb 50% condensation temperatures - real and ideal solution

| log P (bar) | Sn-1 | Sn-2 | Pb-1 | Pb-2 | Pb-3 | Pb-4 | Pb-5 |
|---|---|---|---|---|---|---|---|
| −6 | 1461 | 1478 | 1448 | 1452 | 1448 | 1452 | 1442 |
| −4 | 1791 | 1805 | 1782 | 1789 | 1784 | 1789 | 1778 |
| −2 | 2333 | 2350 | 2324 | 2336 | 2329 | 2337 | 2322 |
| 0 | 3138 | 3148 | 3134 | 3138 | 3140 | 3144 | 3136 |
| 2 | 4254 | 4273 | 4250 | 4272 | 4265 | 4273 | 4258 |

Sn-1: real. Sn-2: ideal. Pb-1: FactSage. Pb-2: Wood and Wade (2013) $\log \gamma = -1265/T$. Pb-3: Sossi et al. (2019) $\log \gamma = -689/T$. Pb-4: Sossi et al. (2019) $\log \gamma = -1320/T$. Pb-5: ideal.

Table 29. Group 15 (P, As, Sb, Bi) 50% condensation temperatures - real solution

| log P (bar) | $P_2O_5$ – 1 | $P_2O_5$ – 2 | $P_2O_5$ - 3 | $As_2O_3$ | $Sb_2O_3$ | $Bi_2O_3$ |
|---|---|---|---|---|---|---|
| −6 | 1583 | 1737 | 1761 | 1544 | 1552 | 1461 |
| −4 | 1948 | 1996 | 2028 | 1804 | 1804 | 1781 |
| −2 | 2474 | 2376 | 2400 | 2302 | 2293 | 2291 |
| 0 | 3176 | 3082 | 3103 | 2958 | 2870 | 3072 |
| 2 | 4322 | 4084 | 4126 | 3831 | 3619 | 4211 |

Real solution models for $P_2O_5$:(1) Ban-Ya (1993). (2) FactSage. (3) Suito et al. (1981).

Table 30. As, Sb, Bi 50% condensation temperatures - ideal solution

| log P (bar) | $As_2O_3$ | $Sb_2O_3$ | $Bi_2O_3$ |
|---|---|---|---|
| −6 | 1200 | 1237 | 1109 |
| −4 | 1407 | 1439 | 1336 |
| −2 | 1702 | 1720 | 1691 |
| 0 | 2159 | 2131 | 2336 |
| 2 | 2949 | 2796 | 3770 |

Table 31. Oxygen 50% condensation temperatures - real and ideal solution

| log P (bar) | Real | Ideal |
|---|---|---|
| −6 | 1858 | 1819 |
| −4 | 2173 | 2134 |
| −2 | 2617 | 2581 |
| 0* | 3294 | 3260 |
| 2 | 4463 | 4427 |

*Ivanov et al. (2022) give 3409 K for 50% condensation of oxygen at one bar total pressure.

Table 32. Dry BSE material chemistry at $10^{-4}$ bar

| Z | Symbol | $T_{50}$/K | T-range | Gas | M/MO | M/M$^{+}$ |
|---|---|---|---|---|---|---|
| 3 | Li | 2122 | 1775-2122 | Li | 65 | 4300 |
| 11 | Na | 1820 | 1820-2013 | Na | 1200 | 46300 |
| 19 | K | 1863 | 1863-1954 | K | 1300 | 190 |
| 37 | Rb | 1772 | 1772 | Rb | 400 | 100 |
| 55 | Cs | 1746 | 1702-1746 | Cs | 160 | 20 |
| | | | | | | |
| 4 | Be | 2466 | – | Be | 9 | $1.1\times10^{9}$ |
| 12 | Mg | 2198 | 2197-2228 | Mg | 20 | $1.8\times10^{7}$ |
| 20 | Ca | 2498 | 2411-2498* | Ca | 50 | 260 |
| 38 | Sr | 2203 | – | Sr | 8 | 35 |
| 56 | Ba | 2124 | – | BaO | $7.9\times10^{-4}$ | 330 |
| | | | | | | |
| 21 | Sc | 2352 | – | ScO | $5.5\times10^{-5}$ | 9700 |
| 39 | Y | 2350 | – | YO | $6.2\times10^{-6}$ | 2,800 |
| 57 | La | 2147 | – | LaO | $5.4\times10^{-9}$ | 1,400 |
| | | | | | | |
| 22 | Ti | 2213 | 2213-2318 | $TiO_2$ | $1.5\times10^{-5}$ | $1.8\times10^{5}$ |
| 40 | Zr | 2498 | – | $ZrO_2$ | $2.4\times10^{-5}$ | $8.0\times10^{3}$ |
| 72 | Hf | 2498 | – | HfO‡ | $1.4\times10^{-6}$ | $1.3\times10^{4}$ |
| | | | | | | |
| 23 | V | 2072 | – | $VO_2$ | $5.2\times10^{-5}$ | $1.4\times10^{7}$ |
| 41 | Nb | 2181 | – | $NbO_2$ | $8.4\times10^{-8}$ | $1.3\times10^{6}$ |
| 73 | Ta | 2124 | – | $TaO_2$ | $3.7\times10^{-10}$ | $1.3\times10^{9}$ |
| | | | | | | |
| 24 | Cr | 2121 | – | $CrO_2$ | 0.2 | $5.0\times10^{6}$ |
| 42 | Mo | 1921 | – | $MoO_3$ | $1.9\times10^{-4}$ | $3.1\times10^{9}$ |
| 74 | W | 2160 | – | $WO_3$ | $1.8\times10^{-6}$ | $4.7\times10^{8}$ |
| | | | | | | |
| 25 | Mn | 2134 | 2134-2149 | Mn | 5 | $7.4\times10^{7}$ |
| 75 | Re | 1559 | – | $Re_2O_7$ | $6.8\times10^{-9}$ | $1.3\times10^{13}$ |
| | | | | | | |
| 26 | Fe | 2138 | 2138-2208 | Fe | 6 | $5.6\times10^{8}$ |
| 44 | Ru | 1089 | – | $RuO_3$ | $4.4\times10^{-9}$ | $8.6\times10^{16}$ |
| 76 | Os | 1404 | – | $OsO_4$ | $9.4\times10^{-8}$ | $7.5\times10^{16}$ |
| | | | | | | |
| 27 | Co | 2112 | – | Co | 30 | $2.9\times10^{9}$ |
| 45 | Rh | 1681 | – | $RhO_2$ | 0.6 | $9.7\times10^{11}$ |
| 77 | Ir | 1761 | – | $Ir_2O_3$ | 0.8 | – |
| | | | | | | |
| 28 | Ni | 2051 | – | Ni | 120 | $1.4\times10^{10}$ |
| 46 | Pd | 1197 | – | PdO | $6.8\times10^{-3}$ | $1.3\times10^{19}$ |
| 78 | Pt | 1772 | – | Pt | 25 | – |
| 29 | Cu | 1783 | 1752-1783 | Cu | 1,500 | $1.3\times10^{12}$ |
| 47 | Ag | 1283 | 1104-1283 | Ag | $5.6\times10^{4}$ | $8.0\times10^{15}$ |
| 79 | Au | 1245 | – | Au | $5.6\times10^{4}$ | $1.1\times10^{23}$ |
| | | | | | | |
| 30 | Zn | 1785 | 1785-1793 | Zn | $1.1\times10^{7}$ | $8.5\times10^{15}$ |
| 48 | Cd | 1733 | – | Cd | $3.0\times10^{4}$ | $3.2\times10^{15}$ |
| 80 | Hg | 393 | – | Hg | $1.5\times10^{8}$ | $1.7\times10^{86}$ |
| | | | | | | |
| 5 | B | 1784 | – | $NaBO_2$ | $1.2\times10^{-11}$ | $1.1\times10^{14}$ |
| 13 | Al | 2498* | 2495-2498 | AlO | 0.6 | 1,500 |
| 31 | Ga | 2033 | – | Ga | 32 | $1.3\times10^{6}$ |
| 49 | In | 1789 | – | In | 32 | $5.5\times10^{6}$ |
| 81 | Tl | 1729 | – | Tl | 1,900 | $5.4\times10^{7}$ |
| | | | | | | |
| 14 | Si | 2159 | 2159-2221 | SiO | $2.3\times10^{-8}$ | $2.2\times10^{9}$ |
| 32 | Ge | 1781 | 1725-1783 | GeO | $2.4\times10^{-7}$ | $3.2\times10^{12}$ |
| 50 | Sn | 1791 | – | SnO | $4.4\times10^{-4}$ | $4.8\times10^{10}$ |
| 82 | Pb | 1782 | 1782-1789 | Pb | 4 | $4.5\times10^{10}$ |
| | | | | | | |
| 15 | P | 1948 | 1948-2028 | $PO_2$ | $3.6\times10^{-5}$ | $2.6\times10^{17}$ |
| 33 | As | 1804 | – | AsO | $4.8\times10^{-2}$ | $9.8\times10^{17}$ |
| 51 | Sb | 1804 | – | SbO | 0.25 | $5.2\times10^{14}$ |
| 83 | Bi | 1781 | – | Bi | 110 | $1.5\times10^{11}$ |
| | | | | | | |
| 8 | O | 2173 | – | $O_2$ | 0.5 | $9.9\times10^{21}$ |
| | | | | | | |
| 58 | Ce | 2122 | – | $CeO_2$ | $3.1\times10^{-9}$ | – |
| 59 | Pr | 2141 | – | PrO | $3.5\times10^{-8}$ | |
| 60 | Nd | 2159 | – | NdO | $4.7\times10^{-7}$ | |
| 62 | Sm | 2167 | – | SmO | $8.1\times10^{-4}$ | |
| 63 | Eu | 2293† | – | EuO | 0.14 | |
| 64 | Gd | 2207 | – | GdO | $7.1\times10^{-7}$ | |
| 65 | Tb | 2256 | – | TbO | $3.5\times10^{-6}$ | |
| 66 | Dy | 2242 | – | DyO | $2.7\times10^{-4}$ | |
| 67 | Ho | 2236 | – | HoO | $1.2\times10^{-4}$ | |
| 68 | Er | 2276 | – | ErO | $1.0\times10^{-4}$ | |
| 69 | Tm | 2288 | – | TmO | $2.5\times10^{-2}$ | |
| 70 | Yb | 2264 | – | Yb | 2.9 | |
| 71 | Lu | 2286 | – | LuO | $1.7\times10^{-5}$ | |
| | | | | | | |
| 90 | Th | 2371 | – | $ThO_2$ | $6.4\times10^{-9}$ | 960 |
| 92 | U | 2129 | – | $UO_3$ | $1.6\times10^{-8}$ | $8.4\times10^{4}$ |
| 94 | Pu | 2253 | – | $PuO_2$ | $2.8\times10^{-6}$ | 250 |

*Greater than 50% condensed where melt initially forms.

†Ideal solution of EuO + $Eu_2O_3$ in the BSE melt

‡$HfO/HfO_2$ = 1.01, see text.

Notes: Columns list atomic number (Z), element symbol, nominal 50% condensation temperature, the range of 50% condensation temperatures for the available activity coefficients, the major gas at the 50% condensation temperature, the molar ratio of monatomic gas to monoxide gas, and the molar ratio of monatomic gas to singly ionized monatomic gas. Dashes indicate missing data; for example, the T-range column is blank for many elements where ideal solution calculations were done or where only one activity coefficient estimate was available. There are no data in the CONDOR code for computing the $M/M^+$ ratios for Ce to Lu.

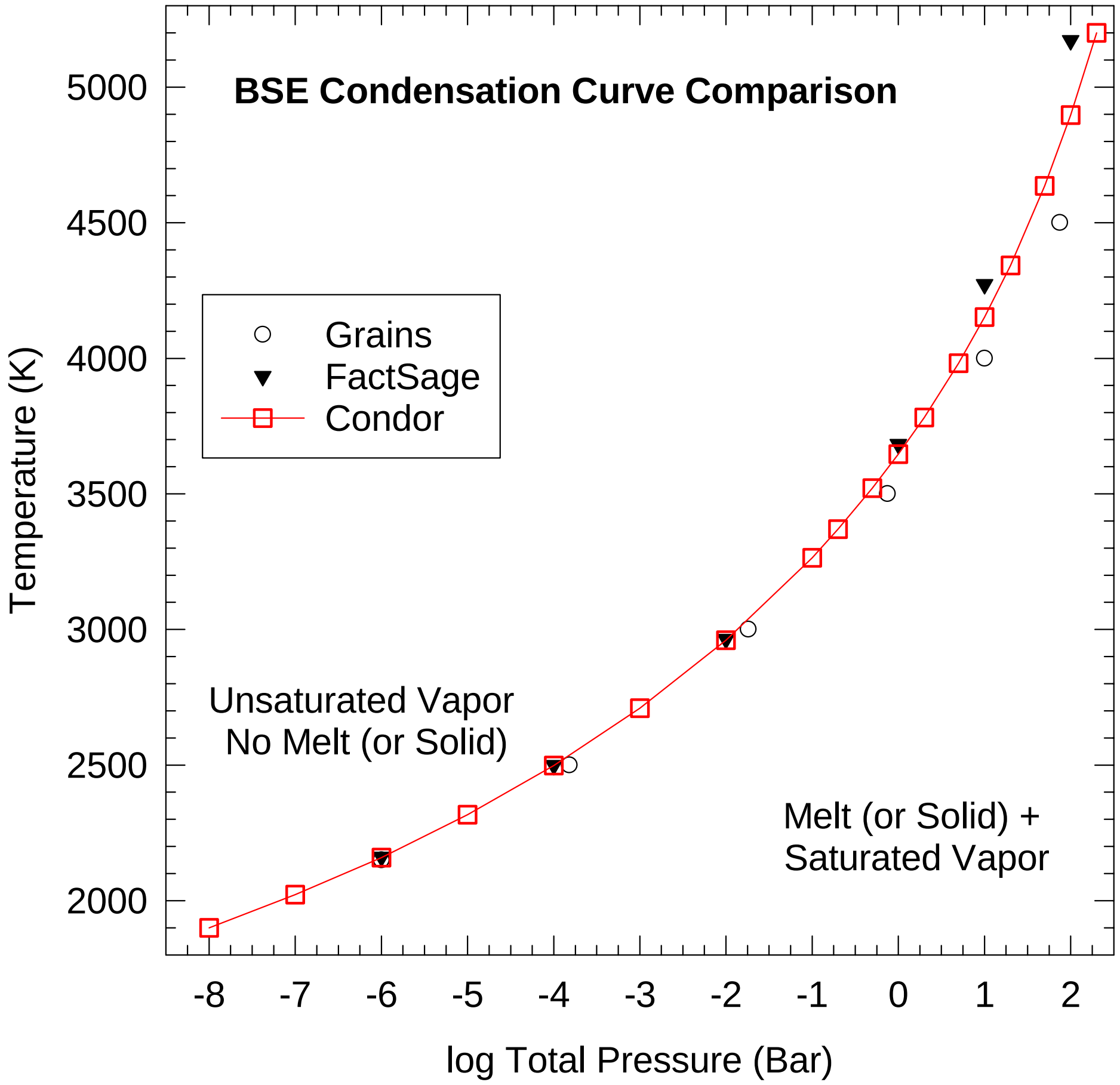

Figure 1

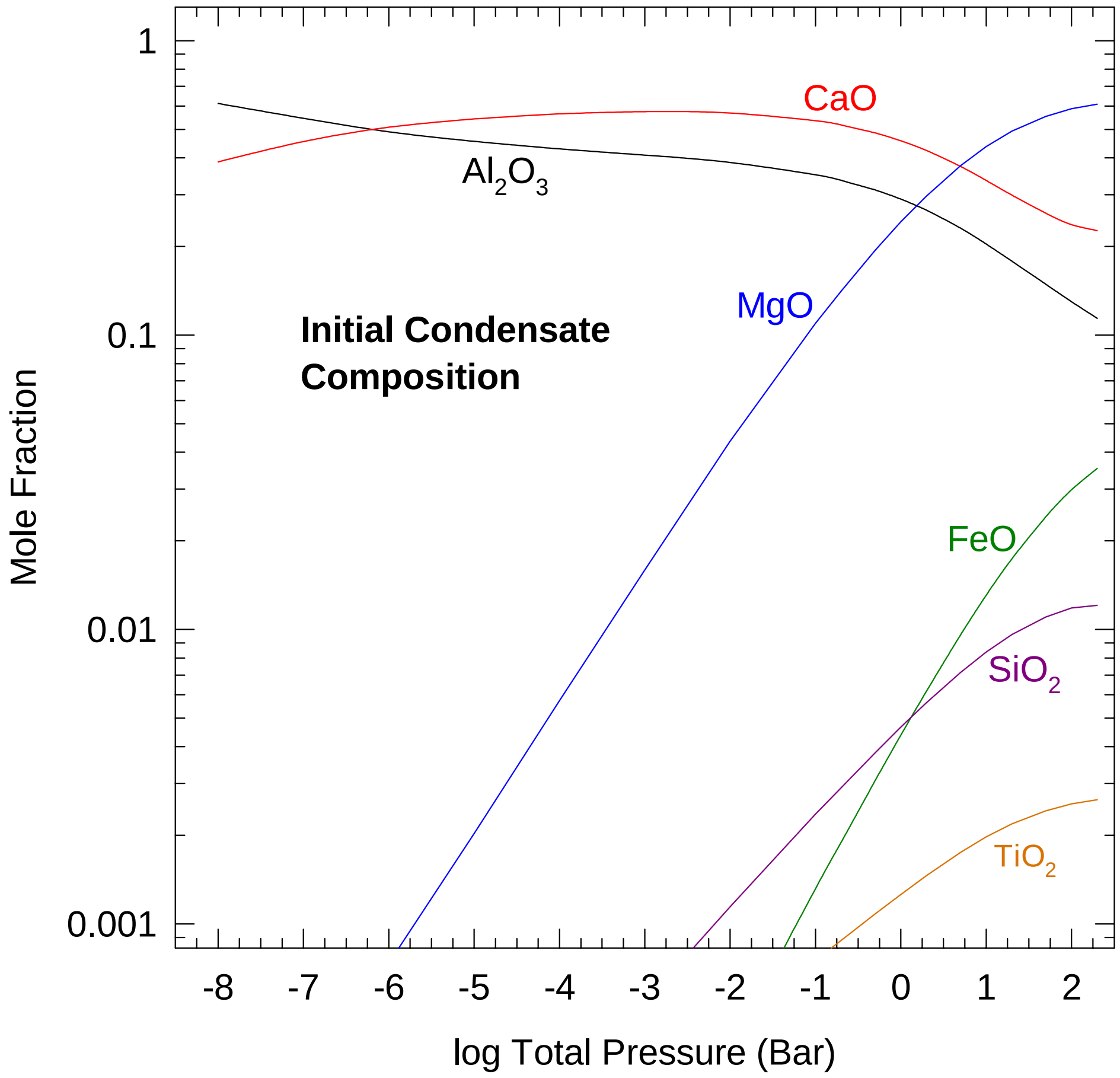

Figure 2

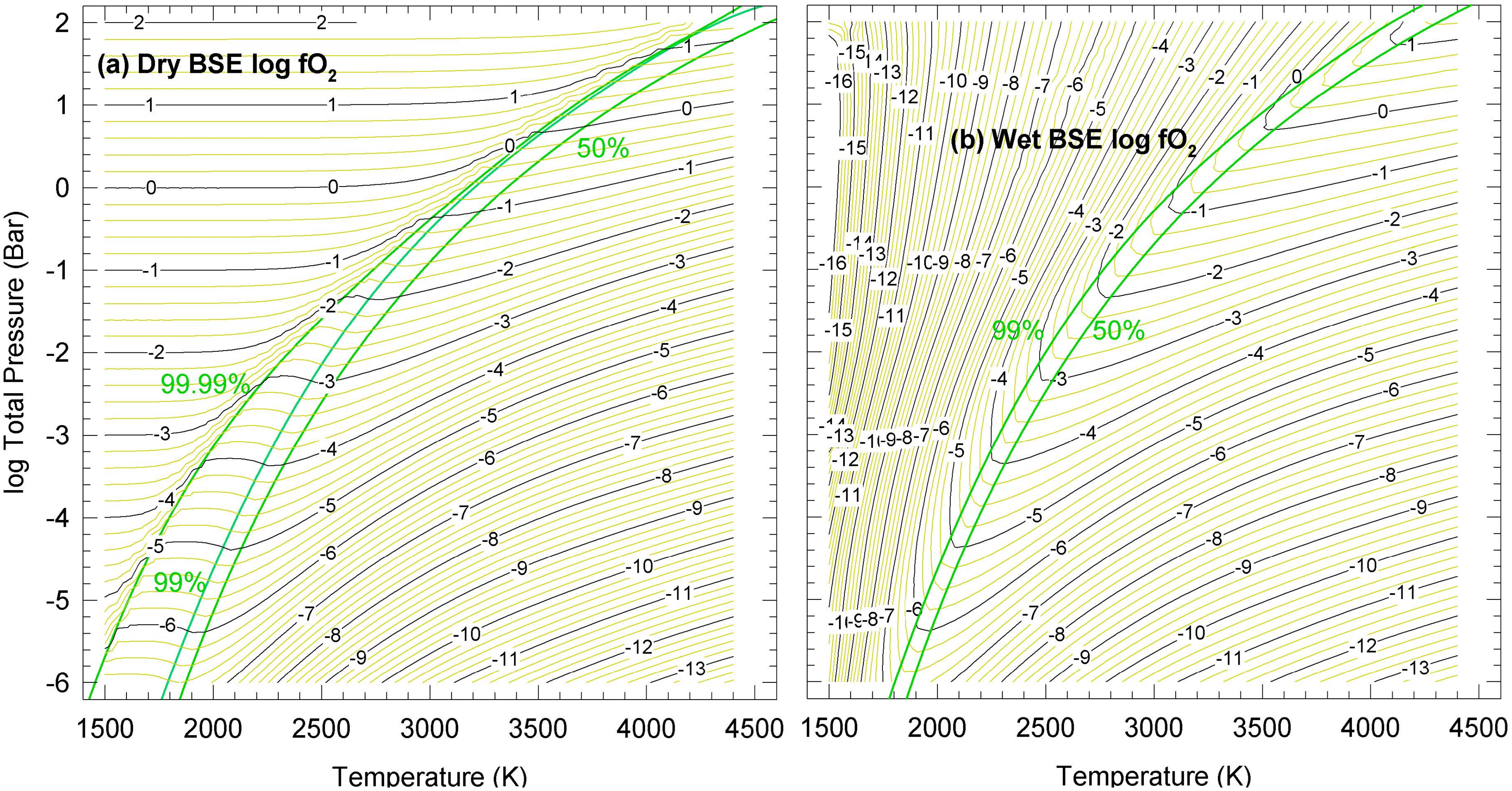

(a) Dry BSE log fO2
(b) Wet BSE log fO2
99.99%
99%
50%
log Total Pressure (Bar)
Temperature (K)

Figure 3a,b

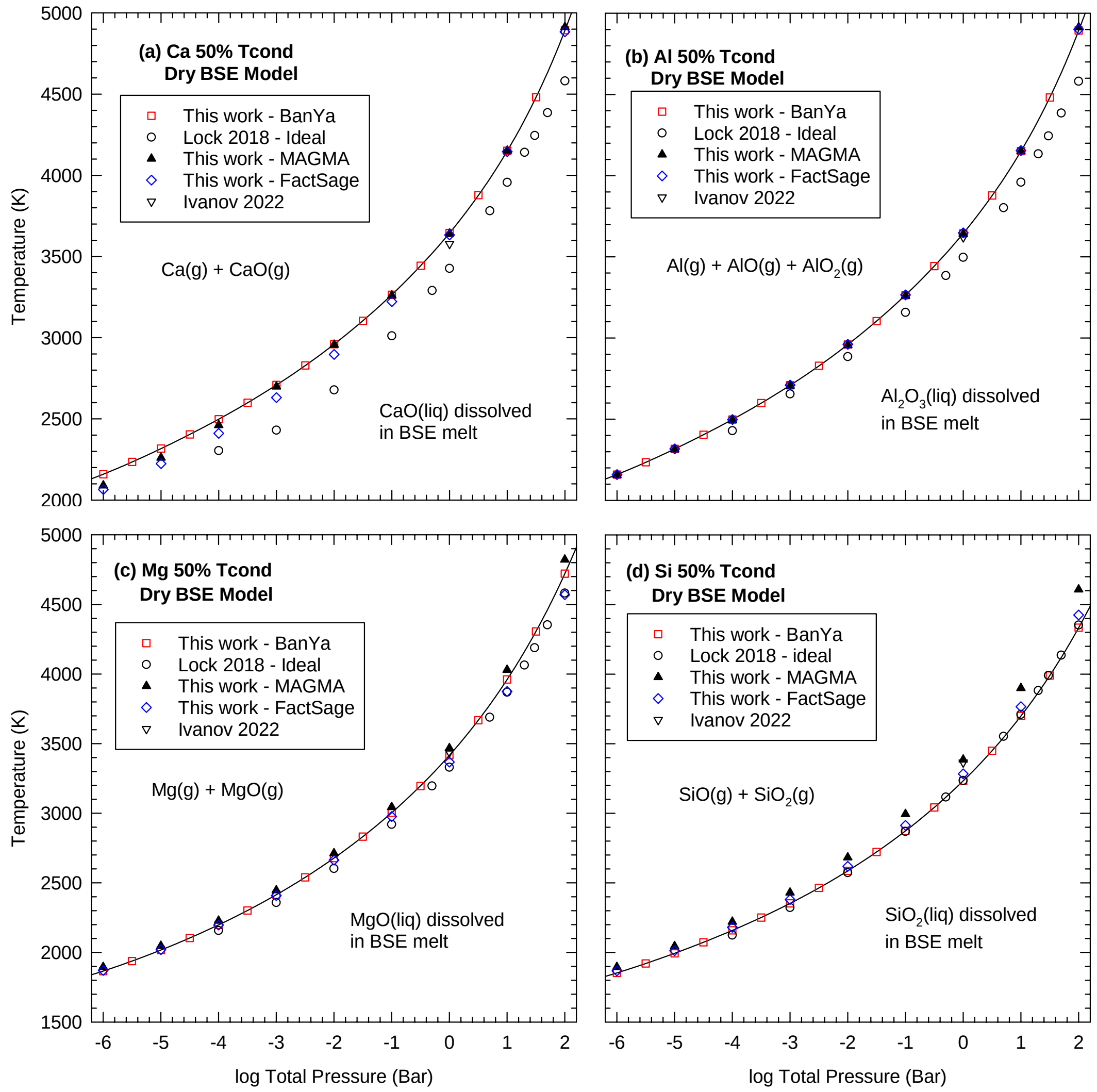
(a) Ca 50% Tcond
Dry BSE Model
This work - BanYa
Lock 2018 - Ideal
This work - MAGMA
This work - FactSage
Ivanov 2022
Ca(g) + CaO(g)
CaO(liq) dissolved in BSE melt
(b) Al 50% Tcond
Dry BSE Model
This work - BanYa
Lock 2018 - Ideal
This work - MAGMA
This work - FactSage
Ivanov 2022
Al(g) + AlO(g) + AlO2(g)
Al2O3(liq) dissolved in BSE melt
(c) Mg 50% Tcond
Dry BSE Model
This work - BanYa
Lock 2018 - Ideal
This work - MAGMA
This work - FactSage
Ivanov 2022
Mg(g) + MgO(g)
MgO(liq) dissolved in BSE melt
(d) Si 50% Tcond
Dry BSE Model
This work - BanYa
Lock 2018 - ideal
This work - MAGMA
This work - FactSage
Ivanov 2022
SiO(g) + SiO2(g)
SiO2(liq) dissolved in BSE melt
Temperature (K)
log Total Pressure (Bar)
5000
4500
4000
3500
3000
2500
2000
1500
-6
-5
-4
-3
-2
-1
0
1
2

Figure 4a-d

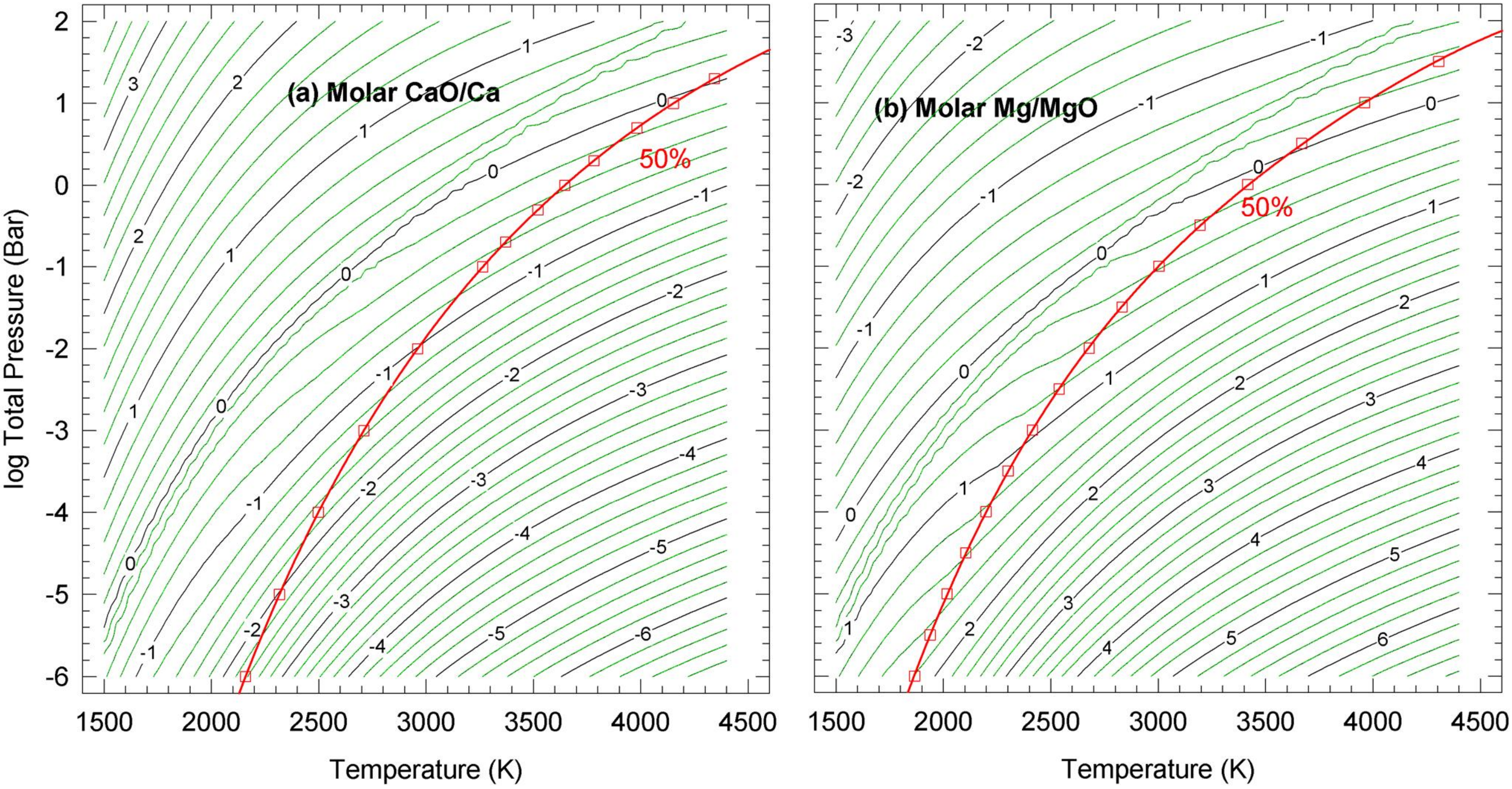

(a) Molar CaO/Ca
(b) Molar Mg/MgO
50%
log Total Pressure (Bar)
Temperature (K)

Figure 5a,b

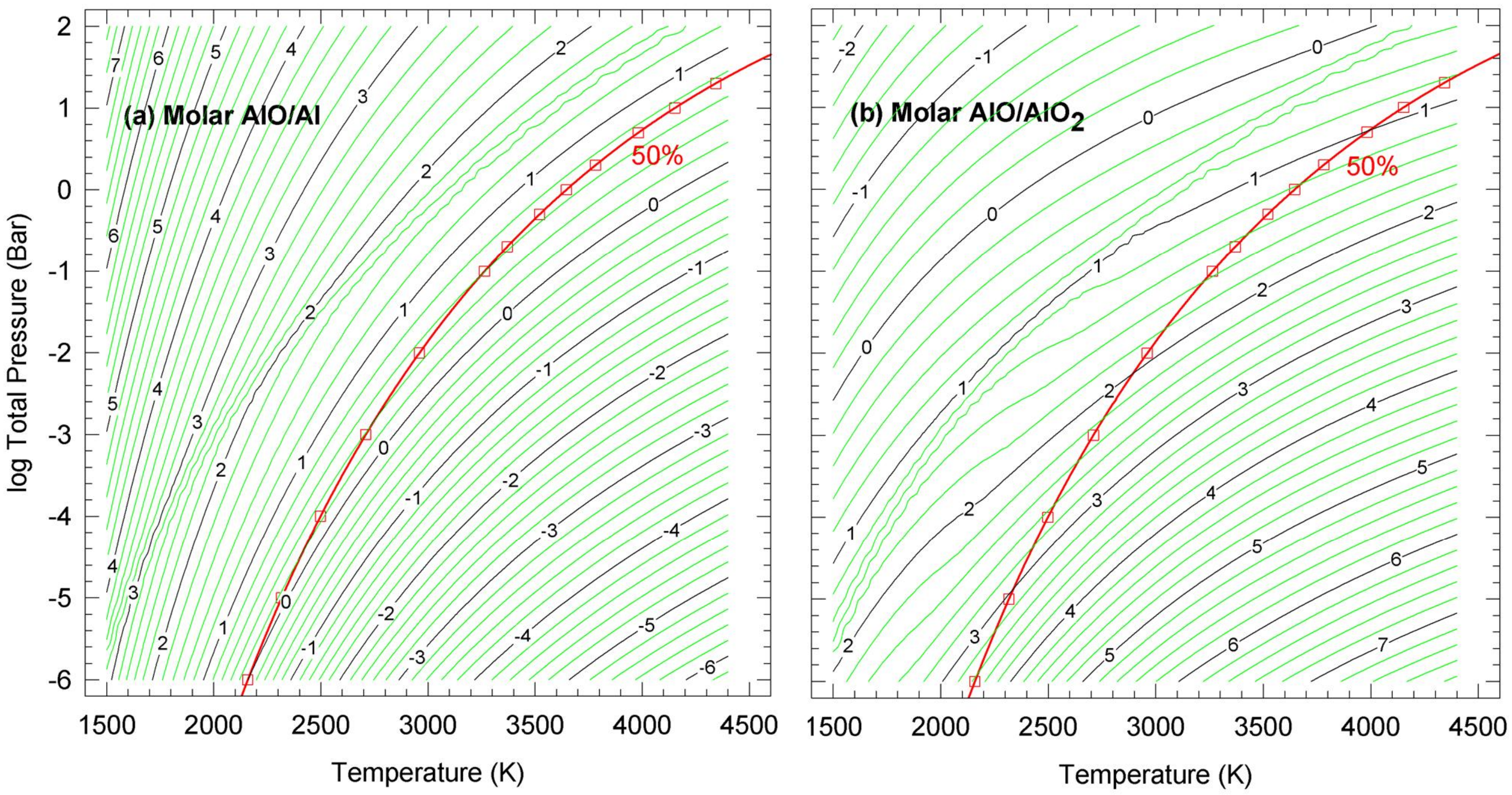

(a) Molar AlO/Al
(b) Molar AlO/AlO2
50%
log Total Pressure (Bar)
Temperature (K)

Figure 6a,b

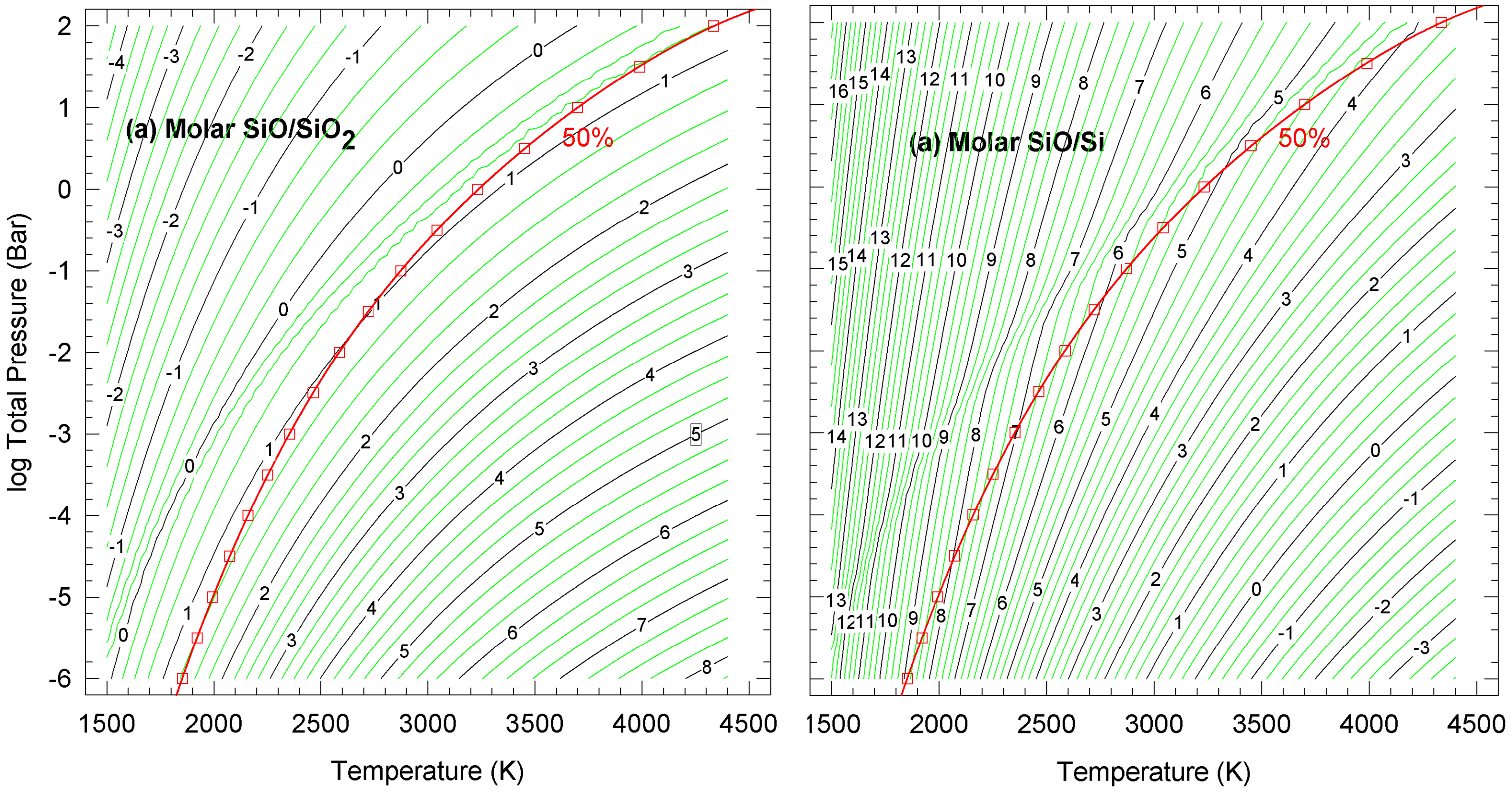

Figure 7a,b

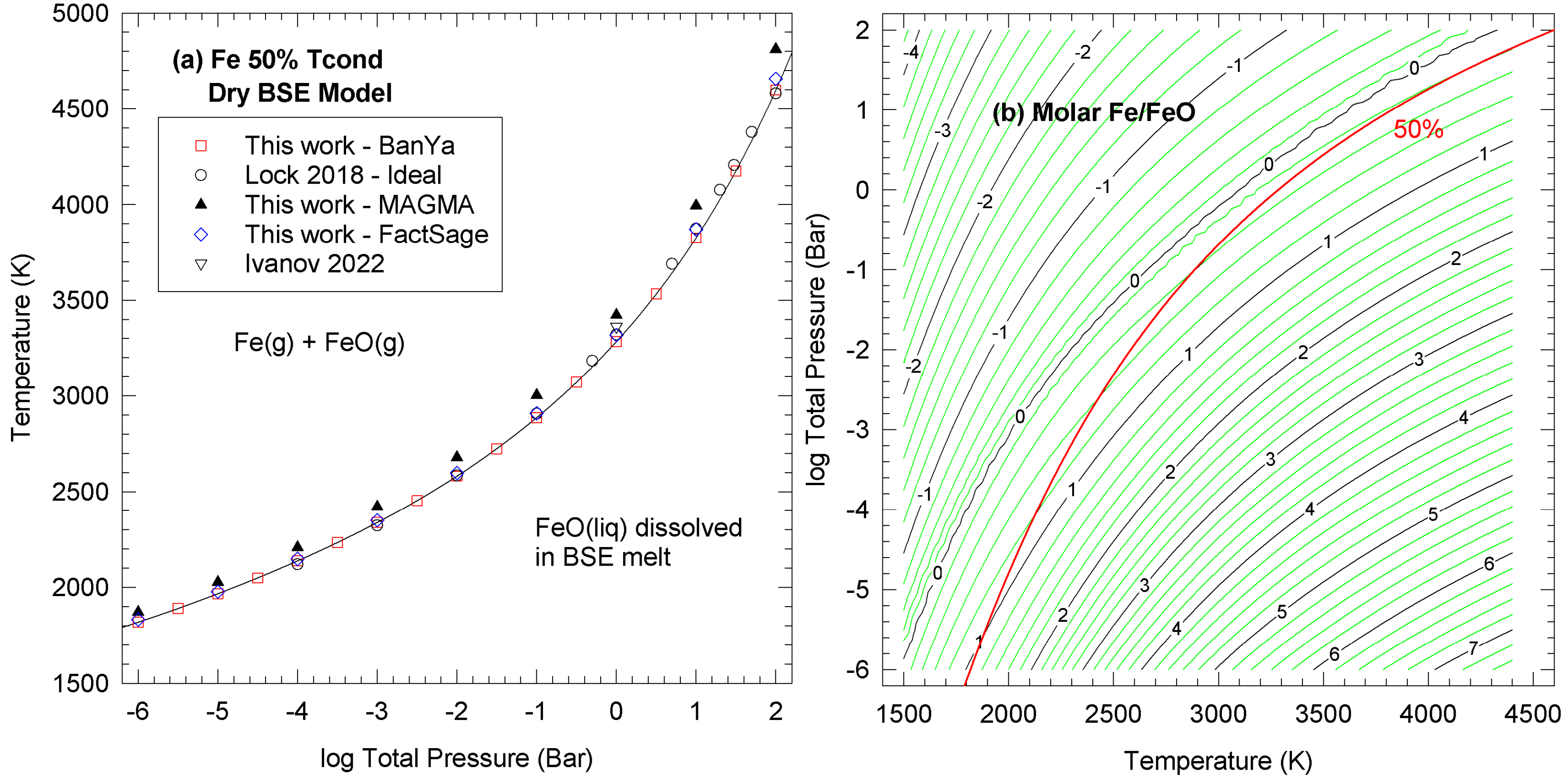

(a) Fe 50% Tcond
Dry BSE Model
This work - BanYa
Lock 2018 - Ideal
This work - MAGMA
This work - FactSage
Ivanov 2022
Fe(g) + FeO(g)
FeO(liq) dissolved in BSE melt
Temperature (K)
log Total Pressure (Bar)
(b) Molar Fe/FeO
50%
log Total Pressure (Bar)
Temperature (K)

Figure 8a,b

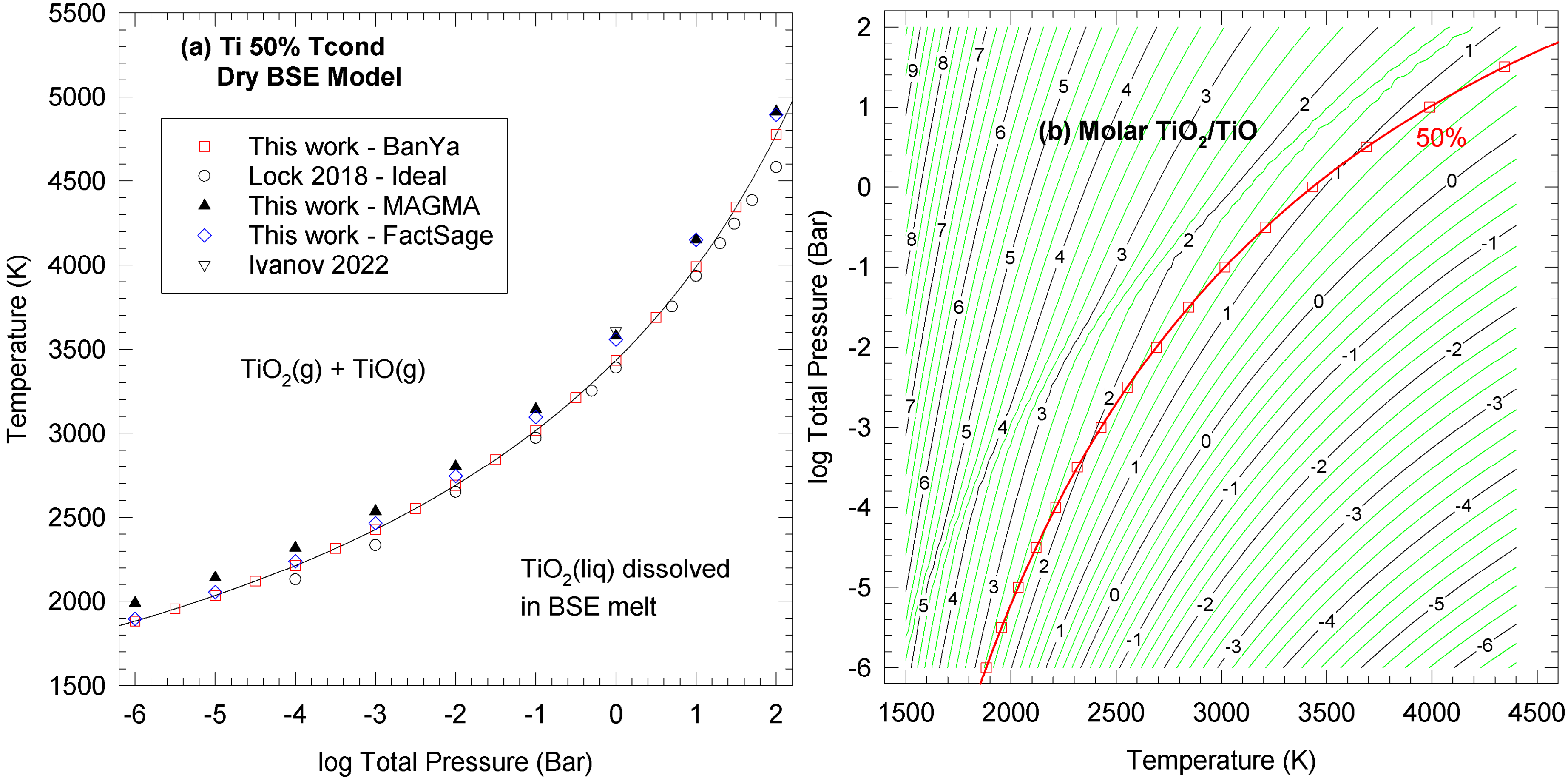

(a) Ti 50% Tcond
Dry BSE Model
This work - BanYa
Lock 2018 - Ideal
This work - MAGMA
This work - FactSage
Ivanov 2022
$TiO_2$(g) + TiO(g)
$TiO_2$(liq) dissolved
in BSE melt
Temperature (K)
log Total Pressure (Bar)
(b) Molar $TiO_2$/TiO
50%
log Total Pressure (Bar)
Temperature (K)

Figure 9a,b

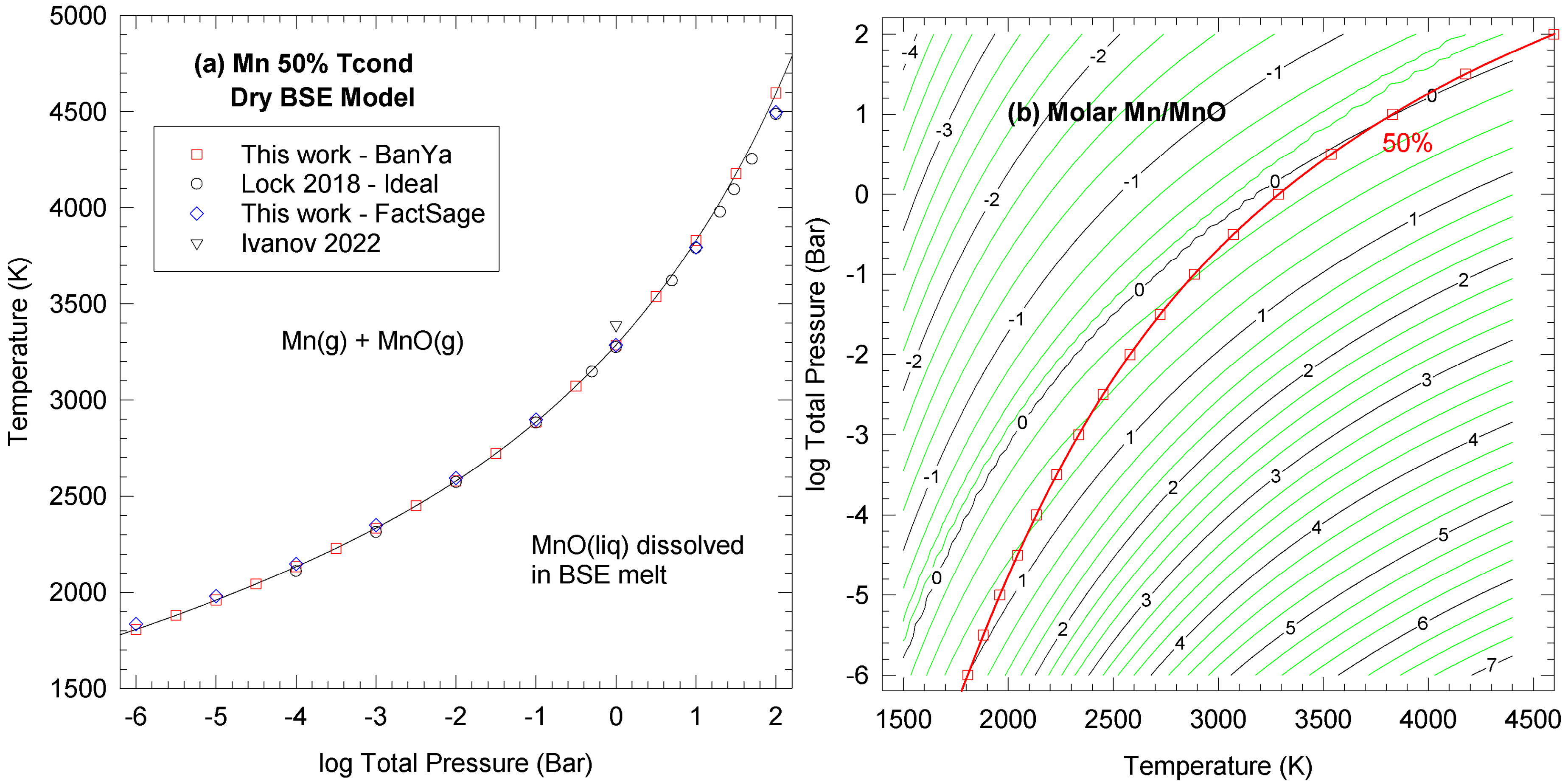

(a) Mn 50% Tcond
Dry BSE Model
This work - BanYa
Lock 2018 - Ideal
This work - FactSage
Ivanov 2022
Mn(g) + MnO(g)
MnO(liq) dissolved in BSE melt
Temperature (K)
log Total Pressure (Bar)
(b) Molar Mn/MnO
50%
log Total Pressure (Bar)
Temperature (K)

Figure 10a,b

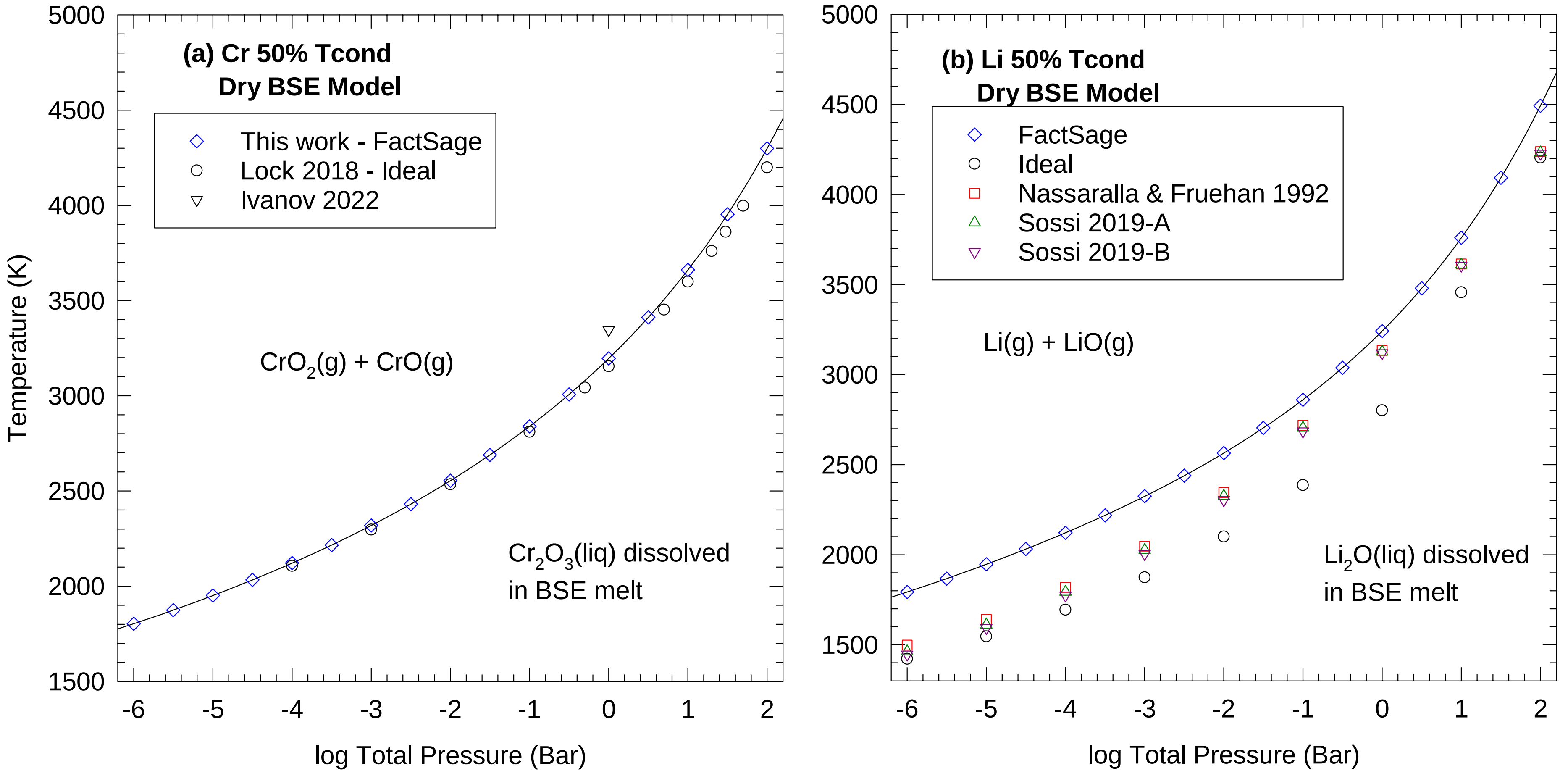

(a) Cr 50% Tcond
Dry BSE Model
This work - FactSage
Lock 2018 - Ideal
Ivanov 2022
CrO2(g) + CrO(g)
Cr2O3(liq) dissolved
in BSE melt
Temperature (K)
log Total Pressure (Bar)
(b) Li 50% Tcond
Dry BSE Model
FactSage
Ideal
Nassaralla & Fruehan 1992
Sossi 2019-A
Sossi 2019-B
Li(g) + LiO(g)
Li2O(liq) dissolved
in BSE melt
log Total Pressure (Bar)

Figure 11a,b

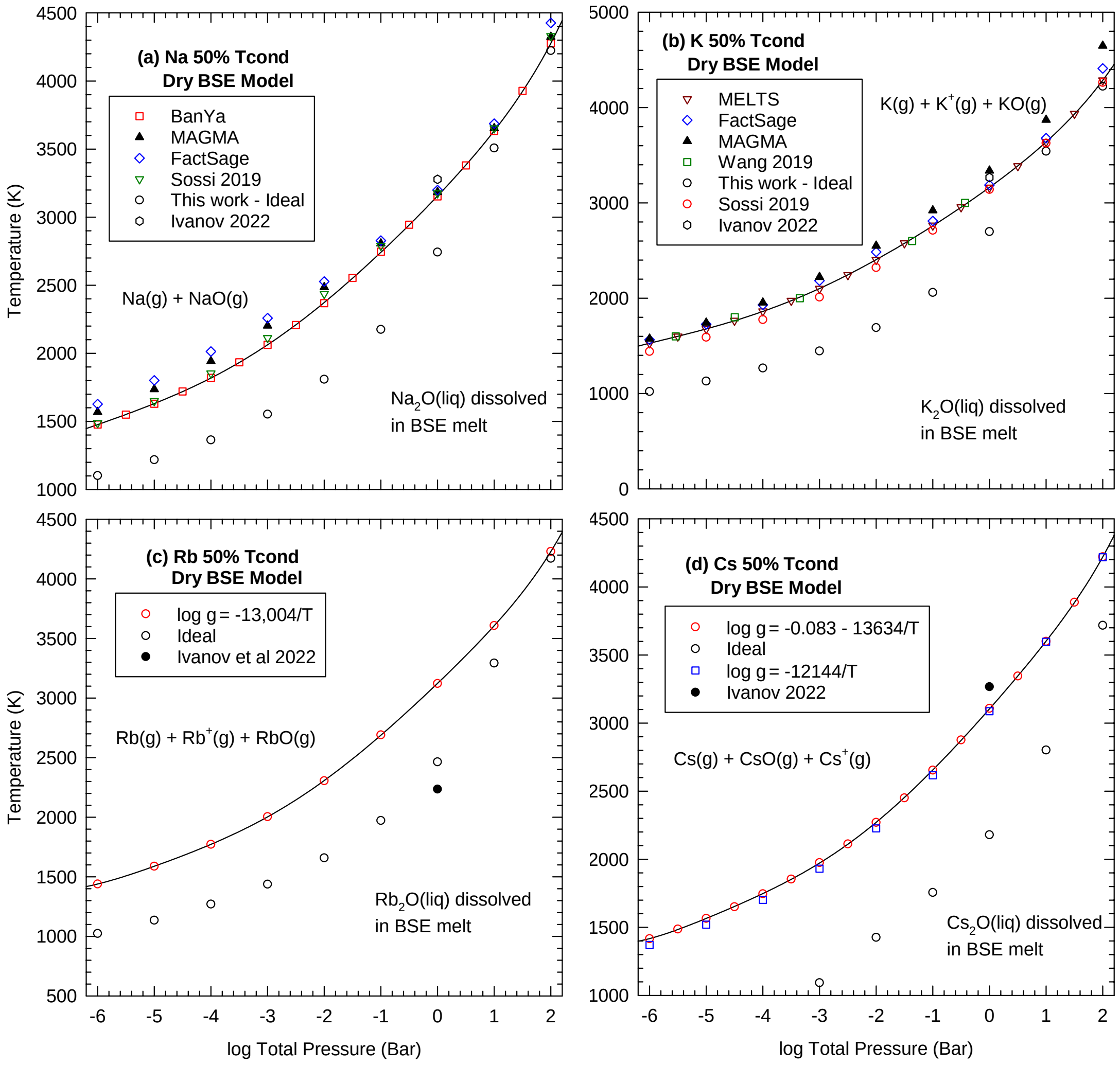
(a) Na 50% Tcond
Dry BSE Model
BanYa
MAGMA
FactSage
Sossi 2019
This work - Ideal
Ivanov 2022
Na(g) + NaO(g)
Na2O(liq) dissolved in BSE melt
(b) K 50% Tcond
Dry BSE Model
MELTS
FactSage
MAGMA
Wang 2019
This work - Ideal
Sossi 2019
Ivanov 2022
K(g) + K+(g) + KO(g)
K2O(liq) dissolved in BSE melt
(c) Rb 50% Tcond
Dry BSE Model
log g = -13,004/T
Ideal
Ivanov et al 2022
Rb(g) + Rb+(g) + RbO(g)
Rb2O(liq) dissolved in BSE melt
(d) Cs 50% Tcond
Dry BSE Model
log g = -0.083 - 13634/T
Ideal
log g = -12144/T
Ivanov 2022
Cs(g) + CsO(g) + Cs+(g)
Cs2O(liq) dissolved in BSE melt
Temperature (K)
log Total Pressure (Bar)

Figure 12a-d

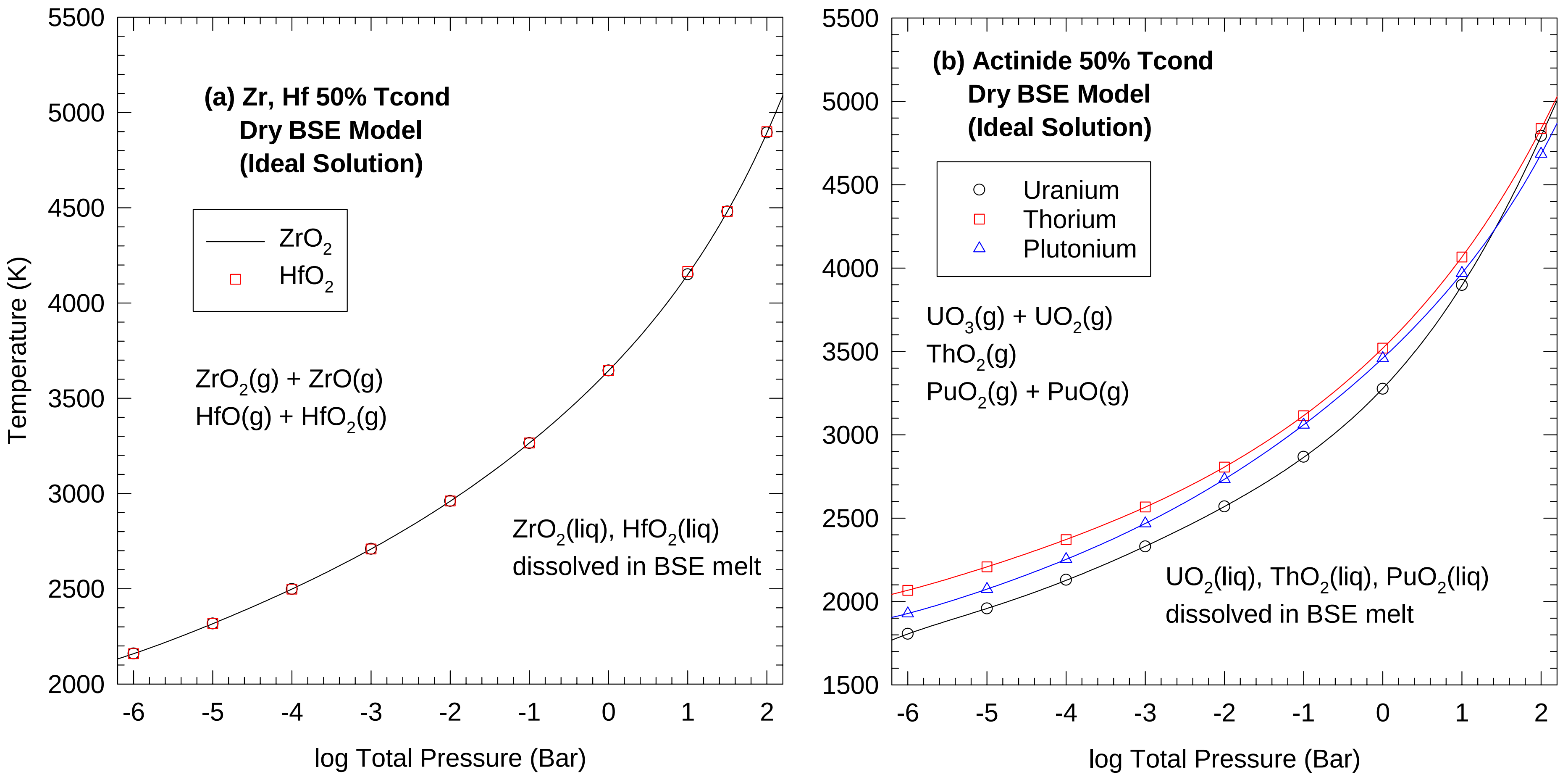

(a) Zr, Hf 50% Tcond
Dry BSE Model
(Ideal Solution)
ZrO2
HfO2
ZrO2(g) + ZrO(g)
HfO(g) + HfO2(g)
ZrO2(liq), HfO2(liq)
dissolved in BSE melt
Temperature (K)
log Total Pressure (Bar)
(b) Actinide 50% Tcond
Dry BSE Model
(Ideal Solution)
Uranium
Thorium
Plutonium
UO3(g) + UO2(g)
ThO2(g)
PuO2(g) + PuO(g)
UO2(liq), ThO2(liq), PuO2(liq)
dissolved in BSE melt
log Total Pressure (Bar)

Figure 13a,b

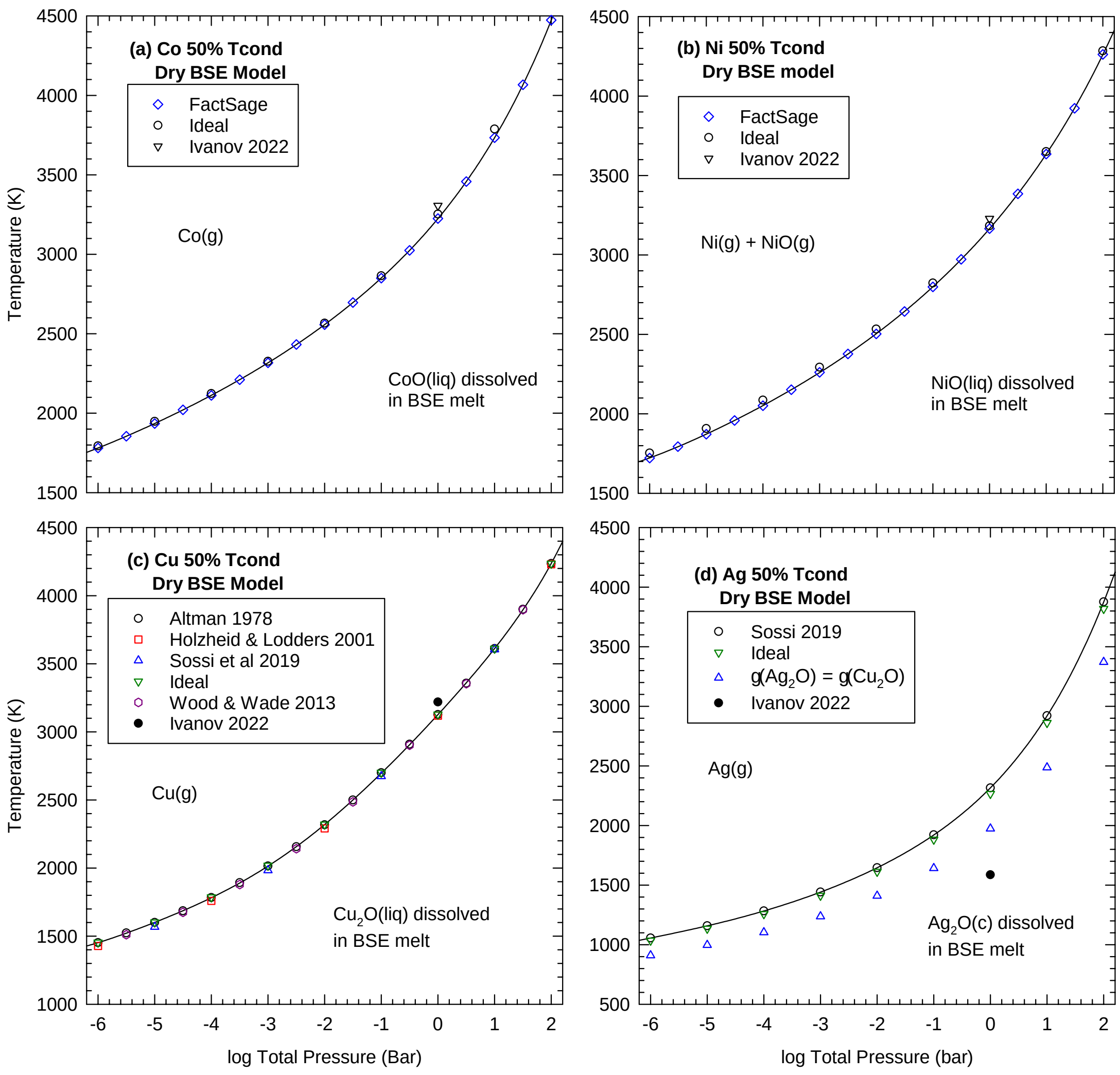

(a) Co 50% Tcond
Dry BSE Model
FactSage
Ideal
Ivanov 2022
Co(g)
CoO(liq) dissolved
in BSE melt
Temperature (K)
(b) Ni 50% Tcond
Dry BSE model
FactSage
Ideal
Ivanov 2022
Ni(g) + NiO(g)
NiO(liq) dissolved
in BSE melt
(c) Cu 50% Tcond
Dry BSE Model
Altman 1978
Holzheid & Lodders 2001
Sossi et al 2019
Ideal
Wood & Wade 2013
Ivanov 2022
Cu(g)
Cu2O(liq) dissolved
in BSE melt
log Total Pressure (Bar)
(d) Ag 50% Tcond
Dry BSE Model
Sossi 2019
Ideal
g(Ag2O) = g(Cu2O)
Ivanov 2022
Ag(g)
Ag2O(c) dissolved
in BSE melt
log Total Pressure (bar)

Figure 14a-d

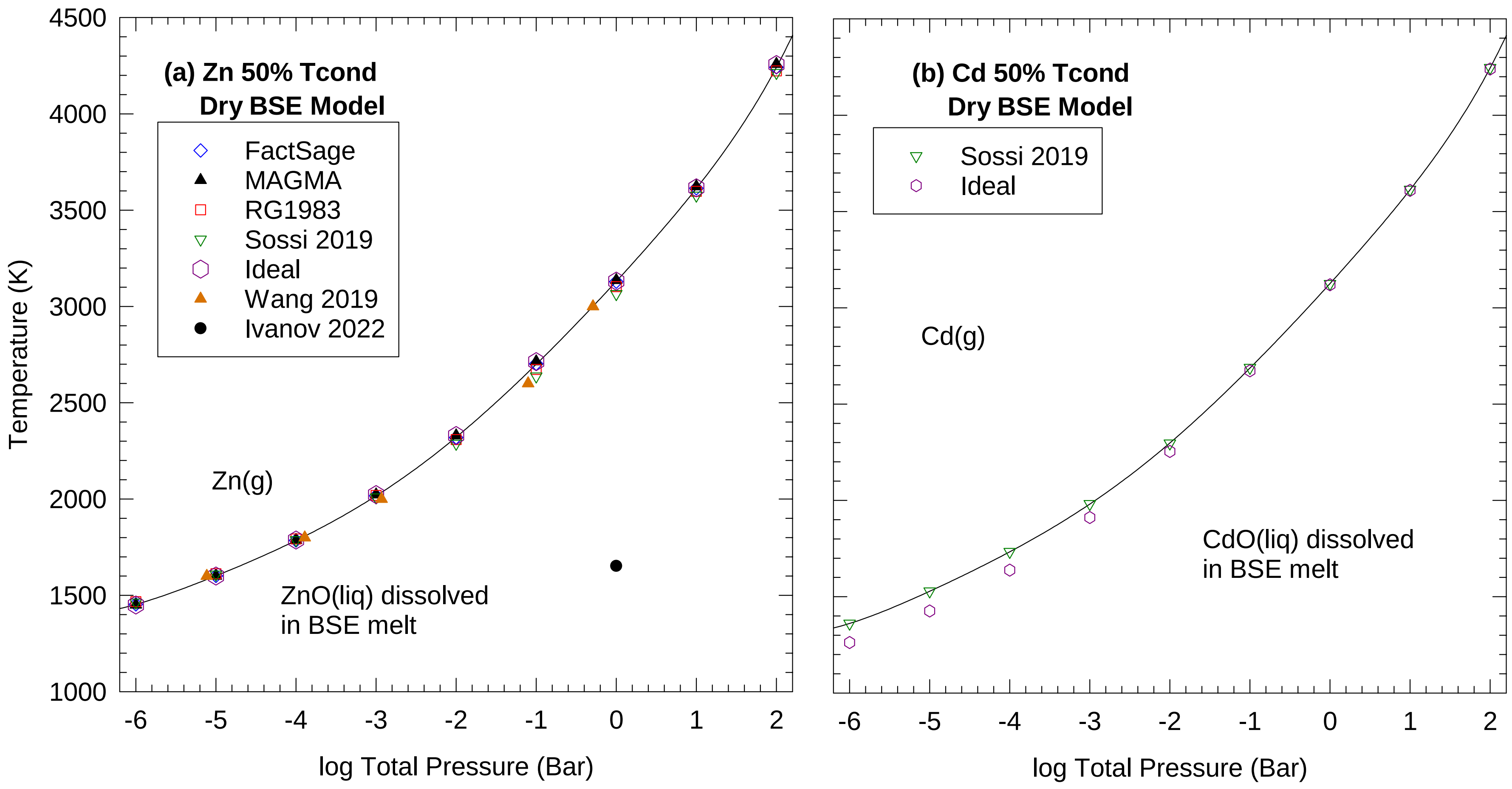

Figure 15a,b

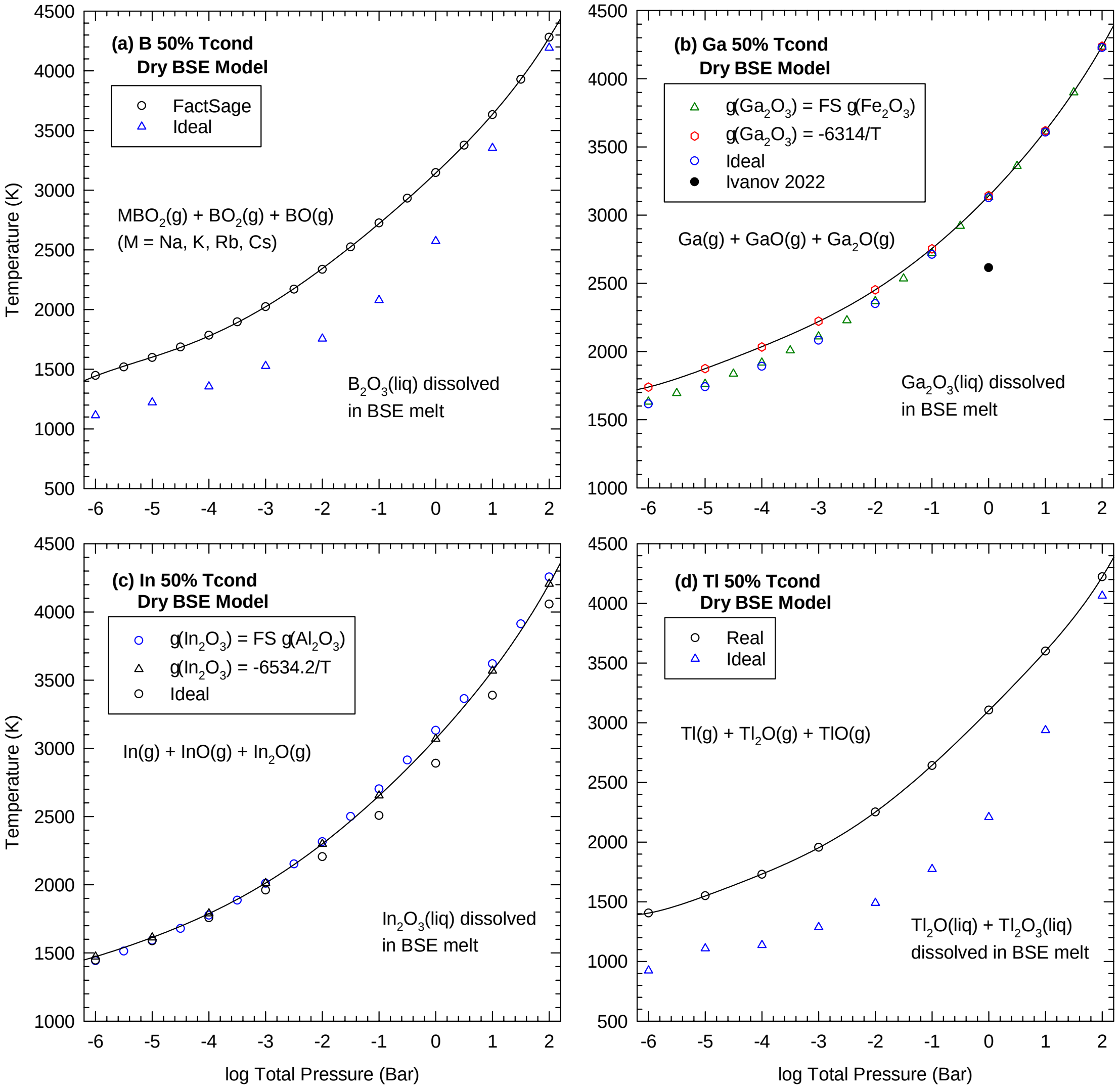

(a) B 50% Tcond
Dry BSE Model
FactSage
Ideal
MBO2(g) + BO2(g) + BO(g)
(M = Na, K, Rb, Cs)
B2O3(liq) dissolved
in BSE melt
(b) Ga 50% Tcond
Dry BSE Model
g(Ga2O3) = FS g(Fe2O3)
g(Ga2O3) = -6314/T
Ideal
Ivanov 2022
Ga(g) + GaO(g) + Ga2O(g)
Ga2O3(liq) dissolved
in BSE melt
(c) In 50% Tcond
Dry BSE Model
g(In2O3) = FS g(Al2O3)
g(In2O3) = -6534.2/T
Ideal
In(g) + InO(g) + In2O(g)
In2O3(liq) dissolved
in BSE melt
(d) Tl 50% Tcond
Dry BSE Model
Real
Ideal
Tl(g) + Tl2O(g) + TlO(g)
Tl2O(liq) + Tl2O3(liq)
dissolved in BSE melt
Temperature (K)
log Total Pressure (Bar)

Figure 16a-d

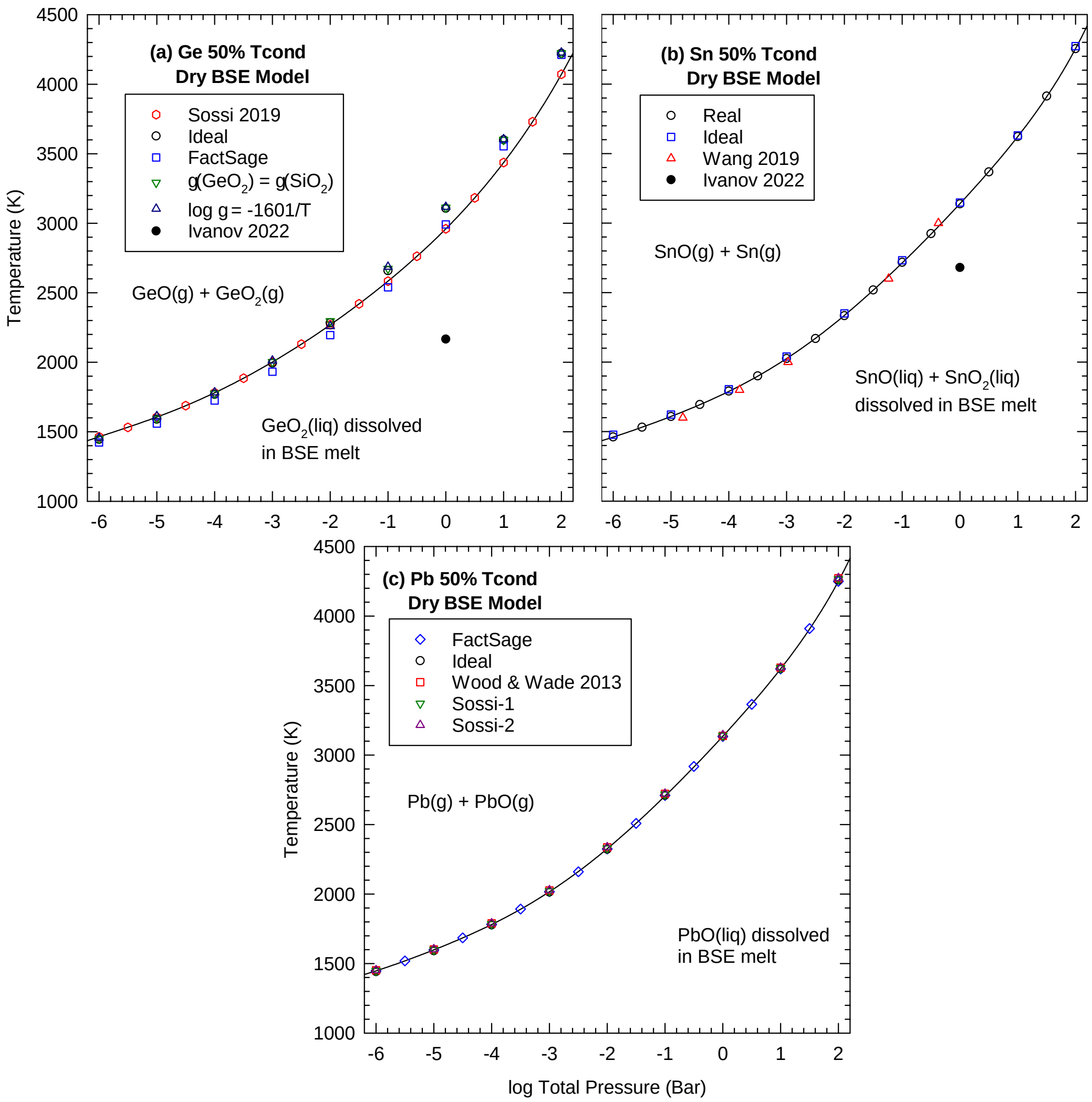

(a) Ge 50% Tcond
Dry BSE Model
Sossi 2019
Ideal
FactSage
g(GeO2) = g(SiO2)
log g = -1601/T
Ivanov 2022
GeO(g) + GeO2(g)
GeO2(liq) dissolved
in BSE melt
Temperature (K)
(b) Sn 50% Tcond
Dry BSE Model
Real
Ideal
Wang 2019
Ivanov 2022
SnO(g) + Sn(g)
SnO(liq) + SnO2(liq)
dissolved in BSE melt
(c) Pb 50% Tcond
Dry BSE Model
FactSage
Ideal
Wood & Wade 2013
Sossi-1
Sossi-2
Pb(g) + PbO(g)
PbO(liq) dissolved
in BSE melt
log Total Pressure (Bar)

Figure 17a-c

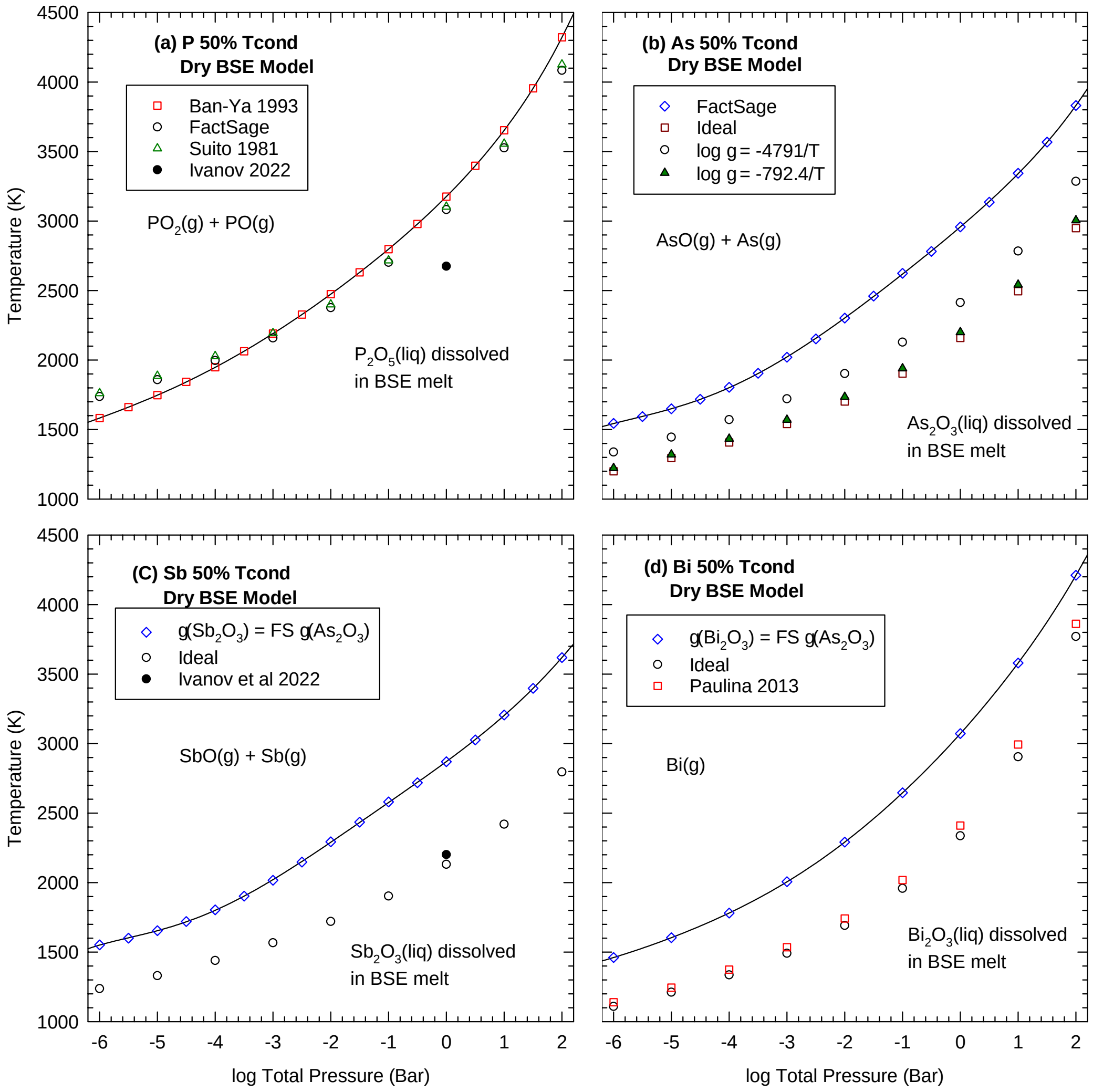

(a) P 50% Tcond
Dry BSE Model
Ban-Ya 1993
FactSage
Suito 1981
Ivanov 2022
PO2(g) + PO(g)
P2O5(liq) dissolved in BSE melt
(b) As 50% Tcond
Dry BSE Model
FactSage
Ideal
log g = -4791/T
log g = -792.4/T
AsO(g) + As(g)
As2O3(liq) dissolved in BSE melt
(C) Sb 50% Tcond
Dry BSE Model
g(Sb2O3) = FS g(As2O3)
Ideal
Ivanov et al 2022
SbO(g) + Sb(g)
Sb2O3(liq) dissolved in BSE melt
(d) Bi 50% Tcond
Dry BSE Model
g(Bi2O3) = FS g(As2O3)
Ideal
Paulina 2013
Bi(g)
Bi2O3(liq) dissolved in BSE melt
Temperature (K)
log Total Pressure (Bar)

Figure 18a-d

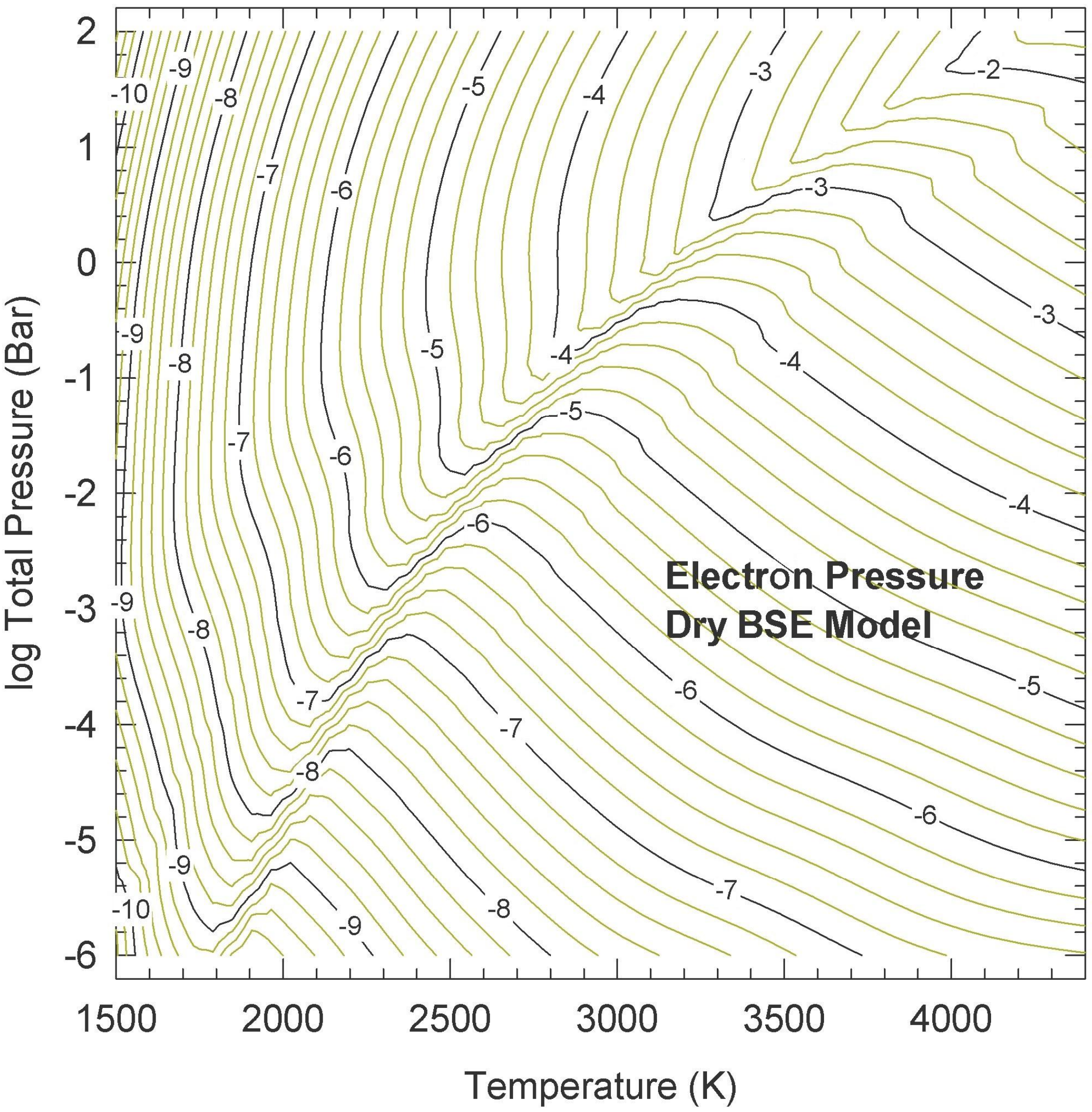

Figure 19